\documentclass[aip,amsmath,amssymb,reprint]{revtex4-1}

\usepackage[dvipsnames]{xcolor}
\usepackage{bm}%
\usepackage{amsmath}
\usepackage{multirow}
\usepackage{soul}
\usepackage[T1]{fontenc}
\usepackage[utf8]{inputenc}
\usepackage{array}
\usepackage{graphicx}
\usepackage{etoolbox}
\usepackage{overpic}
\usepackage{float}
\usepackage{placeins}

\newcommand\Pra{\mbox{\textit{Pr}}} %
\newcommand\Ray{\mbox{\textit{Ra}}} %
\newcommand\Nu{\mbox{\textit{Nu}}} %
\newcommand\Ha{\mbox{\textit{Ha}}} %

\newcommand{\titleput}[3]{
    \put(#1,#2){#3}
}
\newcommand{\axisput}[3]{
    \put(#1,#2){{\footnotesize#3}}
}
\newcommand{\xdataline}{$y = 0$, $z = 0$}
\newcommand{\ydataline}{$x = -0.47$, $z = 0$}
\newcommand{\zdataline}{$x = -0.47$, $y = 0$}
\newcolumntype{L}[1]{>{\raggedright\let\newline\\\arraybackslash\hspace{0pt}}m{#1}}
\newcolumntype{C}[1]{>{\centering\let\newline\\\arraybackslash\hspace{0pt}}m{#1}}
\newcolumntype{R}[1]{>{\raggedleft\let\newline\\\arraybackslash\hspace{0pt}}m{#1}}

\newdimen{\imghgt}
\newdimen{\txthgt}

\makeatletter
\def\@email#1#2{%
 \endgroup
 \patchcmd{\titleblock@produce}
  {\frontmatter@RRAPformat}
  {\frontmatter@RRAPformat{\produce@RRAP{*#1\href{mailto:#2}{#2}}}\frontmatter@RRAPformat}
  {}{}
}%
\makeatother

\begin{document}

\title[Magnetoconvection with walls of finite electrical conductivity]{Magnetoconvection in a long vertical enclosure with walls of finite electrical conductivity} %

\author{Ali Akhtari}
\author{Oleg Zikanov}
\email{zikanov@umich.edu.}
\affiliation{Department of Mechanical Engineering, University of Michigan - Dearborn, 4901 Evergreen Road, Dearborn, 48128, MI, USA}
\author{Dmitry Krasnov}
\affiliation{Institut f\"ur Thermo und Fluiddynamik, Technische Universit\"at Ilmenau, D-98684 Ilmenau, Germany}

\date{\today}

\begin{abstract}
Magnetoconvection in a tall vertical box with vertical hot and cold walls, and an imposed steady uniform magnetic field perpendicular to the temperature gradient, is analyzed numerically. The geometry and the values of the non-dimensional parameters – the Prandtl number of $0.025$, the Rayleigh number of $7.5 \times 10^5$, and the Hartmann number between 0 and 798 – match those of an earlier experiment. A parametric study of the effect of wall electric conductivity, across a wide range of conductance ratio values, on flow properties is performed. Two configurations of electric boundary conditions are explored. In one configuration, all walls have finite electric conductivity, while in the other, only the walls with constant temperature are electrically conducting. The flows are analyzed using their integral properties and distributions of velocity, temperature, and electric currents. It is found that, in general, the convection flow is suppressed by the magnetic field. However, this effect is strongly modified by the wall's electric conductivity and is markedly different for the two wall configurations. The associated changes in flow structure, rate of heat transfer, and flow's kinetic energy are revealed. It is also shown that the assumption of quasi-two-dimensionality may not be valid under some conditions, even at high Hartmann numbers.  
\end{abstract}

\keywords{Convection; Magnetohydrodynamics; Magnetoconvection}

\maketitle %

\section{Introduction}
\label{sec:intro}
Magnetoconvection is the convection motion of an electrically conducting fluid, e.g., a liquid metal, in the presence of a magnetic field. It is a complex process, in which the flow is influenced, at the same time, by buoyancy and electromagnetic forces. Magnetoconvection plays an important role in many systems in nature (e.g., in generation of planetary dynamos) and technology. The technological examples include manufacturing of high-quality steels\cite{Cukierski:2008} and large semiconductor crystals,\cite{Ozoe:2005} operation of liquid metal batteries,\cite{Shen:2016,kelley2018fluid} and, most relevantly for this study, design of liquid-metal breeding blankets for future nuclear fusion reactors.\cite{Abdou:2015,mistrangelo:2021mhd} A distinctive feature of the latter system is that the convection and magnetic field effects are both very strong.

The blanket system surrounding the plasma chamber is a key part of a fusion reactor. It shields the rest of the reactor from radiation, heat, and high-energy neutrons created by the fusion reaction. It serves as a heat exchanger diverting the energy into an external electric power-generation circuit. It is also used to breed tritium via reaction between the neutrons and lithium contained with the blanket.\cite{coen1985lithium} Different blanket concepts are proposed, such as the helium-cooled lead-lithium (HCLL), self-cooled lead-lithium (SCLL), dual coolant lead-lithium (DCLL) and others.\cite{Abdou:2015,mistrangelo:2021mhd}

The results reported in this paper are particularly relevant to separately cooled blankets, such as HCLL. The use of He circuits for heat transfer limits the role of a liquid metal, most likely Li-Pb eutectic, to shielding and breeding of tritium. The metal can be pumped through the blanket modules at low ($\sim$0.1-1 mm/s) speed, thus alleviating the challenging problem of magnetohydrodynamic (MHD) pressure drop and rendering the blanket design more feasible. At such a low speed, the flow of liquid metal within a module is dominated by magnetoconvection caused by strong non-uniform heating in the presence of a very strong ($\sim$10 T) imposed steady magnetic field.

In the presence of the magnetic field, a flow of a liquid metal induces electric currents, which interact with the  magnetic field to generate a Lorentz body force acting on fluid particles. If the Lorentz force is strong, the resulting flow is markedly different from its non-MHD counterpart. The notable effects are the suppression of turbulent fluctuations, development of thin MHD boundary layers at solid walls, and reduction of gradients of flow variables along the magnetic field lines, which results in an anisotropic state. However, such impacts may differ based on how current paths form in the flow field (see, e.g., Ref.~\onlinecite{ZikanovASME:2021} for a review). A major factor affecting the paths of the induced electric currents and, thus, the specific realization of the aforementioned effects and the specific structure of a magnetoconvection flow is the electric conductivity of the walls.\cite{Mueller:2001} This still poorly understood effect of this factor on magnetoconvection flows, in which the Lorentz and buoyancy forces are both strong, is at the focus of this paper.

As discussed in detail in Section \ref{sec:problem}, the effect of wall conductivity on the flow is determined not just by the value of the wall material's conductivity but by the ratio between the total conductances of the wall and the fluid. Usually, high electrical conductivity of liquid metals leads to low values of this ratio, even when the walls are made of well-conducting materials. This consideration has directed many previous studies to the idealized model in which walls are assumed to be perfectly electrically insulating. Several theoretical\cite{buhler1998laminar,Tagawa:2002} and numerical\cite{Authie:2003,Di_P1:2002,burr2002rayleigh,mistrangelo2011numerical,Mas:2012,mistrangelo2014buoyant,liu2019effects,Rhodes:2020} studies of magnetoconvection, however, were completed without this assumption and predicted profound, and sometimes counterintuitive, changes in flow structure even at a small conductance ratio.

\begin{figure}
\centering
\begin{overpic}[trim={1mm 2mm 2mm 0},width=0.48\columnwidth,clip=]{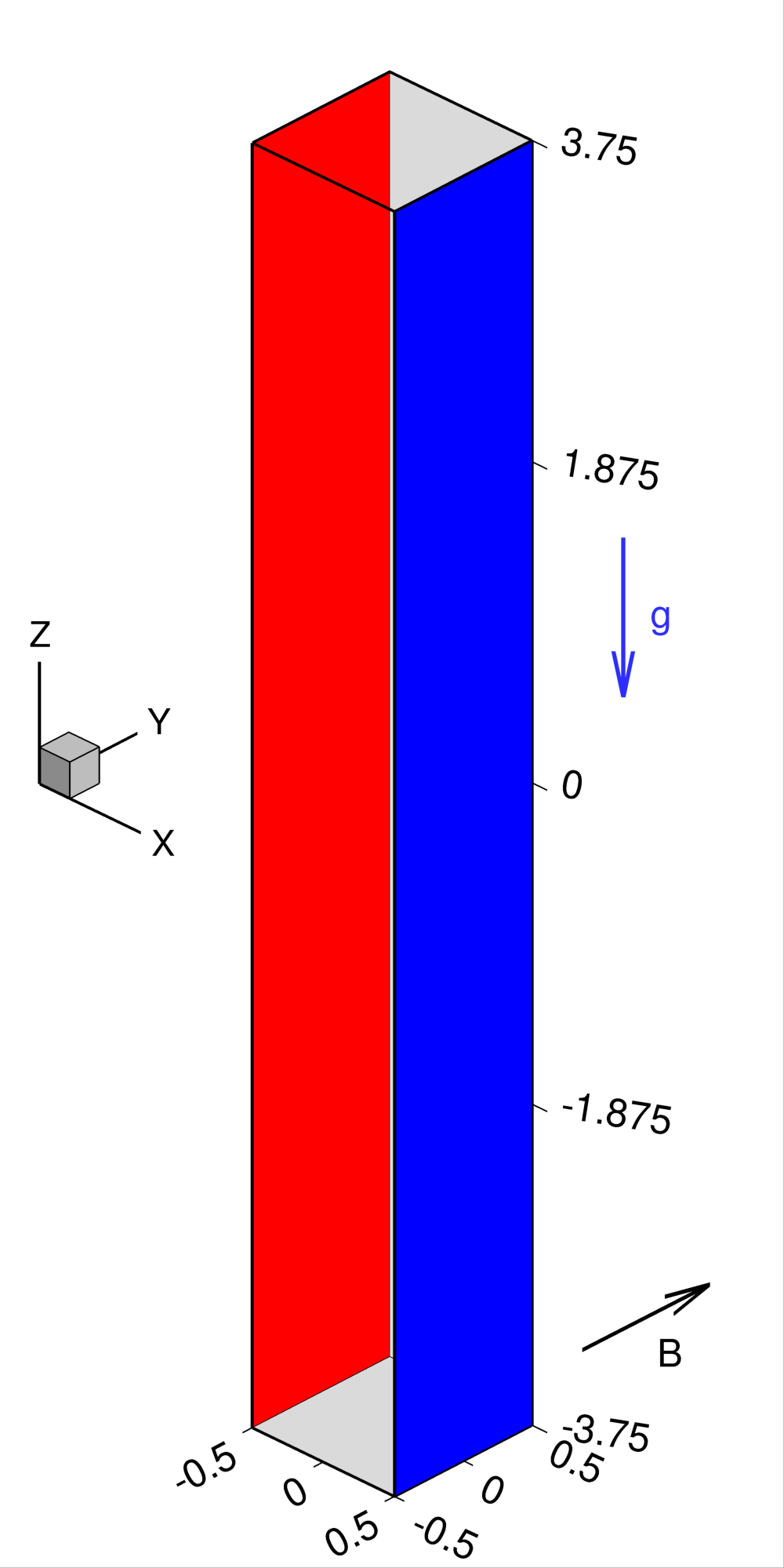}
    \titleput{15}{95}{(a)}
\end{overpic}
\begin{overpic}[trim={1mm 2mm 2mm 0},width=0.24\columnwidth,clip=]{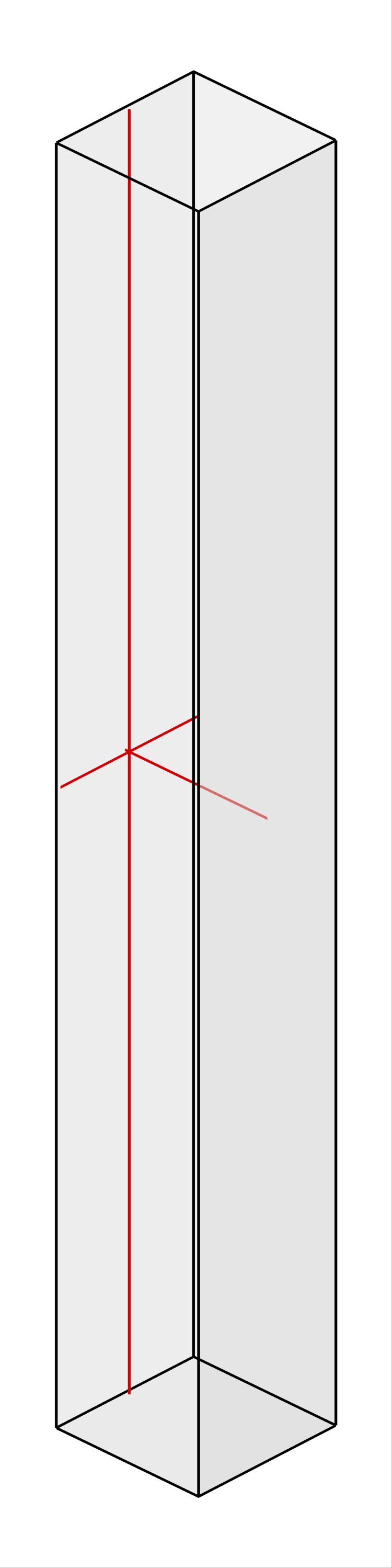}
    \titleput{3}{95}{(b)}
\end{overpic}
\begin{overpic}[trim={1mm 2mm 2mm 0},width=0.24\columnwidth,clip=]{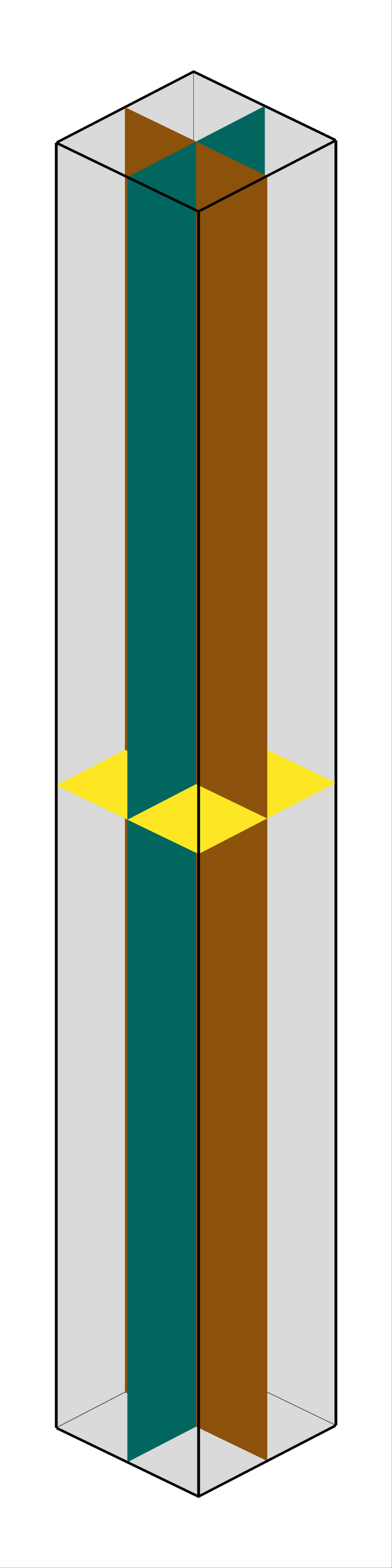}
    \titleput{3}{95}{(c)}
\end{overpic}
\caption{Schematic of the flow. \emph{(a)}, Geometry - a cuboid box with two side walls maintained at constant temperatures: red (hot wall) and blue (cold wall). The other four walls are perfectly thermally insulating. The directions of the applied magnetic field $\bm{B}$ and the gravity acceleration $\bm{g}$ are indicated. \emph{(b)}, Lines, along which velocity, temperature, and induced current profiles are recorded in simulations. Each line passes through the point $(x,y,z) = (-0.47, 0, 0)$ and is parallel to the $x$, $y$ or $z$ direction. (c) Cross-section planes, in which velocity, temperature, and induced current distributions are recorded in simulations. Each surface contains the point $(x,y,z) = (0, 0, 0)$ and has a normal parallel to the  $x$, $y$ or $z$ direction.}
\label{fig:doamin}
\end{figure}

The main goal of this paper is to investigate the effect of wall conductance ratio on the properties of magnetoconvection flow within a specific geometry - a tall cuboid enclosure subjected to heating and cooling at opposite vertical sides in the presence of an external transverse magnetic field parallel to the heated/cooled walls (see Fig.~\ref{fig:doamin}). This is a simplified configuration of a module of the HCLL blanket.\cite{Abdou:2015,Mistrangelo:2014} It is also an example of a system, in which the three major direction -- those of gravity, temperature gradient and magnetic field -- are perpendicular to each other. The configuration was studied before using linearized two-dimensional model,\cite{buhler1998laminar,Tagawa:2002} experiment,\cite{Authie:2002,Authie:2003} and numerical simulations.\cite{Authie:2002,Authie:2003}  Unique features of the flow, such as thin quasi-two-dimensional jets near the hot and cold walls, were found. The configuration was proposed as a benchmark for numerical models of magnetoconvection flows.\cite{Smolentsev:2015verification}

This paper presents results of a parametric study of the configuration based on high-resolution numerical simulations. The imposed magnetic field and wall conductance ratio are varied in broad ranges. To the best knowledge of the authors, this is the first extensive parametric study of such kind.

In what follows, a detailed description of the problem and its parameters, physical model and applied boundary conditions is given in section \ref{sec:problem}. The numerical method is discussed in section \ref{sec:method}. Results are presented in section \ref{sec:results} in two parts. Section \ref{sec:results:allWalls} reports the results obtained for the case in which all walls have equal electric conductance ratios, while section \ref{sec:results:HCWalls} discusses the results for the case, in which only the hot and cold walls are electrically conducting, while other four walls are perfectly insulating. Section \ref{sec:conclusions} provides the concluding remarks.

\section{Problem definition}
\label{sec:problem}

Natural convection flow of a viscous, electrically conducting, single-phase fluid with constant physical properties contained in cuboid box of aspect ratio 7.5 is considered (see Fig.~\ref{fig:doamin}). A transverse constant uniform  magnetic field of strength of $B$ is imposed in the flow domain. The geometry and the non-dimensional flow parameters listed in table \ref{tbl:conv-params} are selected to replicate the experiment.\cite{Authie:2003}.

\begin{table}
    \centering
    \begin{tabular}{l|c|c|c|c|c}
        Property  & $L_x \times L_y \times L_z$ & $\Ray$ & $\Pra$ & $\Ha$ & $C_w$\\
        \hline
        Value & $1 \times 1 \times 7.5$ & {$7.5\times 10^5$} & {0.025} & $0 - 796$ & $0 - 50$\\
    \end{tabular}
    \caption{Non-dimensional parameters of the studied flow, (see text for a detailed discussion and \eqref{eqn:non_dim_para} and \eqref{eqn:Cw} for the definition of $\Ray$, $\Pra$, $\Ha$ and $C_w$).} 
    \label{tbl:conv-params}
\end{table}

Liquid metal flows in typical technical settings satisfy the conditions of small magnetic Reynolds and magnetic Prandtl numbers:
\begin{align}
    Re_m &= \sigma \mu_0 UL \ll 1,\\
    Pr_m &= Re_m/Re = \sigma \mu_0 \nu \ll 1,
    \label{eqn:mag_non_dim_para}
\end{align}
where $\mu_0$, $\sigma$ and $\nu$ are, respectively, magnetic permeability of vacuum, and electrical conductivity and kinematic viscosity of fluid. The conditions are assumed true in our study, which allows us to apply the quasi-static (induction-less) approximation in the description of the electromagnetic effects (see, e.g., Ref.~\onlinecite{Davidson:2016}). Magnetic field fluctuations  caused by the induced currents are neglected in the expressions for the Lorentz force and Ohm's law, which include only the imposed field $\bm{B}$. The full two-way MHD coupling between the flow and the magnetic field is reduced to the one-way influence of the magnetic field on the flow. The induced currents and Lorentz force can be expressed in terms of an electric potential, as shown below.

The other approximation used in this study is the description of the thermal convection effect using the Boussinesq model.

The non-dimensional equations are:
\begin{align}
    \bm{\nabla} \cdot \bm{u} &= 0,\label{eqn:gov:cnt}\\
    \frac{\partial \bm{u}}{\partial t} & +  \left(\bm{u} \cdot \bm{\nabla} \right) \bm{u} \nonumber\\
    &= - \bm{\nabla} p + \sqrt{\frac{Pr}{Ra}} \left( \nabla^2 \bm{u} + Ha^2 \bm{j} \times \bm{e}_B \right) - T \bm{e}_g, \label{eqn:gov:NS}\\
    \frac{\partial T}{\partial t} &+ \bm{u} \cdot \bm{\nabla} T = \sqrt{\frac{1}{Ra Pr}} \nabla^2 T, \label{eqn:gov:heat}\\
    \bm{j} &= - \bm{\nabla} \phi + \bm{u} \times \bm{e}_B, \label{eqn:gov:current}\\
    \nabla^2 \phi &= \bm{\nabla} \cdot \left( \bm{u} \times \bm{e}_B \right) \label{eqn:gov:elpot}
\end{align}
In these equations, $\bm{u}$, $p$,  $T$, $\bm{j}$, and $\phi$  are the dimensionless fields of, respectively, velocity, pressure, temperature, electric current density and electric potential. $\bm{e}_g$ and $\bm{e}_B$ are the unit vectors in the direction of gravity and magnetic field, respectively. The equations are non-dimensionalized using the box width $L_x = L_y = L$, temperature difference between the hot and cold walls $\Delta T = T_{hot} - T_{cold}$, strength of the imposed magnetic field $B_0$ and the free-fall velocity $U \equiv \sqrt{g \beta \Delta T L}$ (here $\beta$ and $g$ are the thermal expansion coefficient and the acceleration of gravity respectively). Electric current density and electric potential are scaled using the combinations $\sigma U B_0$ and $U B_0 L$.

Non-dimensional parameters are the Prandtl, Rayleigh and Hartmann numbers:
\begin{equation}
    \Pra\equiv \frac{\nu}{\alpha}, \hskip3mm
    \Ray\equiv \frac{g \beta \Delta T L^3}{\nu\alpha}, \hskip3mm
    \Ha \equiv B_0 L \sqrt{\frac{\sigma}{\rho \nu}},
    \label{eqn:non_dim_para}
\end{equation}
where $\alpha$ and $\rho$ are, respectively, thermal diffusivity and mass density.

The no-slip boundary conditions $\bm{u}=0$ are applied at all walls. Temperatures at the vertical walls at $x=\pm 0.5$ are constant: $T_{x=0.5}=T_{cold} = -0.5$ and $T_{x=-0.5}=T_{hot} = 0.5$. The other walls are perfectly thermally insulating, so the boundary condition is $\partial T / \partial n = 0$.

The last set of boundary conditions is for the electric potential. We analyze the effect of finite wall electric conductivity on magnetoconvection flows when the walls are thin, i.e., when the wall thickness $\tau_w$ and the typical length scale of the flow $L$ satisfy $\tau_w \ll L$. As a result of this assumption, the distribution of the electric potential across the wall can be ignored. We can use the approximate boundary condition for electric potential at the boundary between fluid and walls instead of solving for a distribution of electric potential in both the fluid domain and the walls. This so-called thin-wall boundary condition initially proposed in Ref.~\onlinecite{Walker:1981} and used in later studies is defined as:
\begin{equation}
    \left.\frac{\partial \phi}{\partial n}\right|_{wall} = \left. C_w \nabla^2_\perp \phi \right|_{wall},
    \label{eqn:elecBND}
\end{equation}
where:
\begin{equation}
    C_w \equiv \frac{\sigma_w \tau_w}{\sigma L}
    \label{eqn:Cw}
\end{equation}
is the wall conductance ratio, i.e., the ratio between the total electric conductances of the wall and the fluid. Here, $\sigma_w$ is the electric conductivity of the wall material, $\bm{n}$ is the outward-facing normal to the boundary, and $\nabla^2_\perp$ is the two-dimensional Laplacian in the plane of the boundary. The physical meaning of \eqref{eqn:elecBND} is that the electric charges carried into a thin wall by the normal component of the electric current are redistributed tangentially in the wall.

The limits $C_w = 0$ and $C_w = \infty$ correspond to the cases of perfectly electrically insulating and perfectly electrically conducting walls. Since the conductivity of the fluid $\sigma$ is high for liquid metals, $C_w$ is typically small, so a perfectly electrically conducting wall is rarely a good approximation. The situation with $C_w\rightarrow 0$ is more complex. On the one hand, this limit is often used in MHD of liquid metal flows as a justification for using the approximation of a perfectly insulating wall. On the other hand, it has been demonstrated in the literature and confirmed by our simulations presented below that in many cases, including that of magnetoconvection, flows with even very small values of $C_w$ can be profoundly different from those with $C_w = 0$.

In this study, two  configurations of  the wall conductance ratios are considered. In the first, called hereafter the configuration A, all walls have the same value of $C_w$. In the configuration B, only the hot and cold walls may have a non-zero value of $C_w$, while the sidewalls perpendicular to the magnetic field and top and bottom walls remain electrically insulating $C_w=0$. Four different values of the wall conductance ratio are considered: $C_w = 0.01, 0.1, 1$ and $50$. The value of $50$ is chosen as a representative of flows with almost perfectly electrically conducting walls.

\section{Method}
\label{sec:method}

The equations \eqref{eqn:gov:cnt} - \eqref{eqn:gov:elpot} are solved numerically using the finite difference method originally described in Refs.~\onlinecite{Krasnov:FD:2011,ZikanovJFM:2013} and extended in Ref.~\onlinecite{krasnov2023tensor} to support direct solution of problems with thin wall boundary conditions \eqref{eqn:elecBND}. The spatial discretization is on a structured non-uniform grid. The discretization scheme is of the second order and nearly fully conservative in the sense that it  conserves  mass, momentum, internal energy and electric charge exactly, and the kinetic energy with a dissipative error of the third order. The time discretization is semi-implicit and of the second order. It is based on the backward differentiation formula and the fully explicit Adams-Bashforth approximation. The diffusive term in \eqref{eqn:gov:heat} is treated implicitly in order to remove the stability limitation on the time step imposed by the small Prandtl number. The standard projection method (see, e.g., Ref.~\onlinecite{Zikanovbook:2019}) is used to compute  pressure and satisfy incompressibility.  Three elliptic equations – for temperature, pressure and electric potential – are solved at every time step using the Tensor-product-Thomas (TPT) method.\cite{Vitoshkin:2013}

In the TPT method, eigenvalue decomposition of the discretized equations is done in two directions. In the third direction, tridiagonal matrix equations are solved using the Thomas algorithm. The thin-wall boundary condition \eqref{eqn:elecBND} affects the separability of the potential equation \eqref{eqn:gov:elpot} thus requiring an unconventional implementation of the TPT algorithm. The procedure was proposed and validated in Ref.~\onlinecite{krasnov2023tensor}. In short, the coefficients of the discretization matrix are modified to incorporate the discretized form of \eqref{eqn:elecBND}. As explained in detail in Ref.~\onlinecite{krasnov2023tensor}, this allows us to solve the problem directly, avoiding iterations used in earlier studies, such as, e.g., Ref.~\onlinecite{Leboucher:1999}.

The total duration of a simulation run is 200 or 400 non-dimensional convective time units. Each simulation starts with the initial conditions consisting of random non-dimensional velocity fluctuations with amplitude $10^{-3}$, and a linear temperature profile representing pure conduction between hot and cold walls. 

While no systematic analysis of the effect of the initial conditions was carried out, tests were conducted for the flow with $\Ha=650$ and all walls electrically conducting. First, the evolution of the flow with $C_w=50$ was computed starting from the initial conditions in the form of the two previously computed fully developed flow fields: one with $\Ha=0$ and one with $\Ha=650$ and $C_w=0$. In the second test, the flows with $C_w=0.1$ and $C_w=1$ were computed starting from the fully developed flow at $\Ha=0$ as an initial condition. It was found in both tests that the change of the initial state did not affect the final fully developed state of the flow.

In order to observe the evolution of the flow, point signals of velocity and temperature, values of the Nusselt number and mean kinetic energy defined as
\begin{eqnarray}
\label{eqn:Nu}
\Nu(t) & \equiv & \frac{1}{V} \int_V \left( u_x T \sqrt{\Ray \Pr} - \frac{\partial T}{\partial x} \right) dV,\\
\label{eqn:TKE}
 E(t) & \equiv & \frac{1}{V}\int_V \left( u_x^2+u_y^2+u_z^2 \right) dV,
\end{eqnarray}
and distributions of solution variables over the lines and planes shown in Fig.~\ref{fig:doamin}b,c are recorded. The flow is considered to be in a fully developed state when $\Nu(t)$ and $E(t)$ are fluctuating around a steady mean. The length of the runs assures that a sufficiently long (never less than 100 non-dimensional time units) evolution of a fully developed flow is calculated. Time-averaged flow properties, such as the mean values of the Nusselt number and kinetic energy
\begin{equation}
\Nu = \left\langle\Nu(t)\right\rangle_t, \: E = \left\langle E(t)\right\rangle_t,
    \label{eqn:meanNuE}
\end{equation}
are computed during the last 100 time units of this period. 

The discretization grid is uniform in the $z$-direction and clustered near the sidewalls in the $x$-$y$-plane. The goal of the clustering is to assure sufficiently fine resolution of the MHD boundary layers at the walls perpendicular to the magnetic field and the thermal boundary layers and near-wall jets at the hot and cold walls parallel to the magnetic  field. Two coordinate transformations are used to generate grid clustering.  Assuming the walls are at $x_i = \pm L_i/2$, they are defined as:
\begin{eqnarray}
\label{eqn:cheb-grid}
x_i & = & \frac{L_i}{2} \left[ \gamma\sin\left(\frac{\pi}{2}\xi \right) +(1-\gamma)\xi \right],\\
\label{eqn:tanh-grid}
x_i & = & \frac{L_i}{2}\frac{\tanh(A \xi)}{\tanh(A)} ,
\end{eqnarray}
where $-1\le\xi \le 1$ is the transformed coordinate, in which the grid is uniform, $\gamma$ is the weight-coefficient determining the mixture of the uniform and Chebyshev-Gauss-Lobatto grids, and $A$ is the constant determining the degree of the $\tanh$ grid clustering. 

\begin{table}
    \centering
    \begin{tabular}{l| c c c}
        $\Ha$ & 0, 85 & 162, 240 & $325 - 796$\\ \hline
        $N_x$ & 64 & 64 & 96\\
        $N_y$ & 64 & 128 & 96\\
        $N_z$ & 480 & 480 & 720\\ \hline
        $x$& $\gamma=0.96$ & $\gamma=0.96$ & $\gamma=0.96$\\
        $y$& $\gamma=0.96$ & $\gamma=0.96$ & $A=3.0$\\
        $z$& $\gamma=0$ & $\gamma=0$ & $\gamma=0$\\ \hline
        $\Delta t$ & $1 \times 10^{-3}$ & $5 \times 10^{-4}$ & $1.6 \times 10^{-4}$
    \end{tabular}
    \caption{Parameters of the computational grid used in simulations of flows at different $\Ha$. $N_x$, $N_y$, $N_z$ are the grid sizes in the $x$, $y$, and $z$ directions. In each direction, the grid is uniform if $\gamma=0$, a mixture of the uniform and Chebyshev-Gauss-Lobatto grids (\ref{eqn:cheb-grid}) if $\gamma>0$, or $\tanh$-clustered  (\ref{eqn:tanh-grid}) if $A>0$. The time step used in the simulations is shown in the last line.} 
    \label{tbl:grid-params}
\end{table}

The parameters of the grid shown in Table \ref{tbl:grid-params} were determined as sufficient for accuracy in the grid-sensitivity study\cite{krasnov2023tensor} performed for the same flow configuration and the same values of $\Ray$, $\Ha$, and $\Pra$ at $C_w=0$. Additional tests conducted in the course of this study have confirmed that further increase of the grid size by the factor of 1.5 in each direction does not change the first two decimal digits of $\Nu$ and $E$ defined by \eqref{eqn:meanNuE}. The grid parameters shown in Table \ref{tbl:grid-params} are used in the simulations irrespective of the value of the wall conductance ratio $C_w$.

A particular challenge in numerical simulations of MHD flows of liquid metals is the need to resolve the Hartmann boundary layers of thickness $\delta_{Ha} \sim 1/\Ha$ forming at the walls perpendicular to the magnetic field. In some flows, resolution by not less than 7-8 grid points is needed.\cite{ZikanovJFM:2013} Our grid sensitivity studies as well as earlier results\cite{Gelfgat:2018} show that the requirements are less stringent in the case of the magnetoconvection in a box. Numerical resolution of the Hartmann boundary layers by 3-4 points is sufficient.

The time step $\Delta t$ is chosen to be not larger than $50\%$ of the value which causes the solution to become numerically unstable.

\section{Results and discussion}
\label{sec:results}
The results obtained for the configurations A and B of the wall boundary conditions are presented separately, in sections \ref{sec:results:allWalls} and  \ref{sec:results:HCWalls}, respectively. The simulations of flows with $C_w=0$ relevant to both cases are discussed only once  in the beginning of section \ref{sec:results:allWalls}. 

\subsection{Configuration A: all walls have the same electric conductivity}
\label{sec:results:allWalls}

\begin{figure}
\centering
\begin{overpic}[width=0.48\textwidth,clip=]{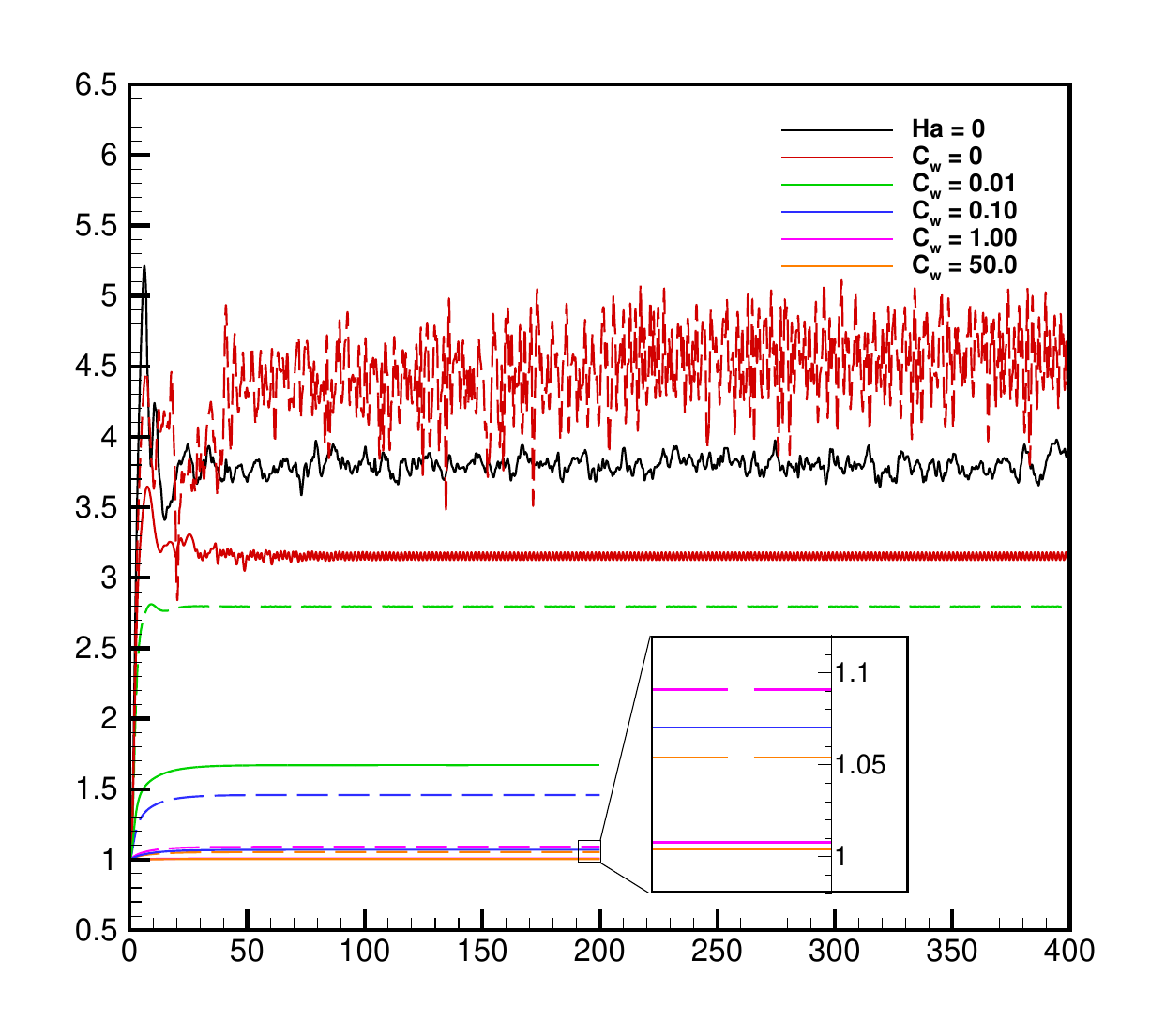}
    \titleput{1}{85}{(a)}
    \axisput{48}{2}{time}
    \axisput{2}{44}{\rotatebox[origin=c]{90}{$\Nu (t)$}}
\end{overpic}\\
\begin{overpic}[width=0.48\textwidth,clip=]{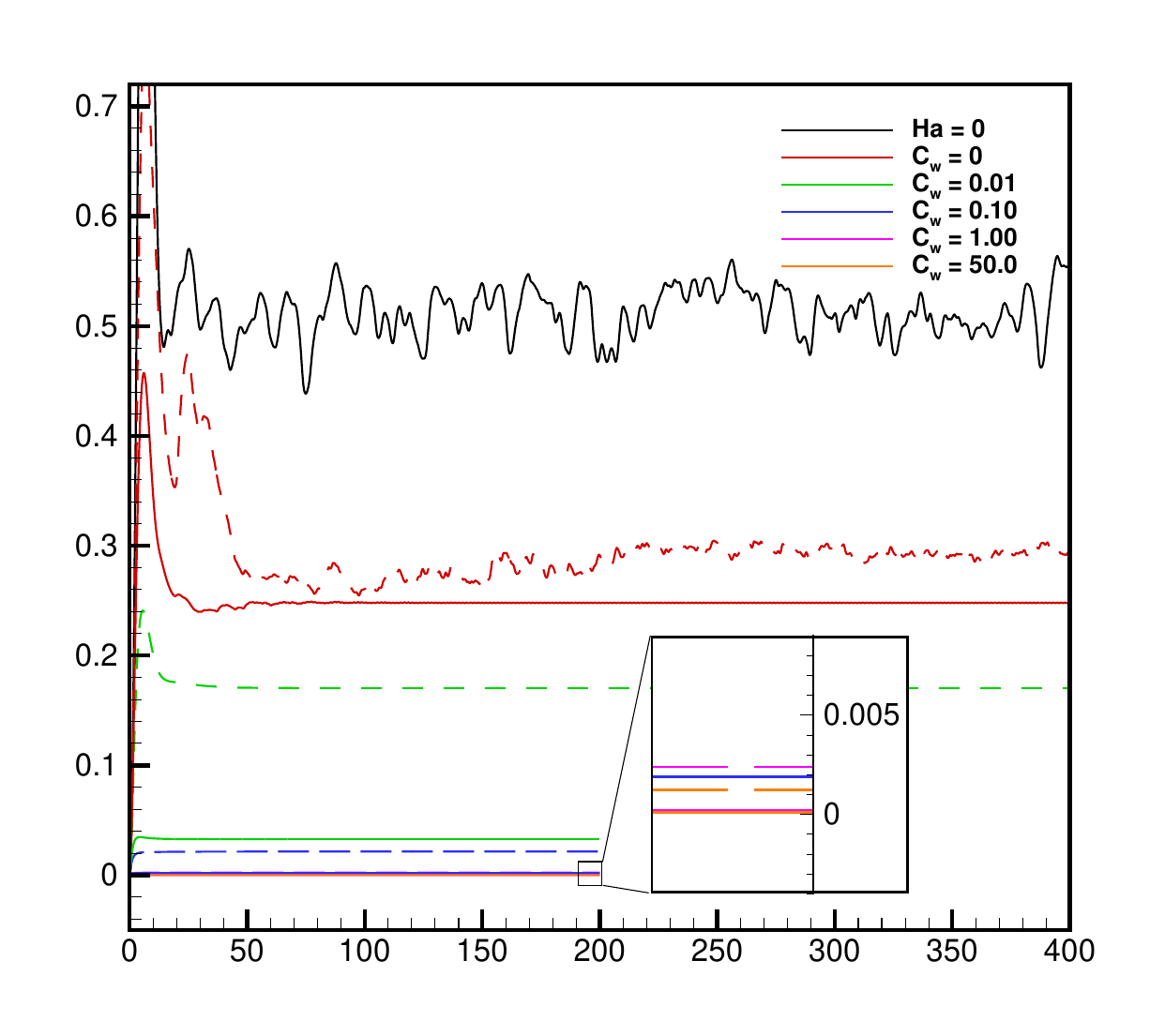}
    \titleput{1}{85}{(b)}
    \axisput{48}{2}{time}
    \axisput{2}{44}{\rotatebox[origin=c]{90}{$E (t)$}}
\end{overpic}
\caption{\emph{(a)} $\Nu(t)$ and \emph{(b)} $E(t)$  (see  \eqref{eqn:Nu} and \eqref{eqn:TKE}) for $\Ha = 325$ (dashed line) and $\Ha = 650$ (solid line) for the configuration A of the wall conductivity conditions. Black solid line represents the case $\Ha = 0$.}
\label{fig:time_NU_E_ac}
\end{figure}

Figure \ref{fig:time_NU_E_ac} shows $\Nu(t)$ and $E(t)$ during the entire simulations for two typical cases, one with moderately high $\Ha = 325$ and the other with high $\Ha = 650$. The results obtained for the pure convection flow at $\Ha = 0$ are included for comparison. A simulation is conducted for 200 time units if the fully developed flow is steady-state, and for 400 time units if this flow is fluctuating. We see in Fig.~\ref{fig:time_NU_E_ac} that the time averaging over the last 100 time units of a simulation used to calculate mean values is sufficient.

\begin{figure}
\centering
\begin{overpic}[width=0.48\textwidth,clip=]{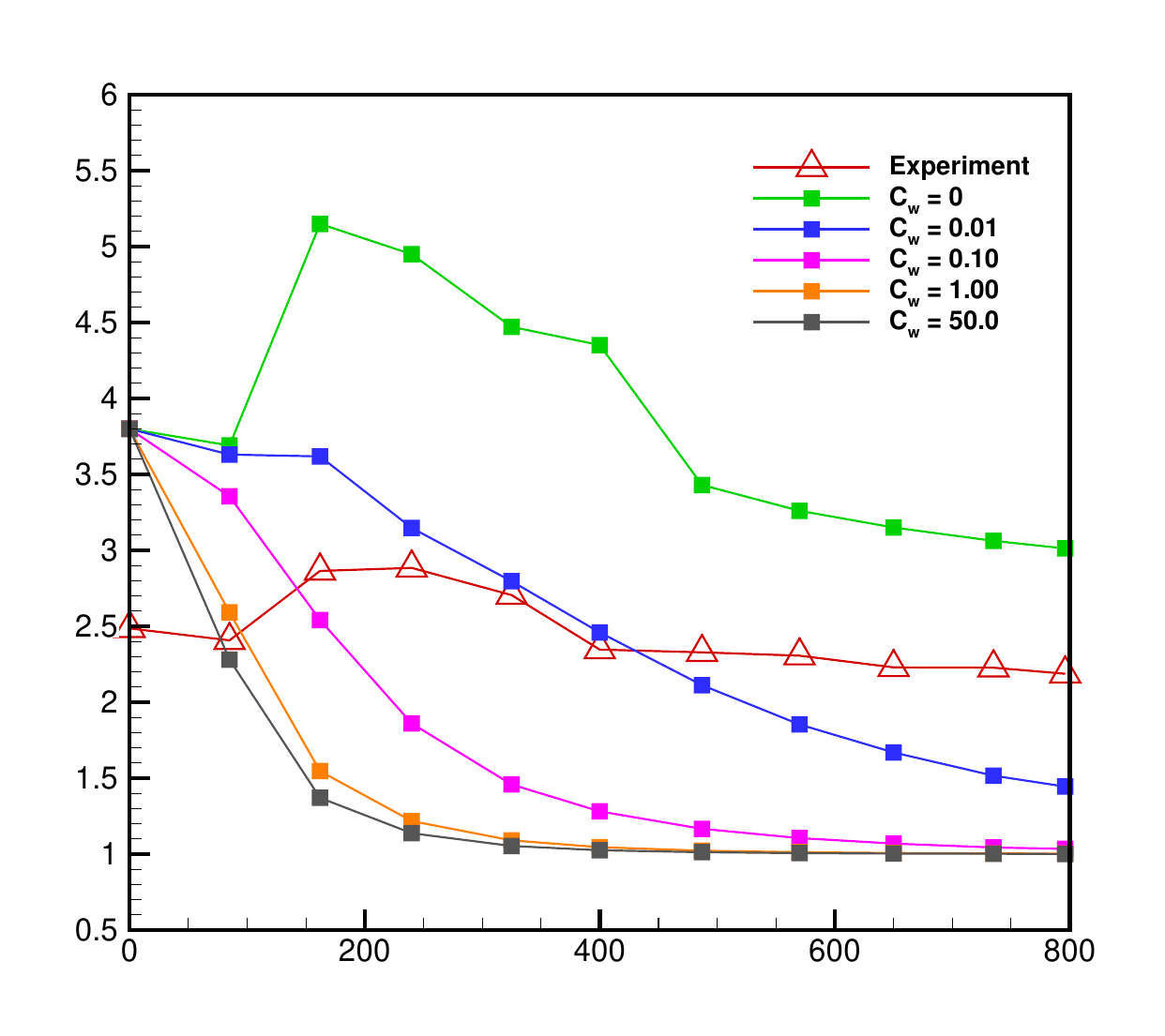}
    \titleput{2}{84}{(a)}
    \axisput{49}{2}{$Ha$}
    \axisput{2}{44}{\rotatebox[origin=c]{90}{$Nu$}}
\end{overpic}\\
\begin{overpic}[width=0.48\textwidth,clip=]{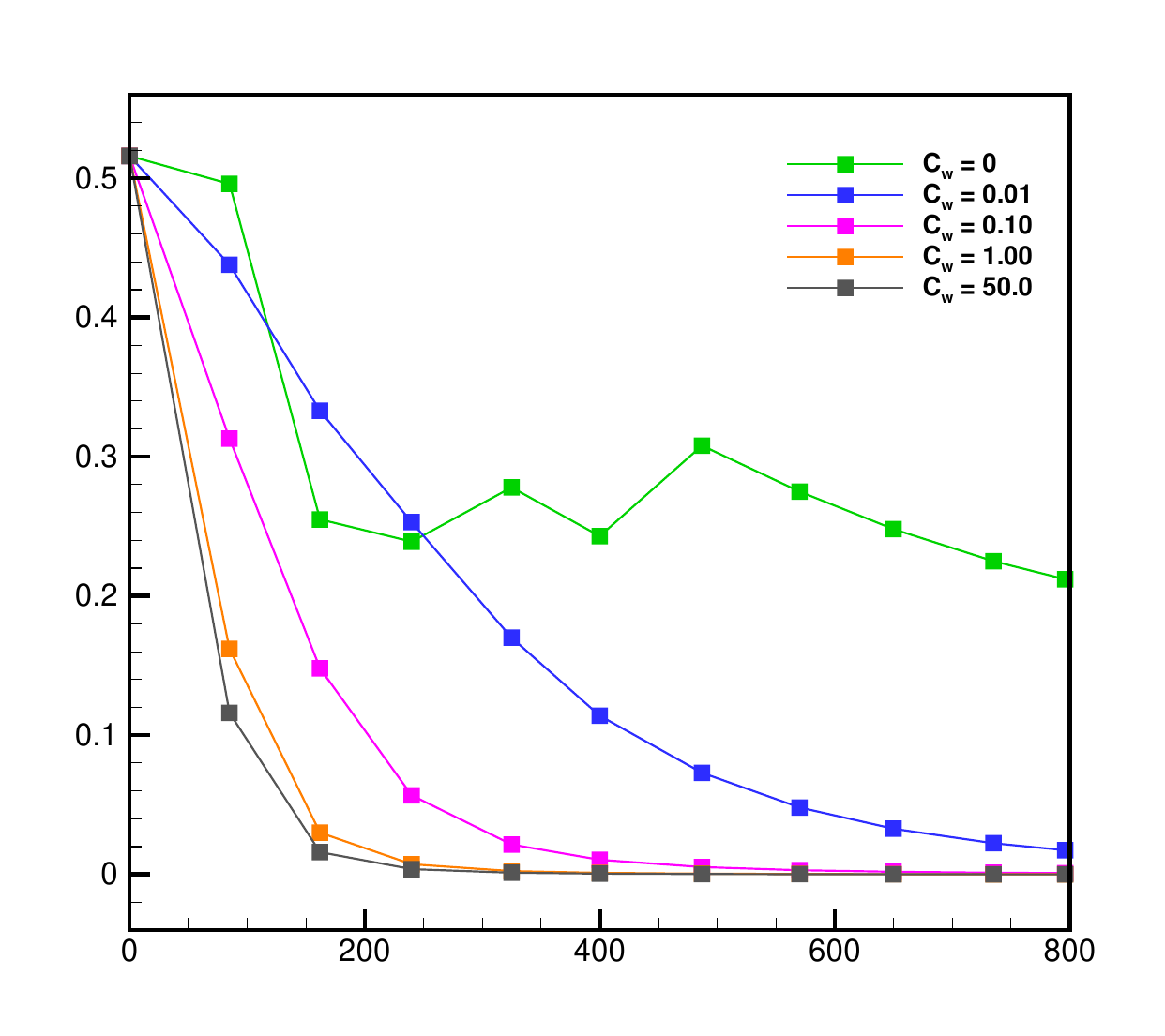}
    \titleput{2}{84}{(b)}
    \axisput{49}{2}{$Ha$}
    \axisput{2}{44}{\rotatebox[origin=c]{90}{$E$}}
\end{overpic}
\caption{Time averaged values of $\Nu$ and $E$ for configuration A (all walls have the same conductance ratio $C_w$). Data for the case of all walls electrically insulating and the experimental data\cite{Authie:2003} (available only for $\Nu$) are shown for comparison.}
\label{fig:NU_E_ac}
\end{figure}

The time averaged values $\Nu$ and $E$ obtained for the configuration A for all the explored values of $\Ha$ and $C_w$ are reported in tables \ref{tbl:Nu_ac} and \ref{tbl:E_ac} and shown in Fig.~\ref{fig:NU_E_ac}. The discussion will also use the distributions of the time-averaged flow fields within the flow domain shown in Figs.~\ref{fig:cont_ac_vert}-\ref{fig:t_ac}.

\begin{table}
    \centering
    \begin{tabular}{c|c|C{1.1cm}C{1.1cm}C{1.1cm}C{1.1cm}C{1.1cm}}
        \multicolumn{2}{c}{} & \multicolumn{5}{c}{$C_w$} \\ \cline{3-7}
        \multicolumn{2}{c}{} & 0 & 0.01 & 0.1 & 1 & 50 \\ \cline{3-7}
        \multirow{10}{*}{$\Ha$} & 0 & \multicolumn{5}{c}{$\overline{3.80}$} \\ %
         & 85 & $\overline{3.69}$ & $\overline{3.63}$ & $\overline{3.36}$ & $\overline{2.59}$ & $\overline{2.28}$ \\ %
         & 162 & $\overline{5.15}$ & $\overline{3.62}$ & $\overline{2.54}$ & ${1.55}$ & ${1.37}$ \\ %
         & 240 & $\overline{4.95}$ & $\overline{3.15}$ & ${1.86}$ & ${1.22}$ & ${1.14}$ \\ %
         & 325 & $\overline{4.47}$ & $\overline{2.80}$ & ${1.46}$ & ${1.09}$ & ${1.05}$ \\ %
         & 400 & $\overline{4.35}$ & ${2.46}$ & ${1.28}$ & ${1.05}$ & ${1.03}$ \\ %
         & 487 & $\overline{3.43}$ & ${2.11}$ & ${1.17}$ & ${1.02}$ & ${1.01}$ \\ %
         & 570 & $\overline{3.26}$ & ${1.85}$ & ${1.11}$ & ${1.01}$ & ${1.01}$ \\ %
         & 650 & $\overline{3.15}$ & ${1.67}$ & ${1.07}$ & ${1.02}$ & ${1.00}$ \\ %
         & 735 & $\overline{3.06}$ & ${1.52}$ & ${1.04}$ & ${1.01}$ & ${1.00}$ \\ %
         & 796 & ${3.01}$ & ${1.45}$ & ${1.04}$ & ${1.00}$ & ${1.00}$ \\
    \end{tabular}
    \caption{Time averaged values of $\Nu$ \eqref{eqn:meanNuE} computed for the configuration A (all walls have the same conductance ratio $C_w$). Values marked with an overline indicate cases in which $\Nu(t)$ fluctuates significantly in a fully developed flow. In other cases, the signal is steady-state after an initial transient.}
    \label{tbl:Nu_ac}
\end{table}

\begin{table*}
    \centering
    \begin{tabular}{c|c|C{2.2cm}C{2.2cm}C{2.2cm}C{2.2cm}C{2.2cm}}
        \multicolumn{2}{c}{} & \multicolumn{5}{c}{$C_w$} \\ \cline{3-7}
        \multicolumn{2}{c}{} & 0 & 0.01 & 0.1 & 1 & 50 \\ \cline{3-7}
        \multirow{10}{*}{$\Ha$} & 0 & \multicolumn{5}{c}{$\overline{5.16 \times 10^{-1}}$} \\
         & 85 & $\overline{4.96 \times 10^{-1}}$ & $\overline{4.38 \times 10^{-1}}$ & $\overline{3.13 \times 10^{-1}}$ & $\overline{1.62 \times 10^{-1}}$ & $\overline{1.16 \times 10^{-1}}$ \\
         & 162 & $\overline{2.55 \times 10^{-1}}$ & $\overline{3.33 \times 10^{-1}}$ & $\overline{1.48 \times 10^{-1}}$ & ${3.00 \times 10^{-2}}$ & ${1.60 \times 10^{-2}}$ \\
         & 240 & $\overline{2.39 \times 10^{-1}}$ & $\overline{2.53 \times 10^{-1}}$ & ${5.68 \times 10^{-2}}$ & ${7.32 \times 10^{-3}}$ & ${3.79 \times 10^{-3}}$ \\
         & 325 & $\overline{2.78 \times 10^{-1}}$ & $\overline{1.70 \times 10^{-1}}$ & ${2.15 \times 10^{-2}}$ & ${2.38 \times 10^{-3}}$ & ${1.21 \times 10^{-3}}$ \\
         & 400 & $\overline{2.43 \times 10^{-1}}$ & ${1.14 \times 10^{-1}}$ & ${1.05 \times 10^{-2}}$ & ${1.10 \times 10^{-3}}$ & ${5.57 \times 10^{-4}}$ \\
         & 487 & $\overline{3.08 \times 10^{-1}}$ & ${7.30 \times 10^{-2}}$ & ${5.28 \times 10^{-3}}$ & ${5.26 \times 10^{-4}}$ & ${2.66 \times 10^{-4}}$ \\
         & 570 & $\overline{2.75 \times 10^{-1}}$ & ${4.81 \times 10^{-2}}$ & ${3.02 \times 10^{-3}}$ & ${2.93 \times 10^{-4}}$ & ${1.48 \times 10^{-4}}$ \\
         & 650 & $\overline{2.48 \times 10^{-1}}$ & ${3.28 \times 10^{-2}}$ & ${1.89 \times 10^{-3}}$ & ${1.80 \times 10^{-4}}$ & ${9.08 \times 10^{-5}}$ \\
         & 735 & $\overline{2.25 \times 10^{-1}}$ & ${2.24 \times 10^{-2}}$ & ${1.22 \times 10^{-3}}$ & ${1.14 \times 10^{-4}}$ & ${5.76 \times 10^{-5}}$ \\
         & 796 & ${2.12 \times 10^{-1}}$ & ${1.74 \times 10^{-2}}$ & ${9.16 \times 10^{-4}}$ & ${8.52 \times 10^{-5}}$ & ${4.21 \times 10^{-5}}$ \\
    \end{tabular}
    \caption{Time averaged values of $E$ \eqref{eqn:meanNuE} computed for the configuration A (all walls have the same conductance ratio $C_w$). Values marked with an overline indicate cases in which $E(t)$ fluctuates significantly in a fully developed flow. In other cases, the signal is steady-state after an initial transient.}
    \label{tbl:E_ac}
\end{table*}

\subsubsection{All walls are perfectly electrically insulating} \label{sec:results:allWalls:c0}
We start with the results obtained for the case of all walls perfectly electrically insulating $C_w=0$. The data were largely presented in our earlier paper.\cite{krasnov2023tensor} A brief discussion is included here for the sake of comparison with the results obtained for conducting walls. We will also use the discussion to introduce the key mechanisms determining the state and properties of the flow.

We see in tables \ref{tbl:Nu_ac} and \ref{tbl:E_ac} and Fig.~\ref{fig:NU_E_ac} that a weak magnetic field ($\Ha=85$) causes a slight decrease of $\Nu$. This is followed by substantial increase at $\Ha=162$ and a gradual decrease at higher $\Ha$. The kinetic energy $E$ decreases substantially with $\Ha$ at weak magnetic field ($\Ha\le 162$), and varies only slightly at higher $\Ha$. 

The variation of $\Nu$ with $\Ha$ is in a  qualitative agreement with the data of the experiment.\cite{Authie:2003} There is, however, a large and consistent quantitative discrepancy. The computed values of $\Nu$ are 50 to 70\% higher than measured. This effect has already been discussed in Ref.~\onlinecite{krasnov2023tensor}, where it was attributed to the challenges associated with conducting thermal convection experiments using liquid metals, particularly the difficulty of reproducing the idealized boundary condition of a constant-temperature wall in an experiment. Another possible explanation, namely the effect of imperfect electrical insulation of the wall in the experiment is invalidated by our results presented below. A further discussion of the challenges faced in experiments can be found in Ref.~\onlinecite{ZikanovASME:2021}.

\begin{figure*}
\centering
\begin{overpic}[width=0.115\textwidth,clip=]{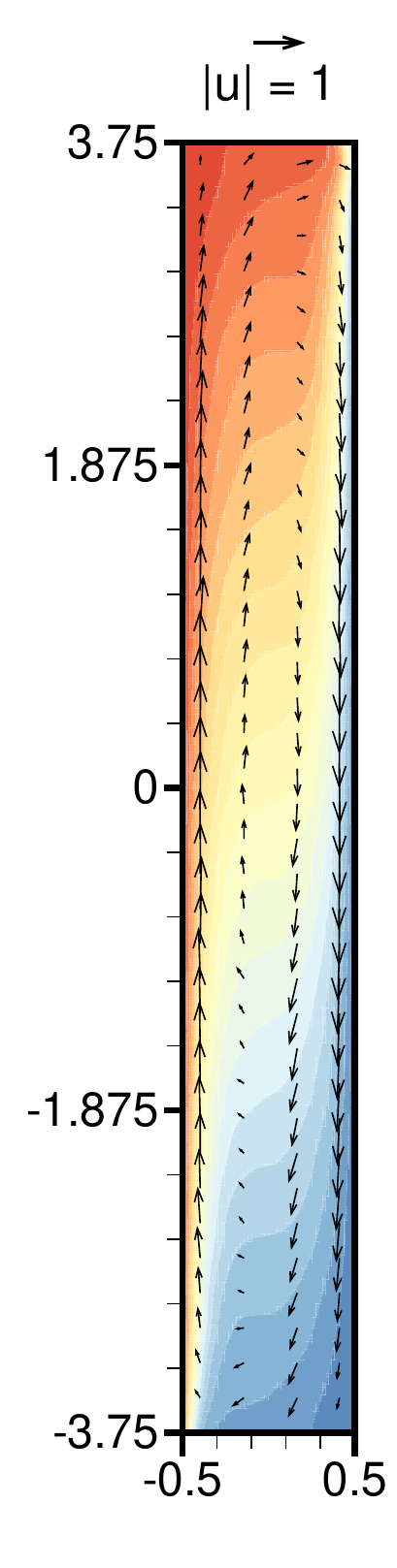}
    \titleput{1}{95}{(a)}
    \axisput{15.5}{1.5}{X}
    \axisput{2}{48}{\rotatebox[origin=c]{90}{Z}}
\end{overpic}
\begin{overpic}[width=0.115\textwidth,clip=]{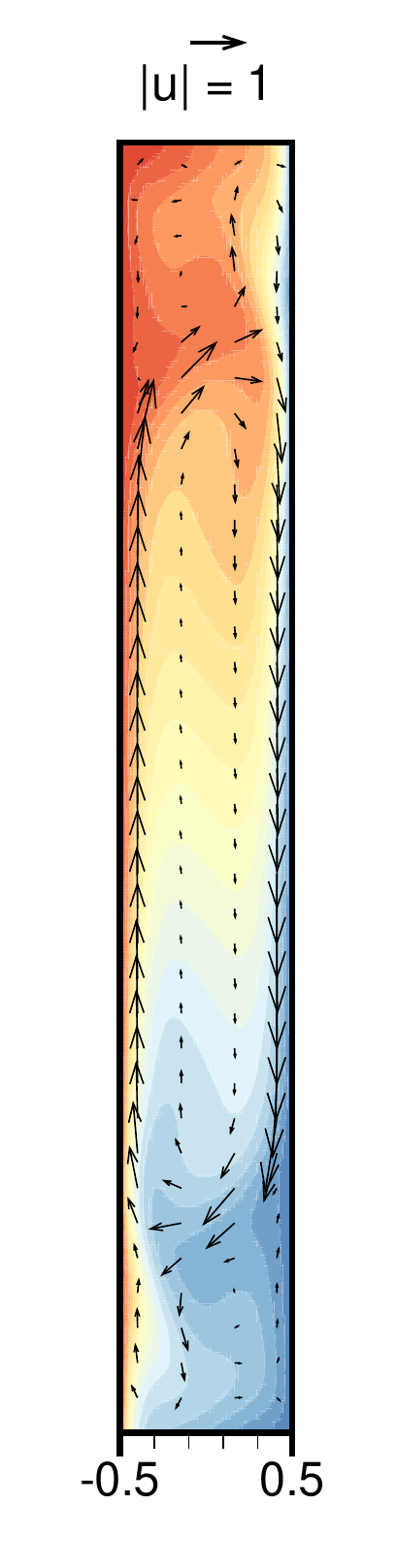}
    \titleput{1}{95}{(b)}
    \axisput{11.5}{1.5}{X}
\end{overpic}
\begin{overpic}[width=0.115\textwidth,clip=]{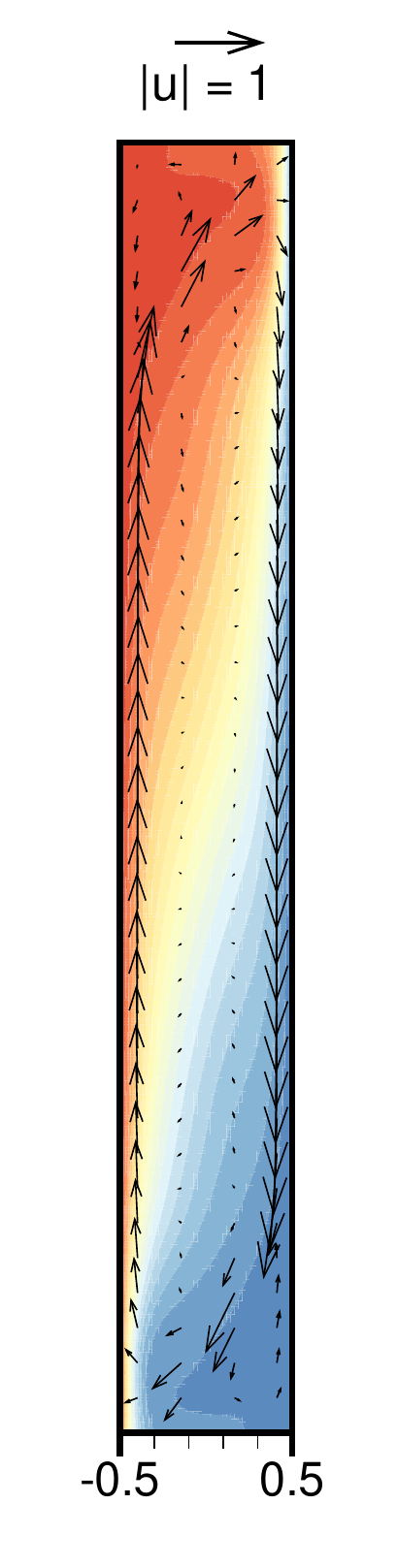}
    \titleput{1}{95}{(c)}
    \axisput{11.5}{1.5}{X}
\end{overpic}
\begin{overpic}[width=0.115\textwidth,clip=]{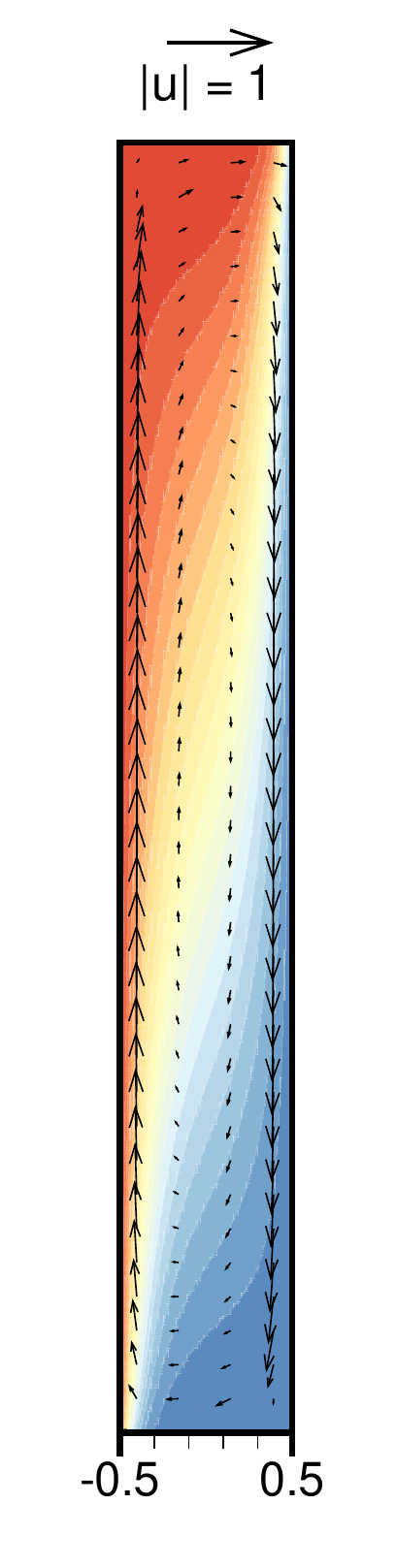}
    \titleput{1}{95}{(d)}
    \axisput{11.5}{1.5}{X}
\end{overpic}
\begin{overpic}[width=0.115\textwidth,clip=]{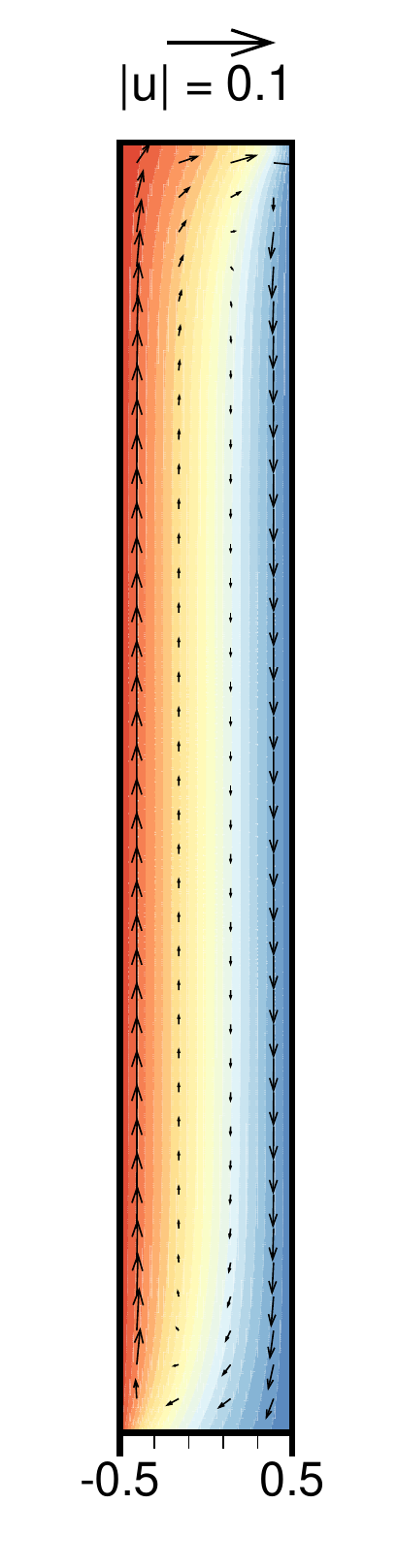}
    \titleput{1}{95}{(e)}
    \axisput{11.5}{1.5}{X}
\end{overpic}
\begin{overpic}[width=0.115\textwidth,clip=]{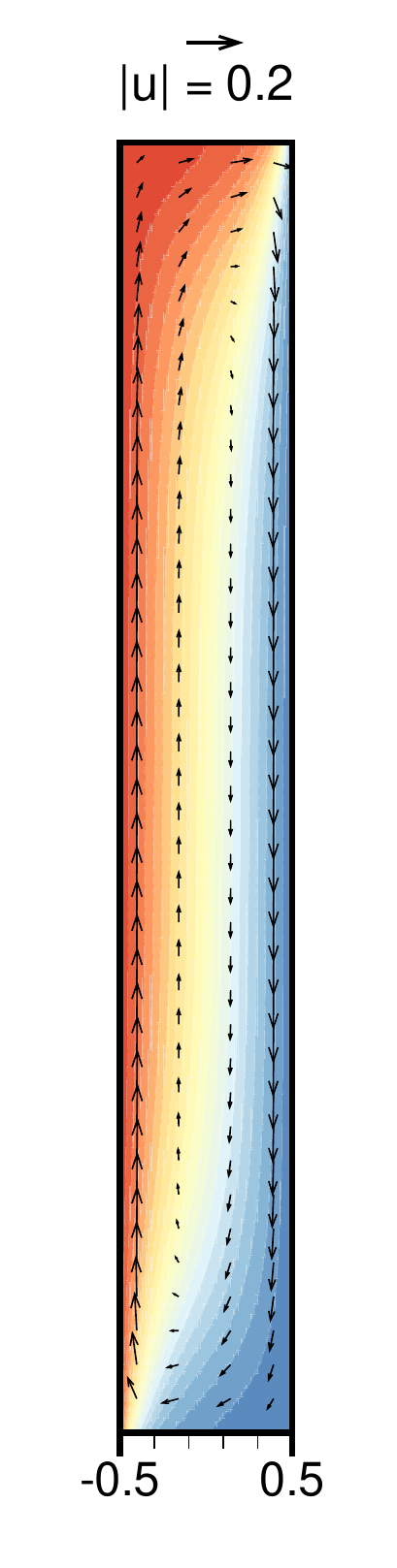}
    \titleput{1}{95}{(f)}
    \axisput{11.5}{1.5}{X}
\end{overpic}
\begin{overpic}[width=0.23\textwidth,clip=]{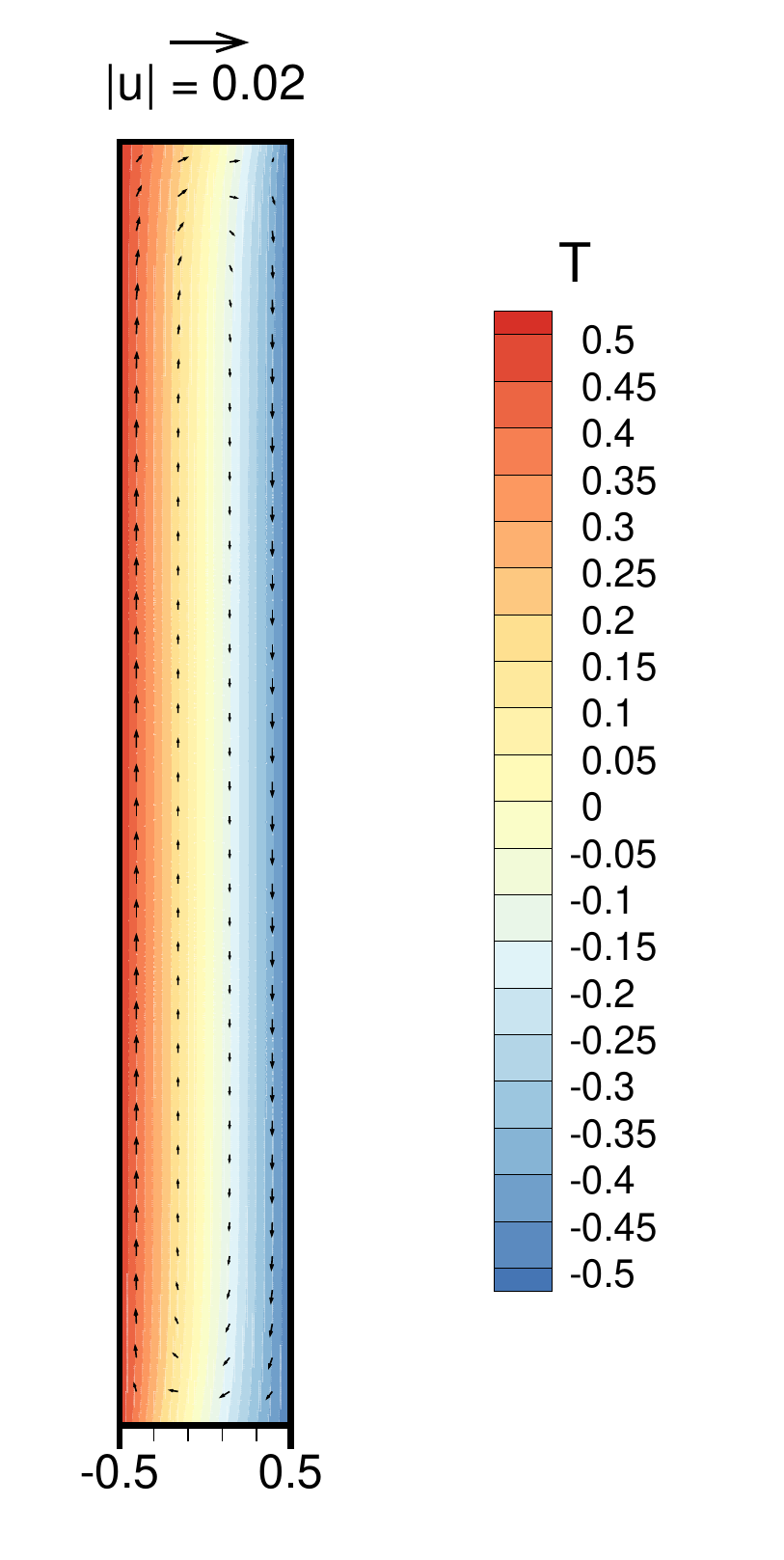}
    \titleput{-0.5}{95}{(g)}
    \axisput{11.5}{1.5}{X}
\end{overpic}
\begin{overpic}[width=0.115\textwidth,clip=]{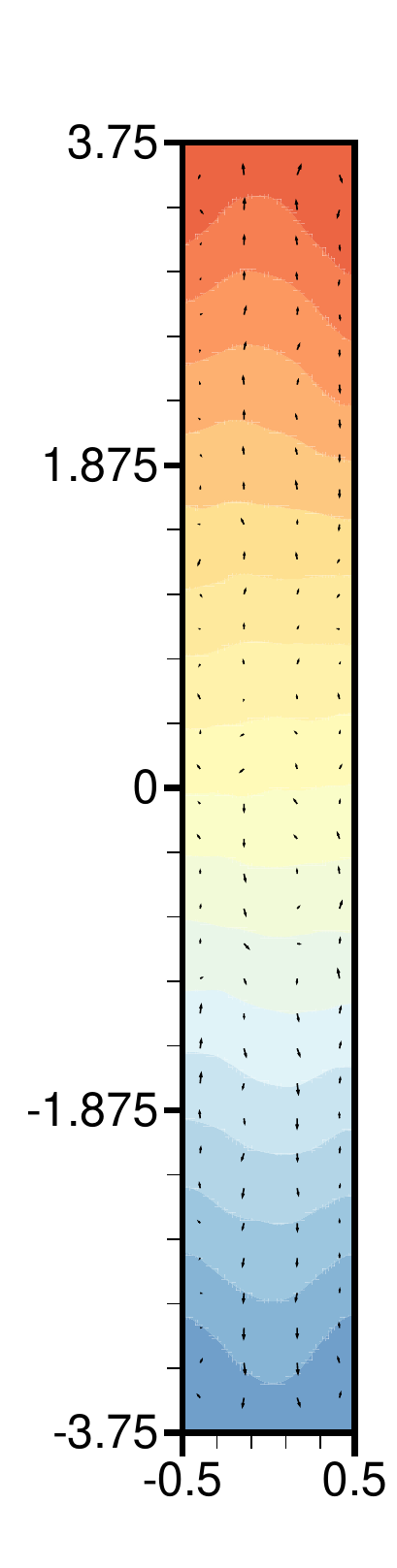}
    \axisput{15.5}{1.5}{Y}
    \axisput{2}{48}{\rotatebox[origin=c]{90}{Z}}
\end{overpic}
\begin{overpic}[width=0.115\textwidth,clip=]{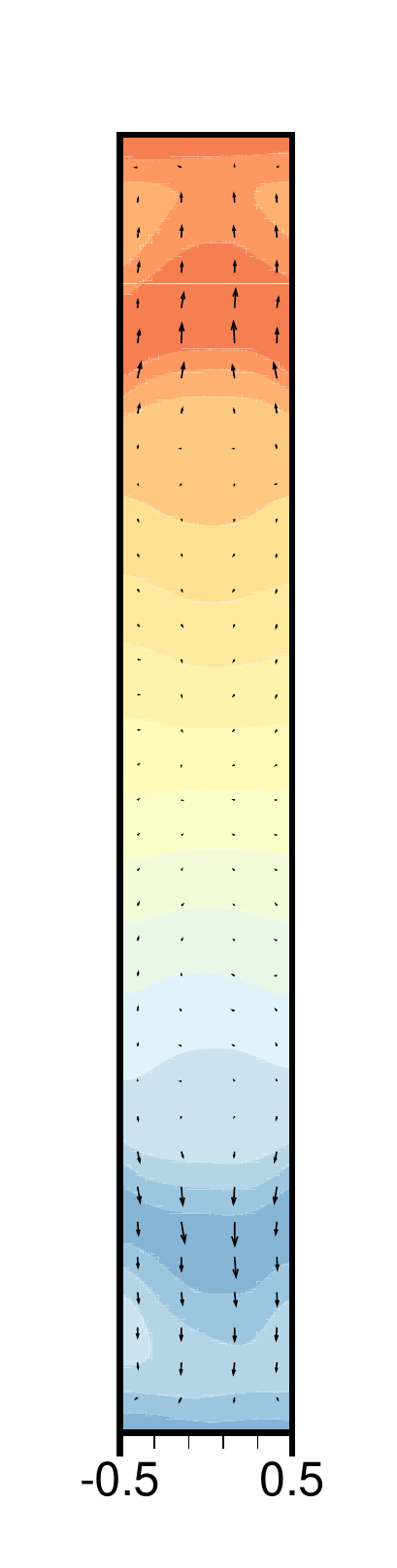}
    \axisput{11.5}{1.5}{Y}
\end{overpic}
\begin{overpic}[width=0.115\textwidth,clip=]{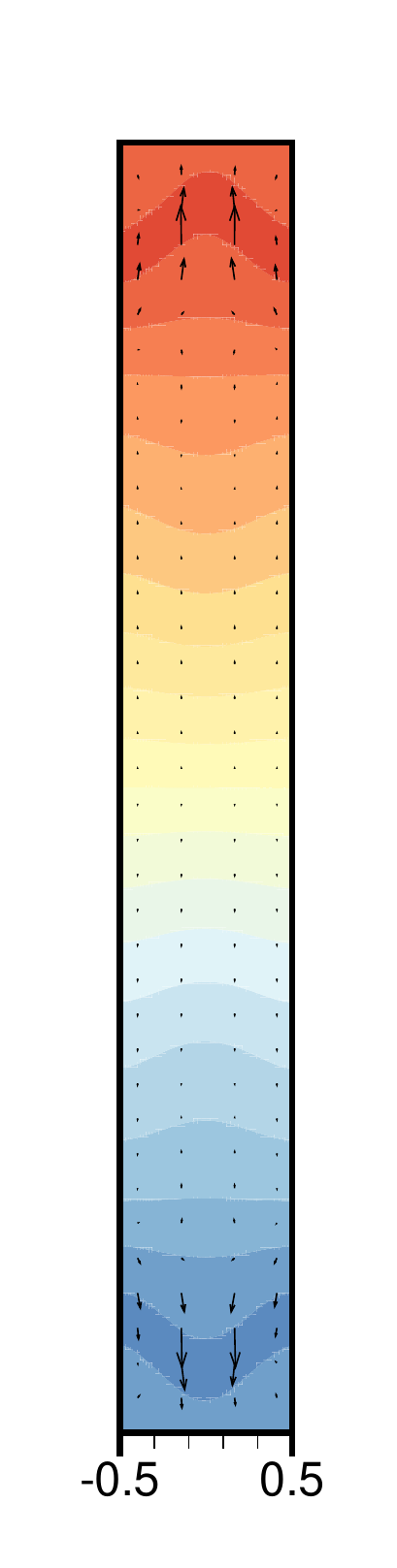}
    \axisput{11.5}{1.5}{Y}
\end{overpic}
\begin{overpic}[width=0.115\textwidth,clip=]{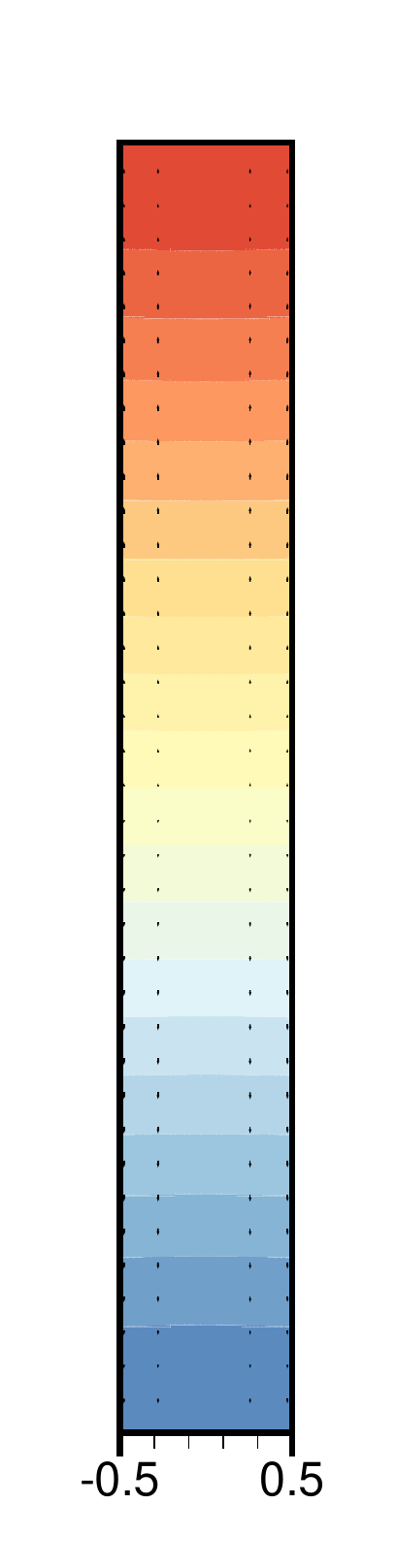}
    \axisput{11.5}{1.5}{Y}
\end{overpic}
\begin{overpic}[width=0.115\textwidth,clip=]{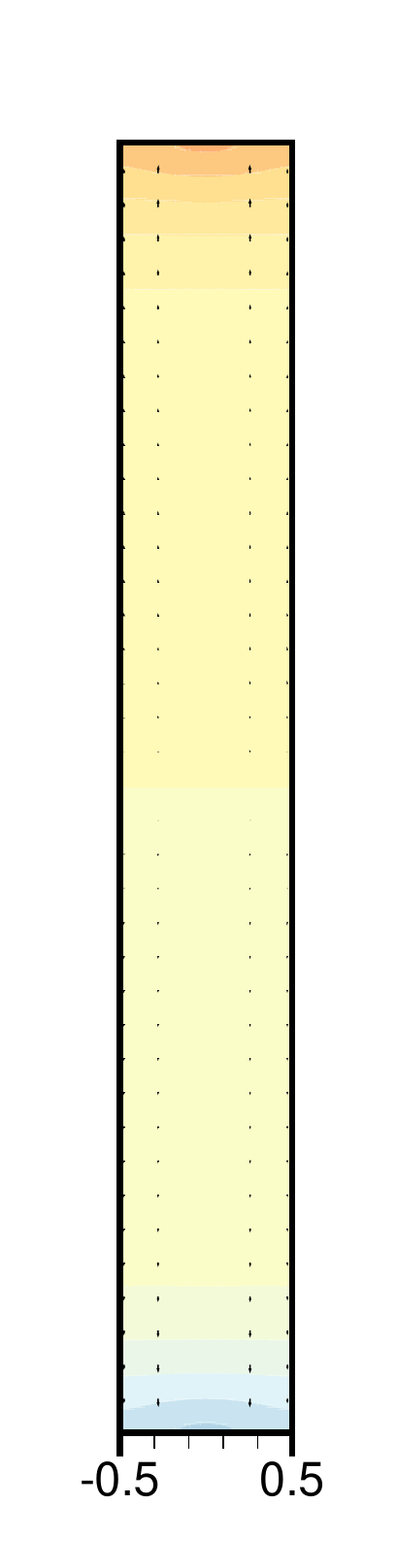}
    \axisput{11.5}{1.5}{Y}
\end{overpic}
\begin{overpic}[width=0.115\textwidth,clip=]{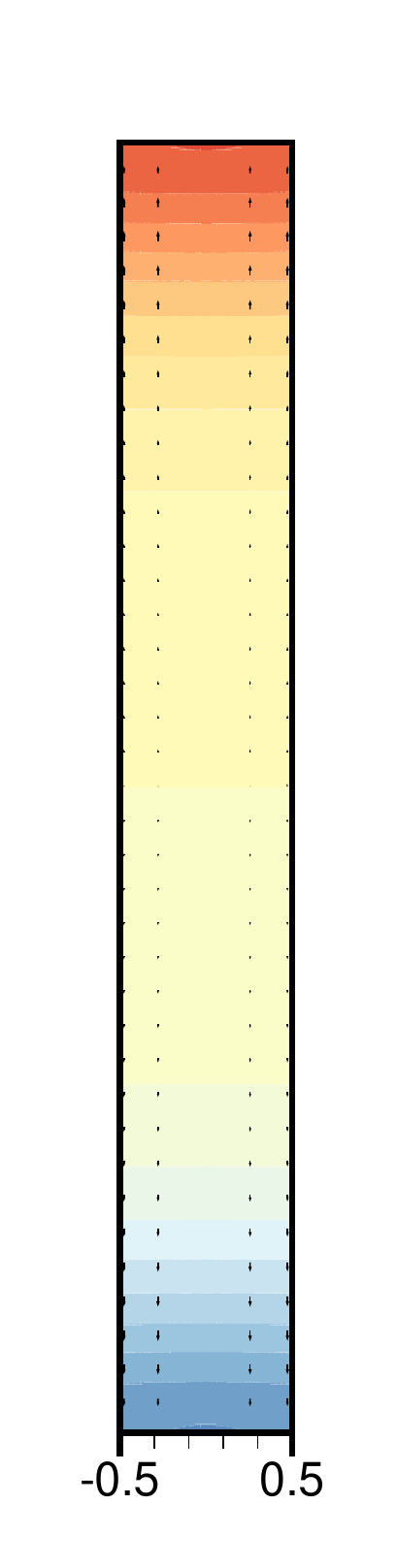}
    \axisput{11.5}{1.5}{Y}
\end{overpic}
\begin{overpic}[width=0.23\textwidth,clip=]{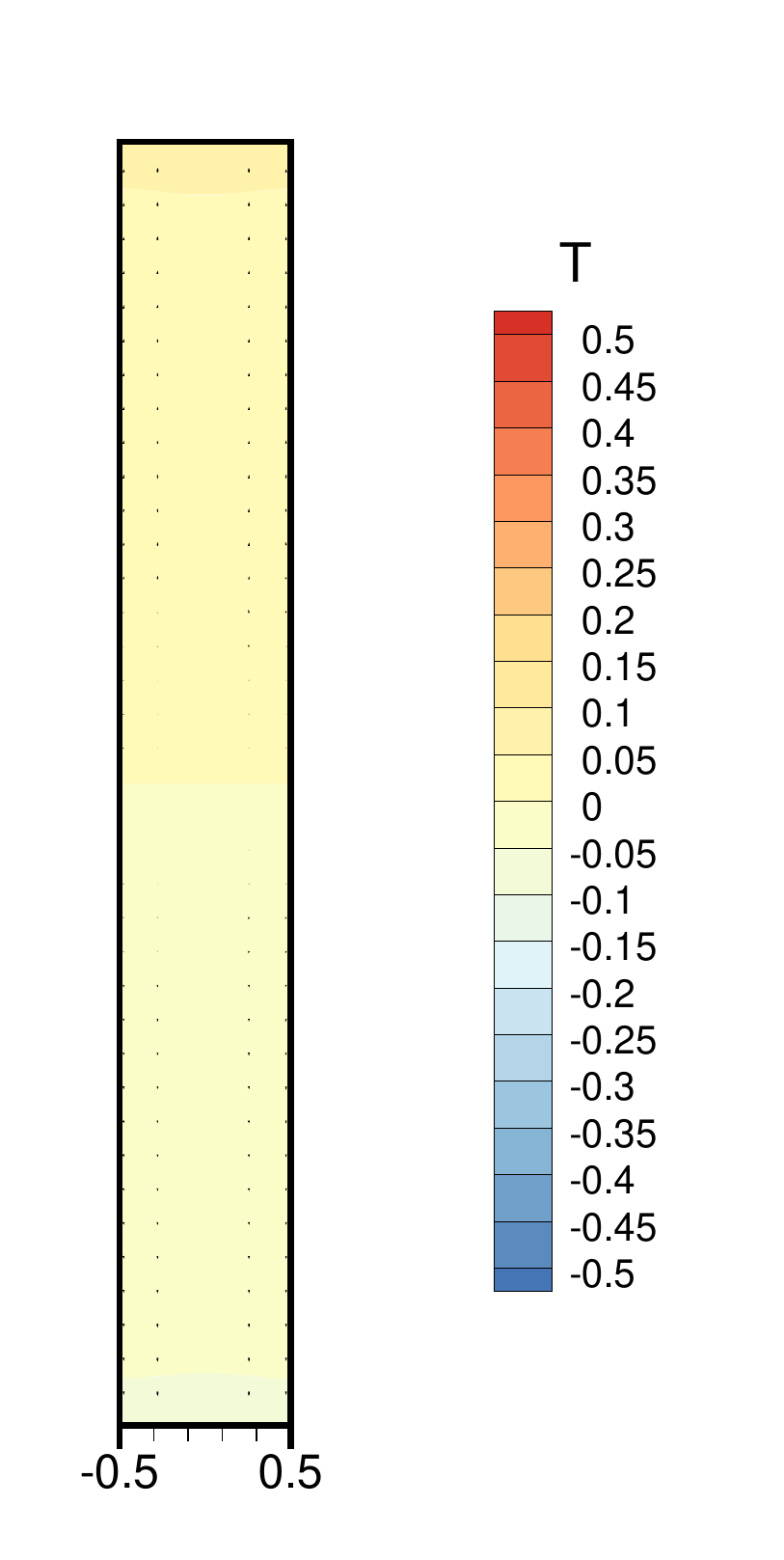}
    \axisput{11.5}{1.5}{Y}
\end{overpic}
\caption{Time-averaged distributions of $T$ and a projection of the velocity on the vertical mid-plane cross-sections $y = 0$ and $x = 0$ (see Fig.~\ref{fig:doamin}c) for configuration A. Results for the flows with $Ha = 0$ \emph{(a)}, $Ha = 85$, $C_w = 0.01$ \emph{(b)}, $Ha = 85$, $C_w = 1.0$ \emph{(c)}, $Ha = 400$, $C_w = 0.01$ \emph{(d)}, $Ha = 400$, $C_w = 1.0$ \emph{(e)}, $Ha = 796$, $C_w = 0.01$ \emph{(f)} and $Ha = 796$, $C_w = 1.0$ \emph{(g)} are shown. The top row shows cross-section $y = 0$ perpendicular to the magnetic field direction. The bottom row shows the cross-section $x = 0$ parallel to the magnetic field direction.}
\label{fig:cont_ac_vert}
\end{figure*}

\begin{figure*}
\centering
\begin{overpic}[width=0.28\textwidth,clip=]{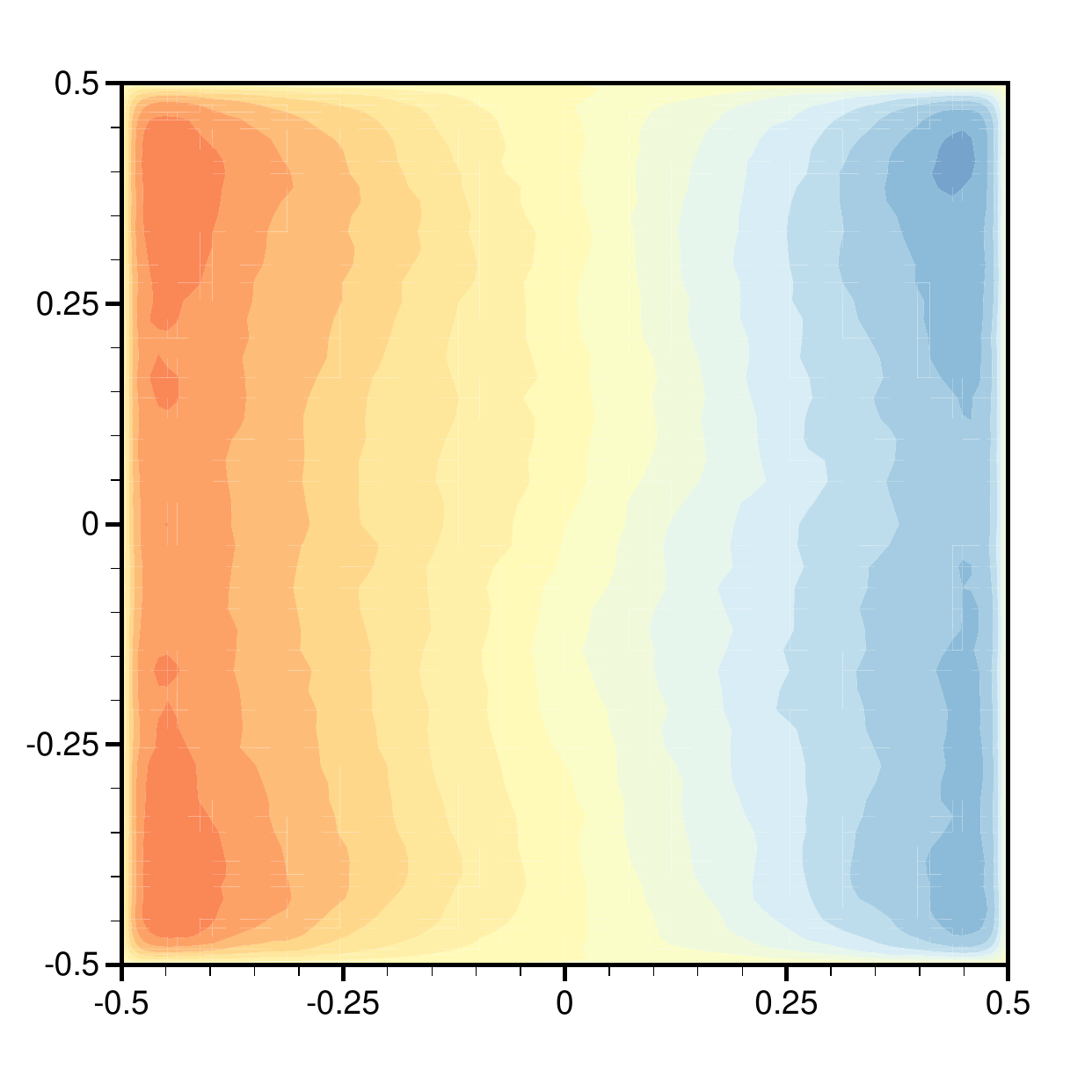}
    \titleput{0}{99}{(a)}
    \axisput{49}{0.5}{X}
    \axisput{0}{49}{\rotatebox[origin=c]{90}{Y}}
\end{overpic}
\begin{overpic}[width=0.28\textwidth,clip=]{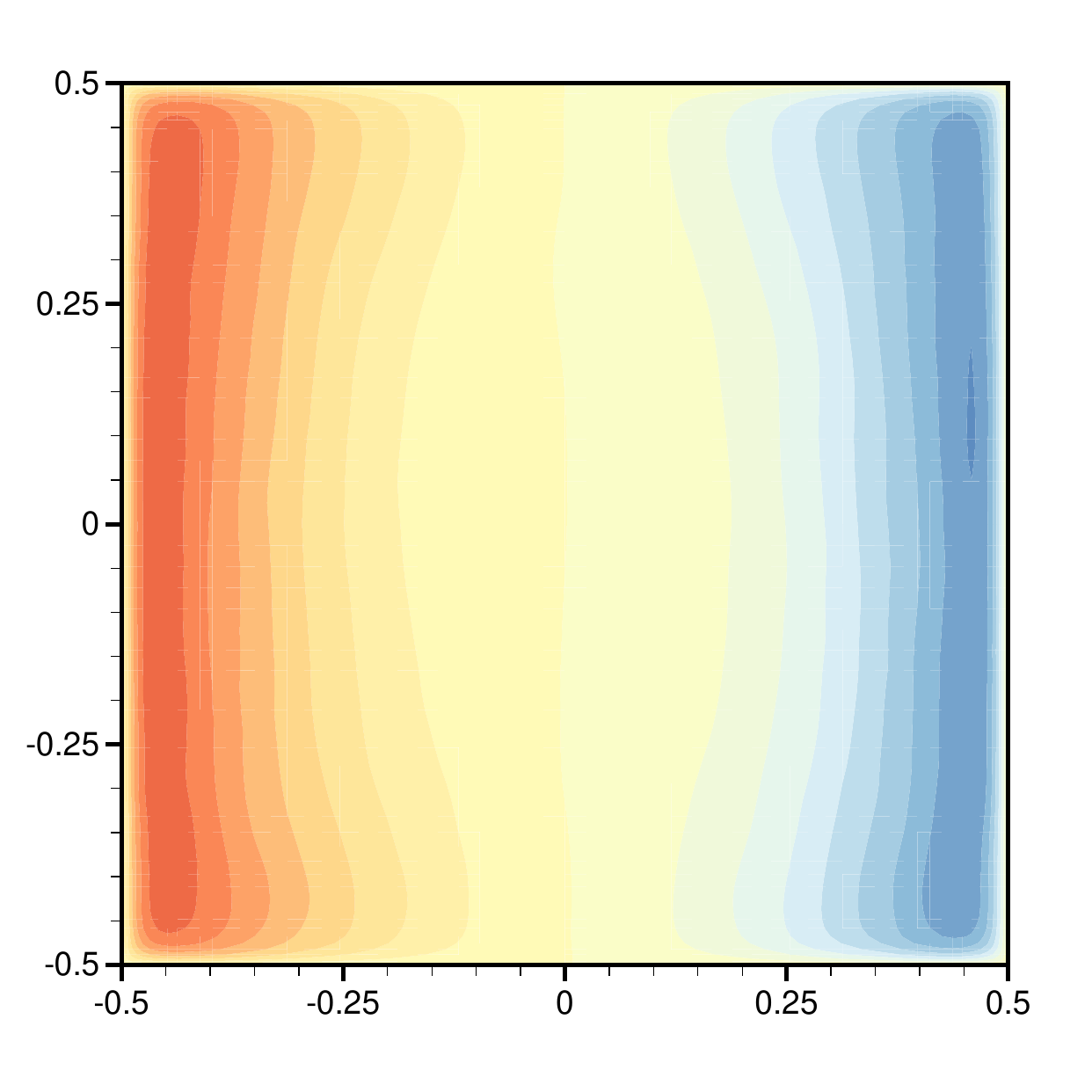}
    \titleput{0}{99}{(b)}
    \axisput{49}{0.5}{X}
    \axisput{0}{49}{\rotatebox[origin=c]{90}{Y}}
\end{overpic}
\begin{overpic}[width=0.315\textwidth,clip=]{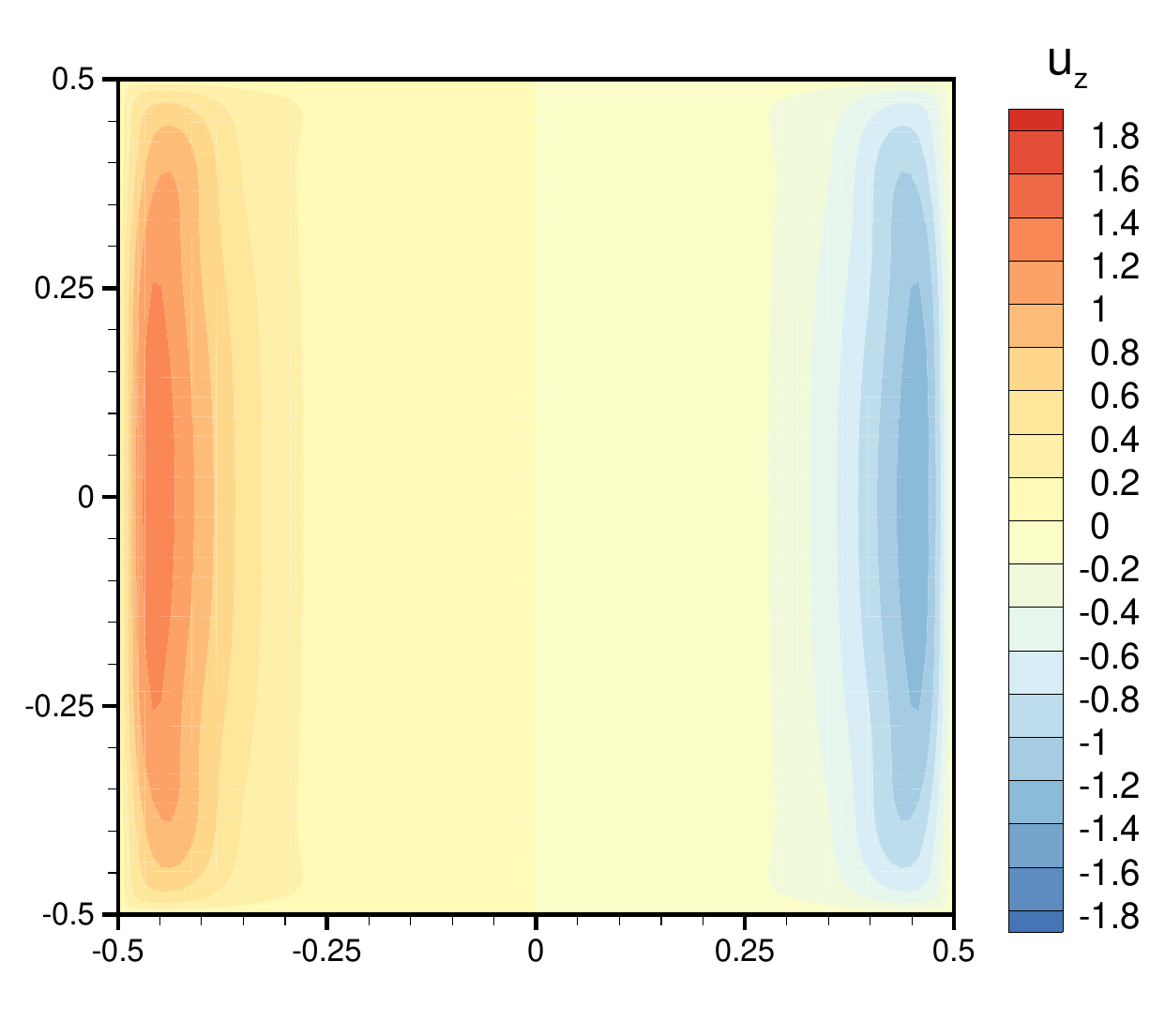}
    \titleput{1}{88}{(c)}
    \axisput{43.5}{0.25}{X}
    \axisput{0}{43.5}{\rotatebox[origin=c]{90}{Y}}
\end{overpic}\vskip1em
\begin{overpic}[width=0.28\textwidth,clip=]{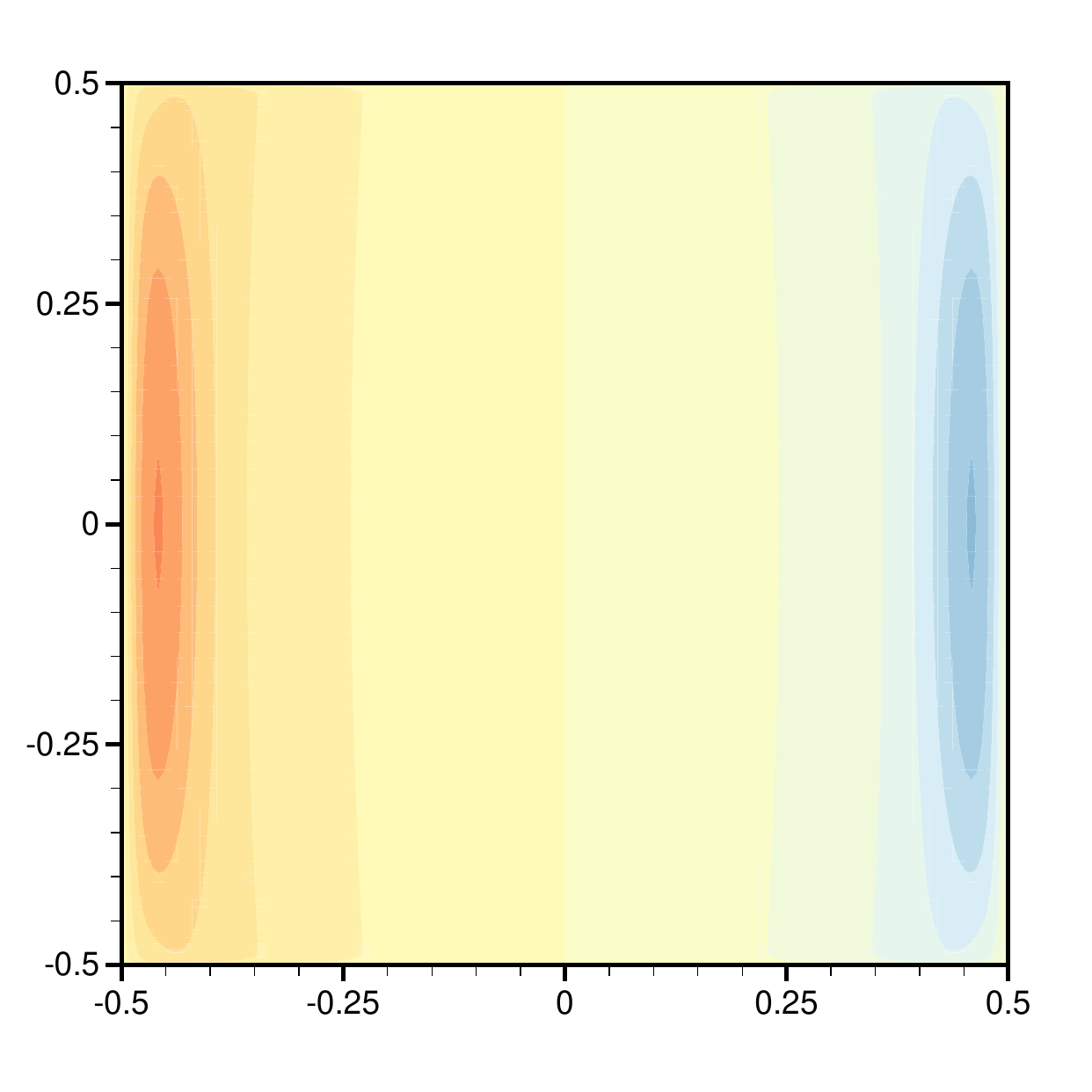}
    \titleput{0}{99}{(d)}
    \axisput{49}{0.5}{X}
    \axisput{0}{49}{\rotatebox[origin=c]{90}{Y}}
\end{overpic}
\begin{overpic}[width=0.28\textwidth,clip=]{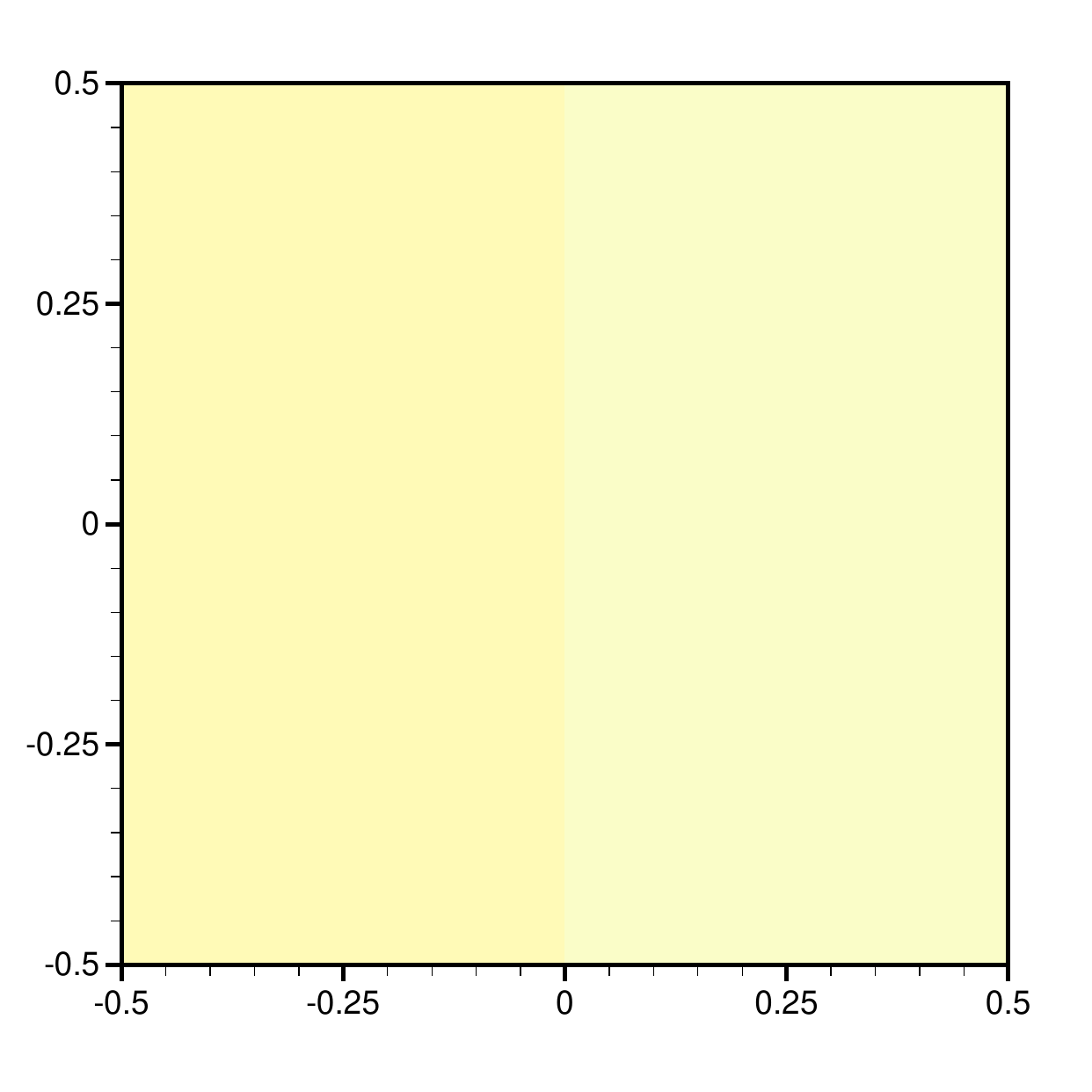}
    \titleput{0}{99}{(e)}
    \axisput{49}{0.5}{X}
    \axisput{0}{49}{\rotatebox[origin=c]{90}{Y}}
\end{overpic}
\begin{overpic}[width=0.315\textwidth,clip=]{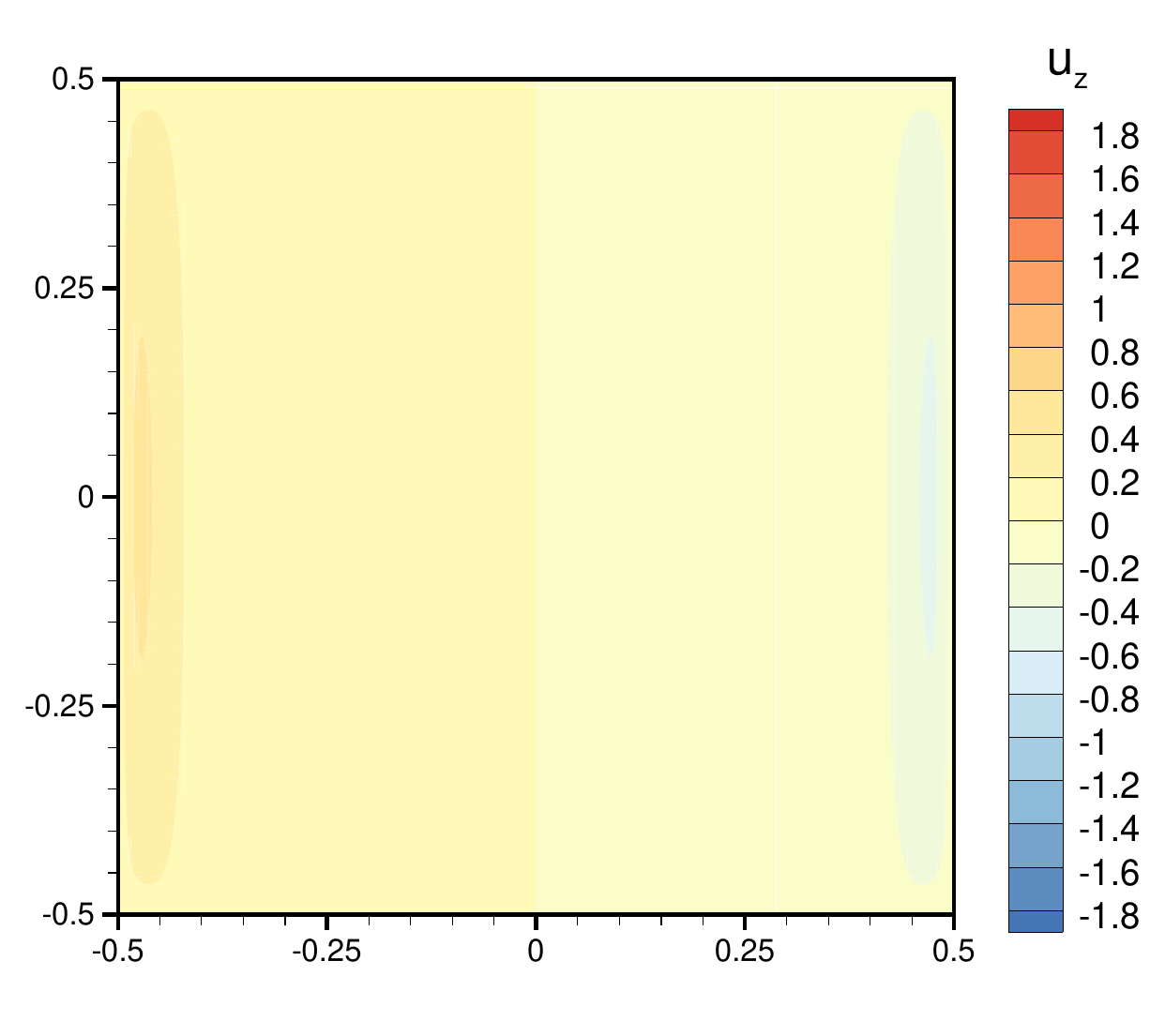}
    \titleput{1}{88}{(f)}
    \axisput{43.5}{0.25}{X}
    \axisput{0}{43.5}{\rotatebox[origin=c]{90}{Y}}
\end{overpic}
\caption{Time-averaged distributions of vertical velocity $u_z$ on the horizontal mid-plane cross-section $z = 0$ (see Fig.~\ref{fig:doamin}c) for configuration A. Results for the flows with $Ha = 0$ \emph{(a)}, $Ha = 85$, $C_w = 0.01$ \emph{(b)}, $Ha = 85$, $C_w = 1.0$ \emph{(c)}, $Ha = 400$, $C_w = 0.01$ \emph{(d)}, $Ha = 400$, $C_w = 1.0$ \emph{(e)} and $Ha = 796$, $C_w = 0.01$ \emph{(f)} are shown.}
\label{fig:cont_ac_hor}
\end{figure*}

\begin{figure*}
\begin{overpic}[width=0.41\textwidth,clip=]{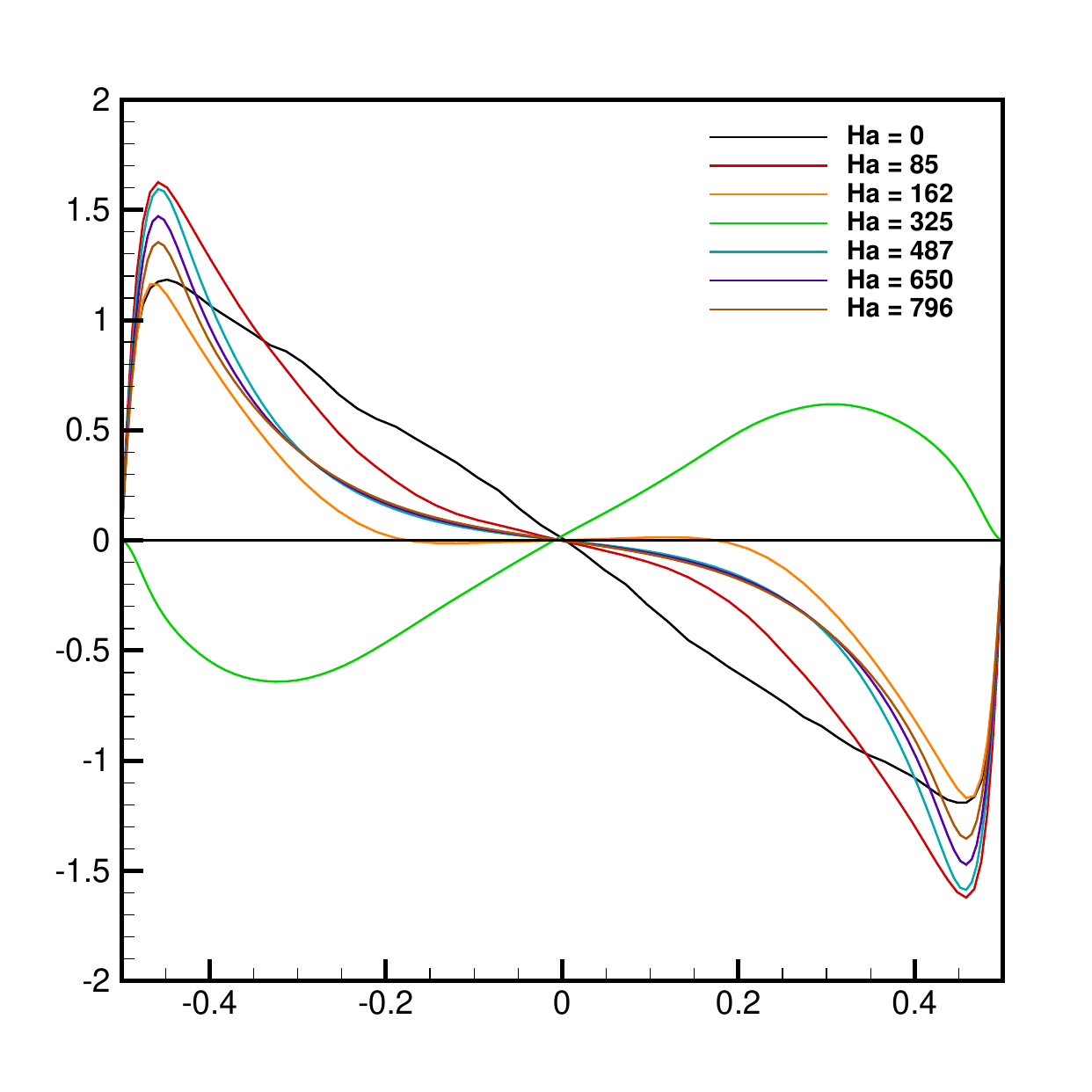}
    \titleput{2}{94}{(a)}
    \axisput{50}{2}{X}
    \axisput{2}{50}{\rotatebox[origin=c]{90}{$u_z$}}
\end{overpic}\hfil\hfil
\begin{overpic}[width=0.41\textwidth,clip=]{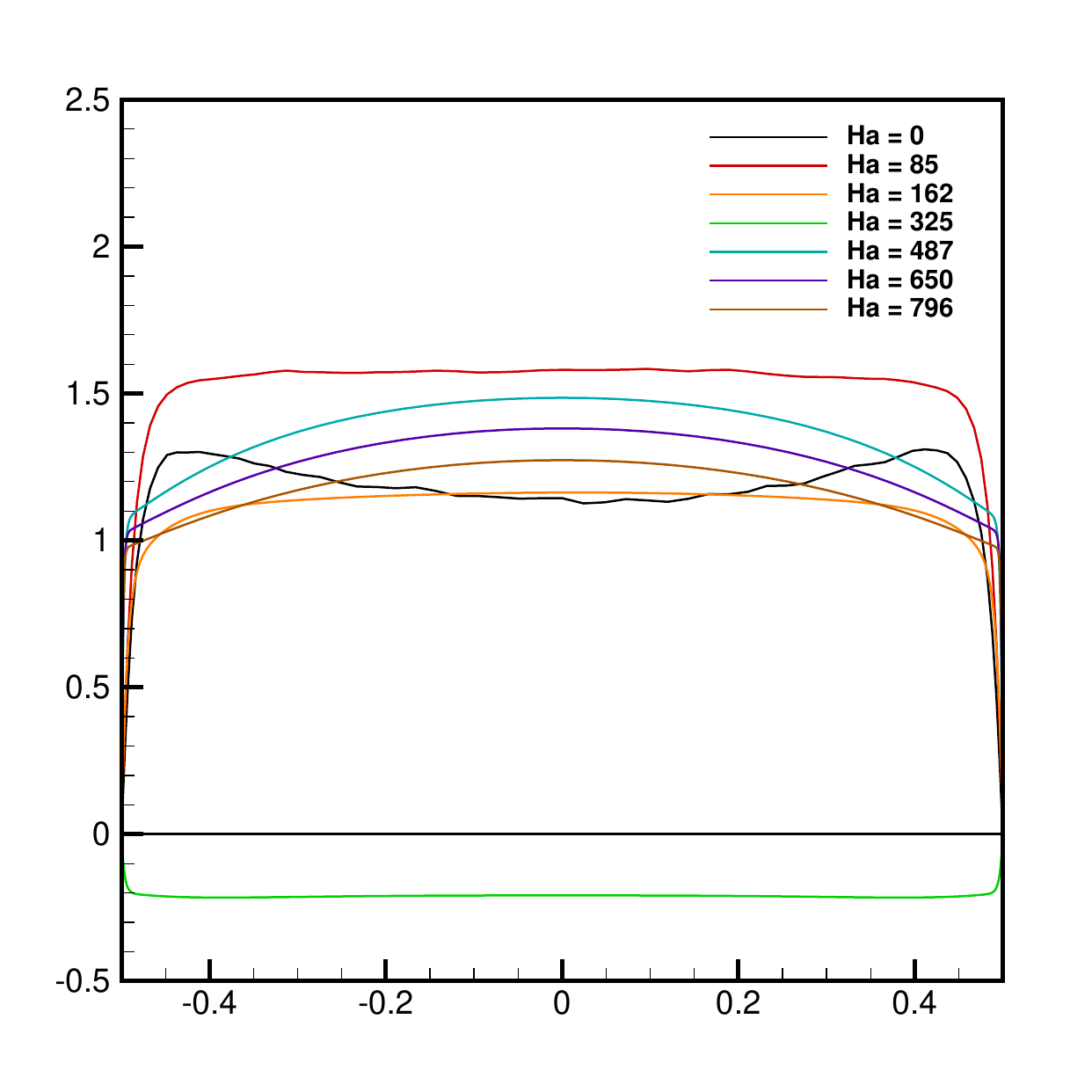}
    \axisput{50}{2}{Y}
    \axisput{2}{50}{\rotatebox[origin=c]{90}{$u_z$}}
\end{overpic}\\
\begin{overpic}[width=0.41\textwidth,clip=]{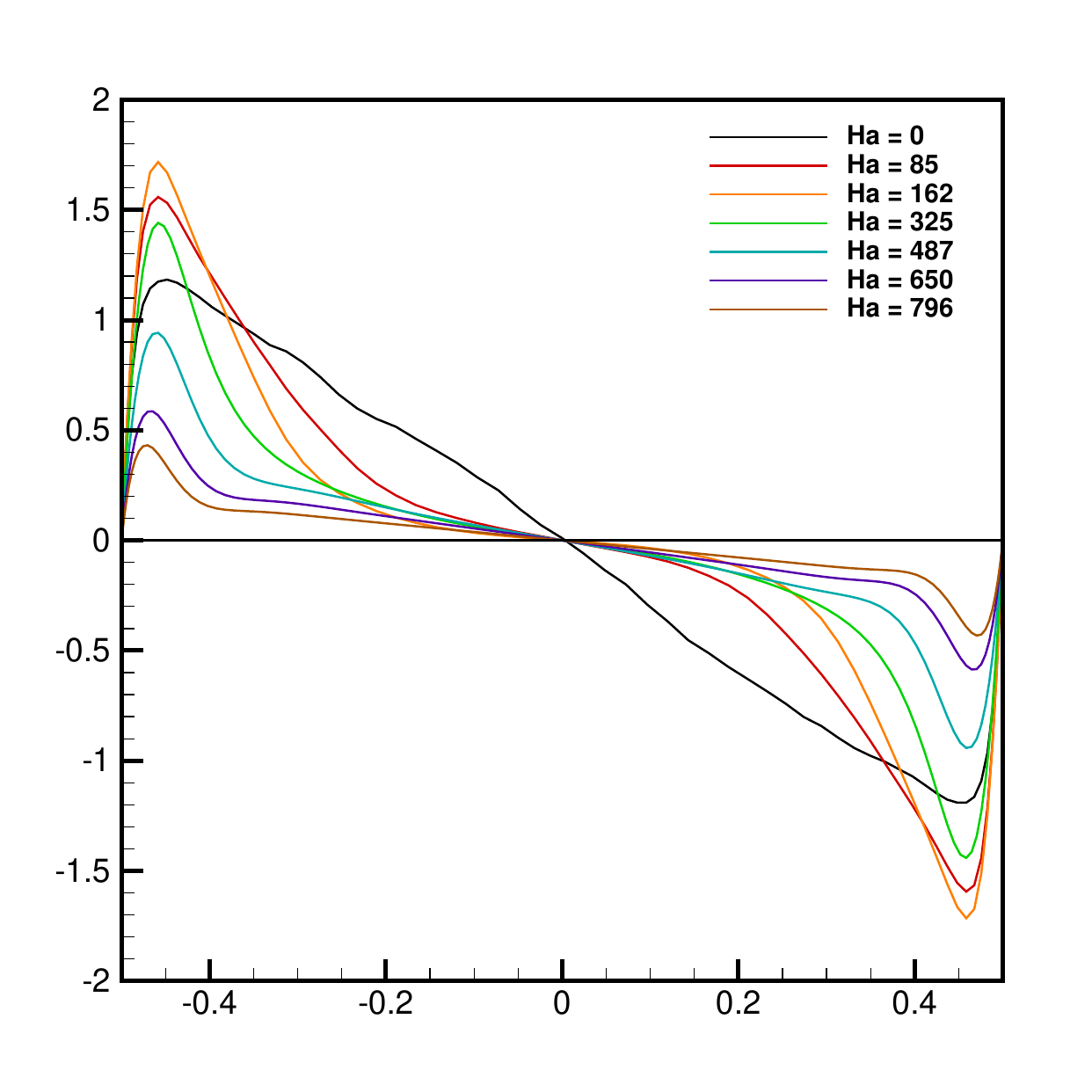}
    \titleput{2}{94}{(b)}
    \axisput{50}{2}{X}
    \axisput{2}{50}{\rotatebox[origin=c]{90}{$u_z$}}
\end{overpic}\hfil\hfil
\begin{overpic}[width=0.41\textwidth,clip=]{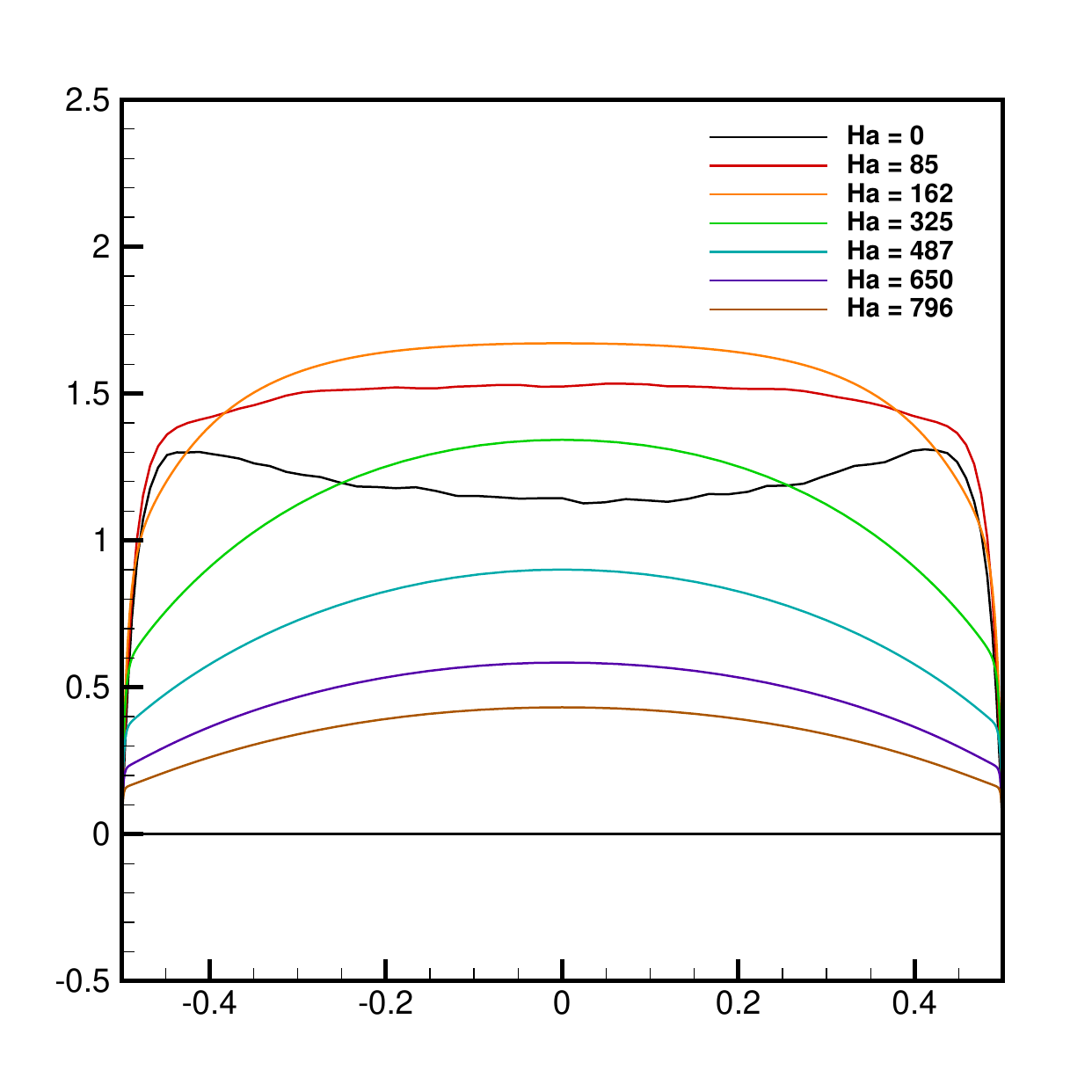}
    \axisput{50}{2}{Y}
    \axisput{2}{50}{\rotatebox[origin=c]{90}{$u_z$}}
\end{overpic}\\
\begin{overpic}[width=0.41\textwidth,clip=]{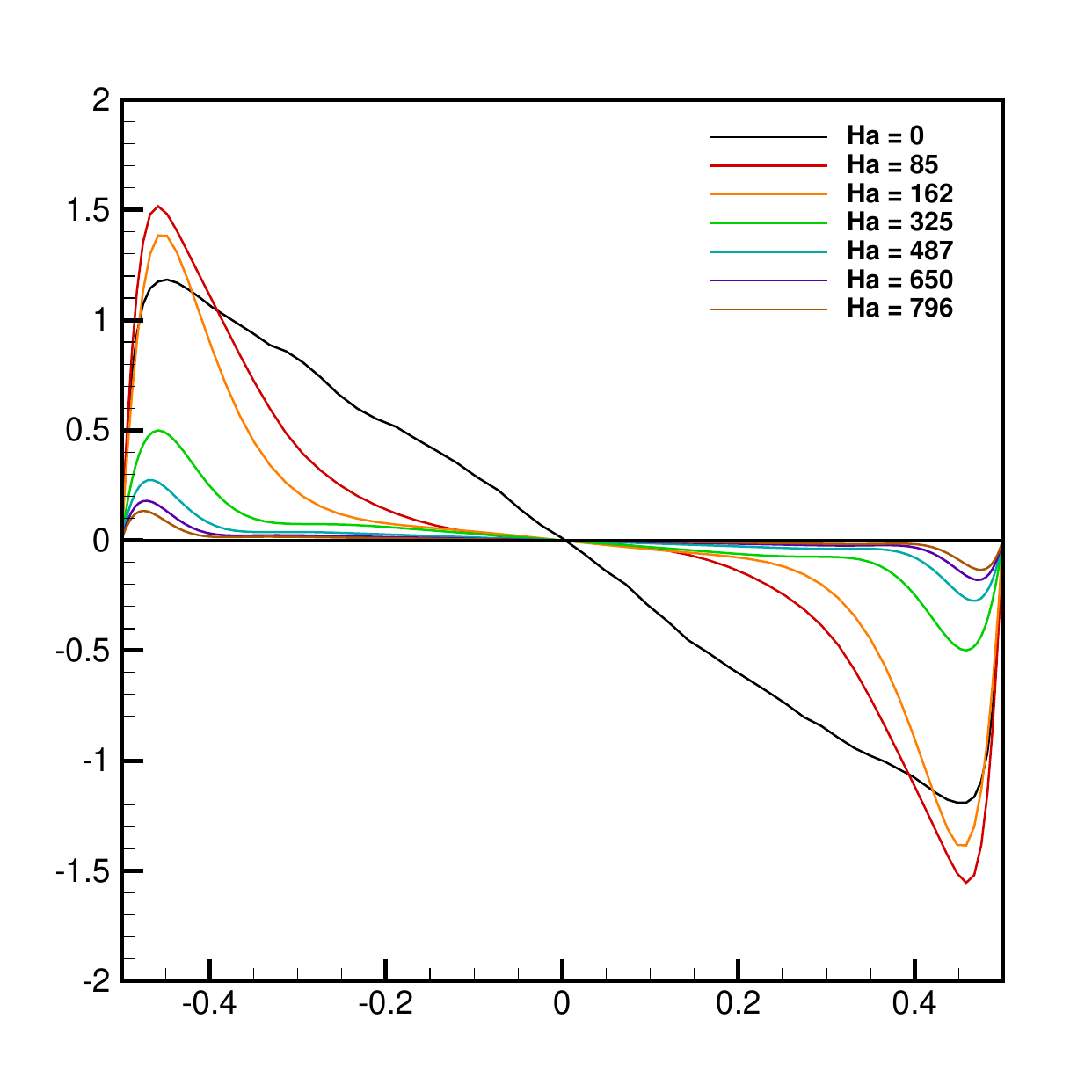}
    \titleput{2}{94}{(c)}
    \axisput{50}{2}{X}
    \axisput{2}{50}{\rotatebox[origin=c]{90}{$u_z$}}
\end{overpic}\hfil\hfil
\begin{overpic}[width=0.41\textwidth,clip=]{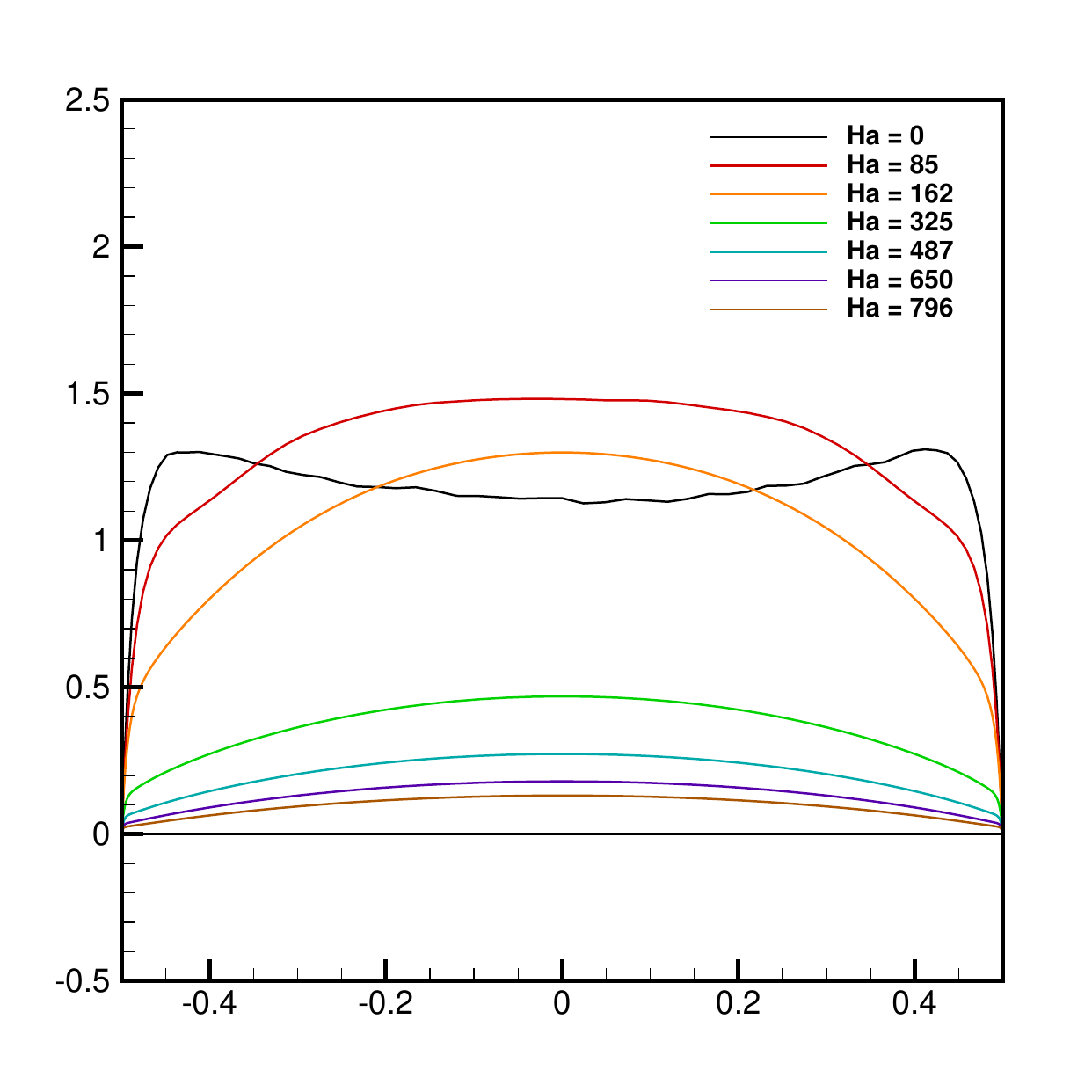}
    \axisput{50}{2}{Y}
    \axisput{2}{50}{\rotatebox[origin=c]{90}{$u_z$}}
\end{overpic}
\caption{Timed-averaged profiles of vertical velocity along the lines \xdataline~(left) and \ydataline~(right) shown in Fig.~\ref{fig:doamin}b. Results obtained for $C_w=0$ \emph{(a)} and the configuration A with $C_w = 0.01$ \emph{(b)} and $C_w = 0.1$ \emph{(c)} are shown for various $\Ha$.}
\label{fig:uz_ac_xy}
\end{figure*}

The observed variation of $\Nu$ and $E$ with $\Ha$ may appear to contradict to the known ability of an imposed magnetic field to suppress fluctuations of velocity. A monotonic decrease of both $\Nu$ and $E$ may appear more natural. An explanation of the effect is based on the role played by large-scale circulation in the flow field and heat transfer. As demonstrated in Figs.~\ref{fig:cont_ac_vert}, \ref{fig:cont_ac_hor} and \ref{fig:uz_ac_xy}, the flow is dominated by planar jets, an ascending one near the hot wall and a descending one near the cold wall. The jets form large-scale eddies in the $x$-$z$-plane. In fluctuating flow regimes, the jets and vortices are unsteady and exist as dominant features of the mean flow. 

The large-scale circulation may consist of just one eddy, with secondary smaller eddies in the corners (see Fig.~\ref{fig:cont_ac_vert}b), or several large eddies. The presence of multiple eddies increases the heat transfer rate via two mechanisms. One is the direct transport of heat by horizontal streams separating the eddies.  The second is that eddies not extending through the entire height of the box disrupt the thermal boundary layers and remove hot fluid from the hot wall and cold fluid from the cold wall early, effectively reducing the average thickness of the boundary layer and, thus, its thermal resistance. Another effect of the formation of multiple eddies is that it limits the range, where the vertical jets near the walls can accelerate, thus resulting in lower jet velocity and lower mean kinetic energy. 

\begin{figure*}
\centering
\begin{overpic}[width=0.41\textwidth,clip=]{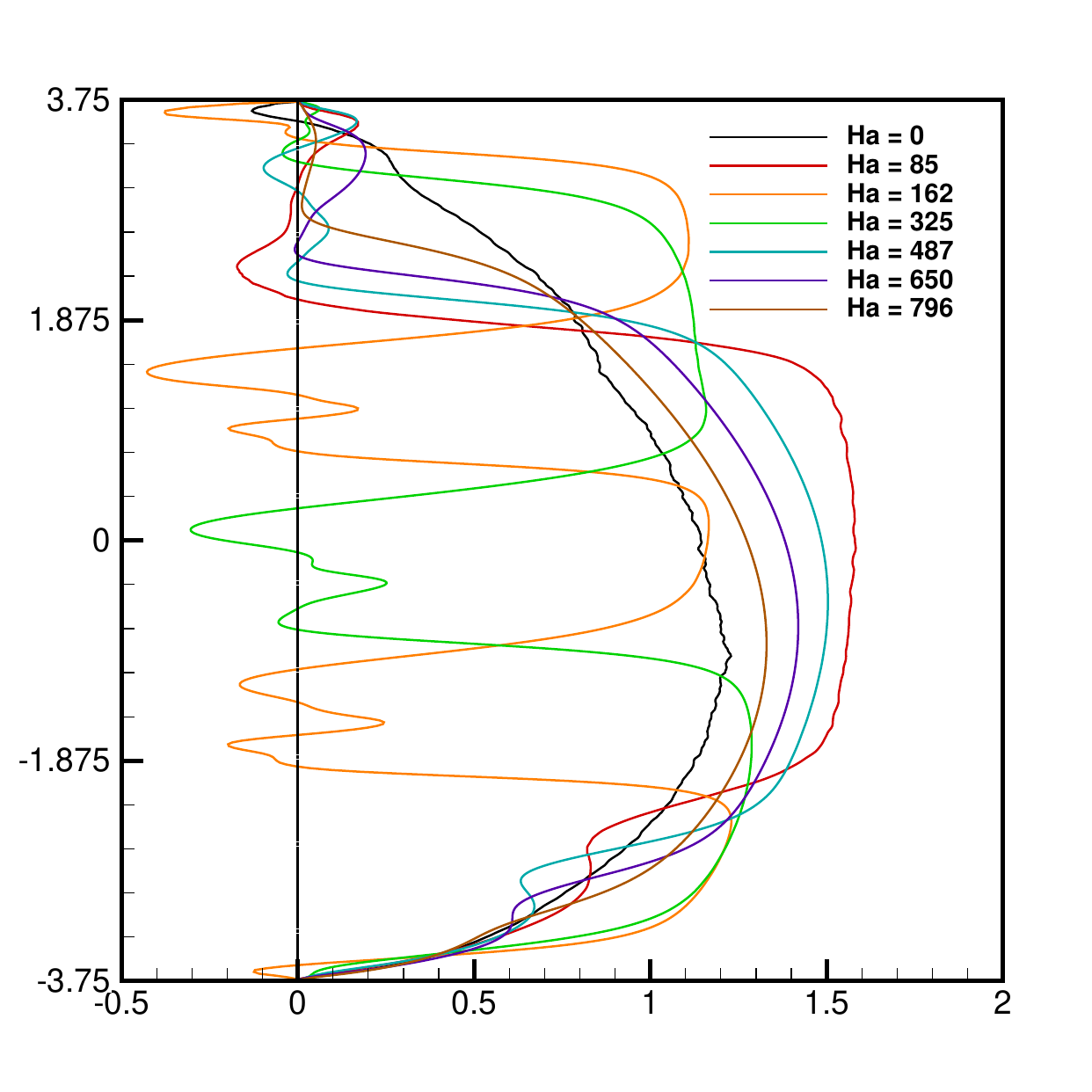}
    \titleput{2}{95}{(a)}
    \axisput{50}{2}{$u_z$}
    \axisput{2}{49}{\rotatebox[origin=c]{90}{Z}}
\end{overpic}\hfil\hfil
\begin{overpic}[width=0.41\textwidth,clip=]{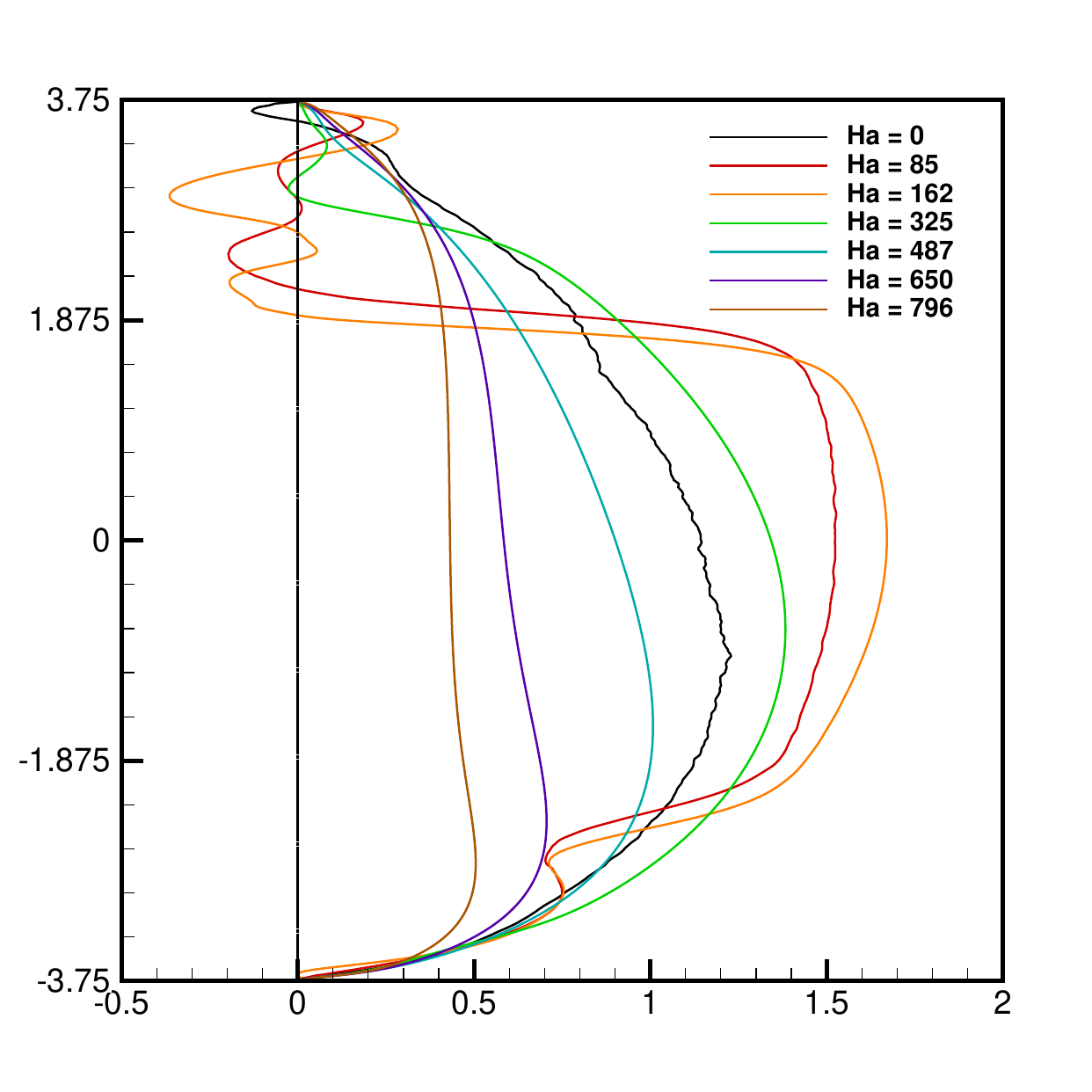}
    \titleput{2}{95}{(b)}
    \axisput{50}{2}{$u_z$}
    \axisput{2}{49}{\rotatebox[origin=c]{90}{Z}}
\end{overpic}\\
\begin{overpic}[width=0.41\textwidth,clip=]{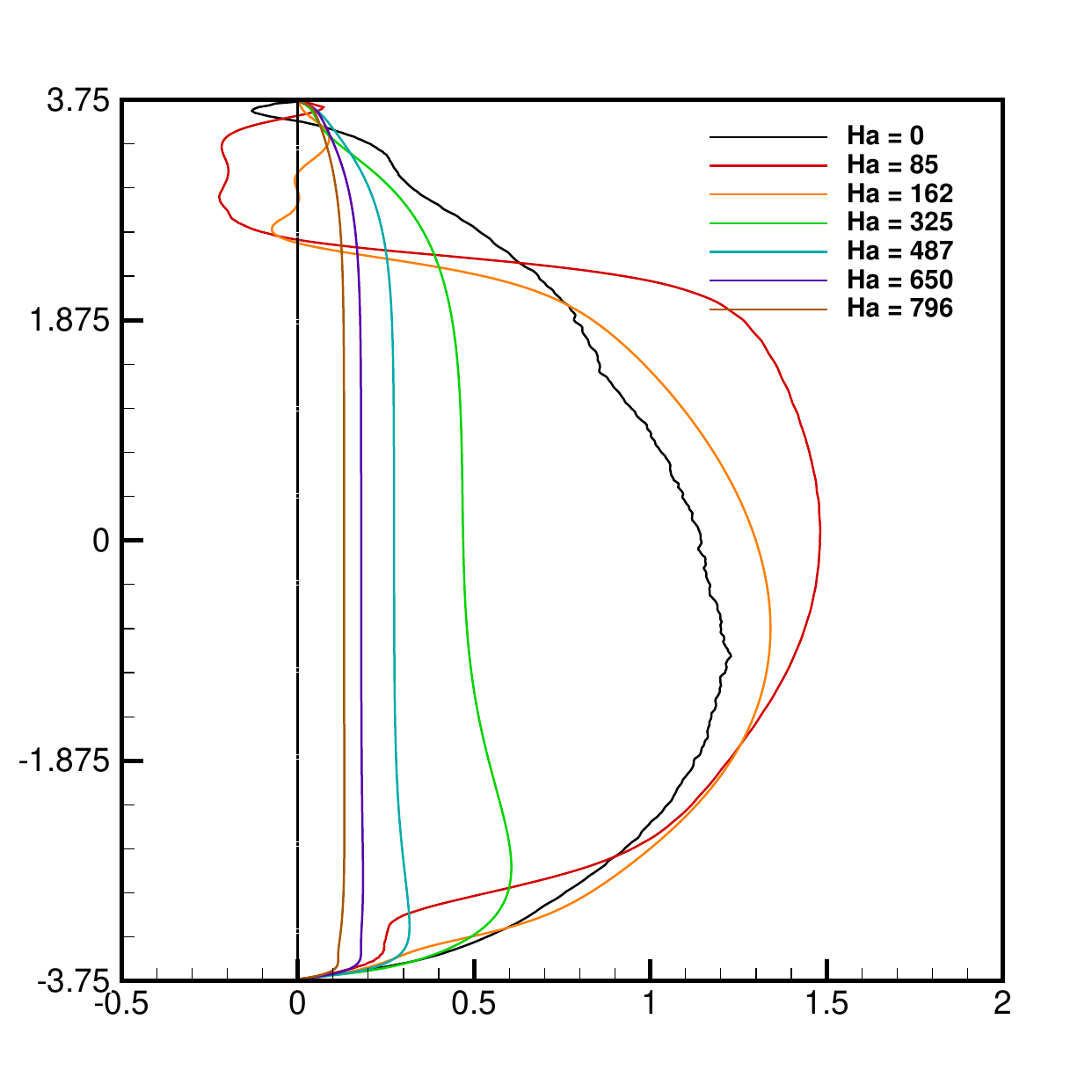}
    \titleput{2}{95}{(c)}
    \axisput{50}{2}{$u_z$}
    \axisput{2}{49}{\rotatebox[origin=c]{90}{Z}}
\end{overpic}\hfil\hfil
\begin{overpic}[width=0.41\textwidth,clip=]{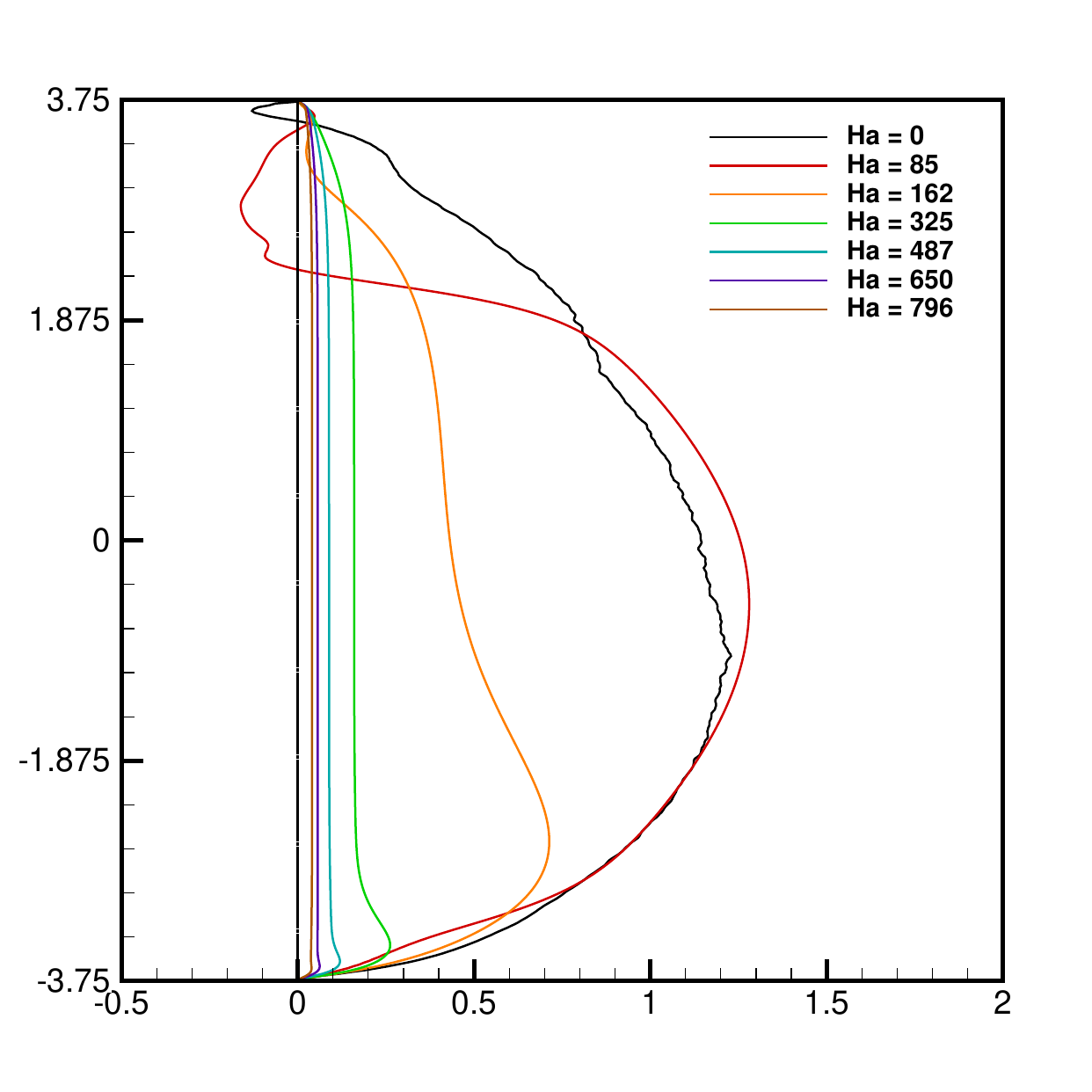}
    \titleput{2}{95}{(d)}
    \axisput{50}{2}{$u_z$}
    \axisput{2}{49}{\rotatebox[origin=c]{90}{Z}}
\end{overpic}
\caption{Timed-averaged profiles of vertical velocity along the line \zdataline~shown in Fig.~\ref{fig:doamin}b. Results obtained for $C_w=0$ \emph{(a)} and the configuration A with  $C_w = 0.01$ \emph{(b)} $C_w = 0.1$ \emph{(c)}, and $C_w = 1.0$ \emph{(d)} are shown for various $\Ha$. A black dotted line indicates $u_z=0$.}
\label{fig:uz_ac_z}
\end{figure*}

\begin{figure*}
\begin{overpic}[width=0.41\textwidth,clip=]{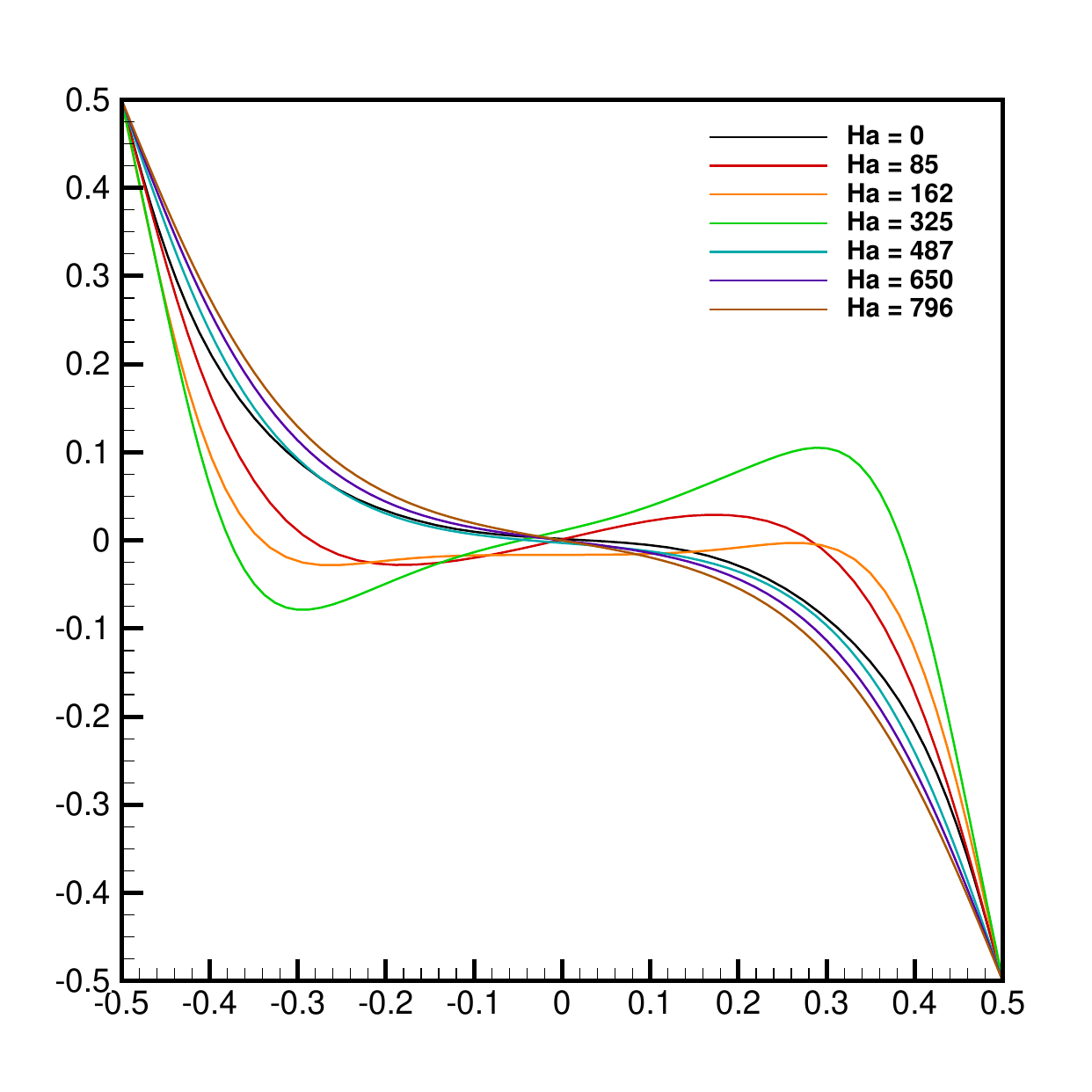}
    \titleput{2}{95}{(a)}
    \axisput{50}{2}{X}
    \axisput{2}{49}{\rotatebox[origin=c]{90}{T}}
\end{overpic}\hfil\hfil
\begin{overpic}[width=0.41\textwidth,clip=]{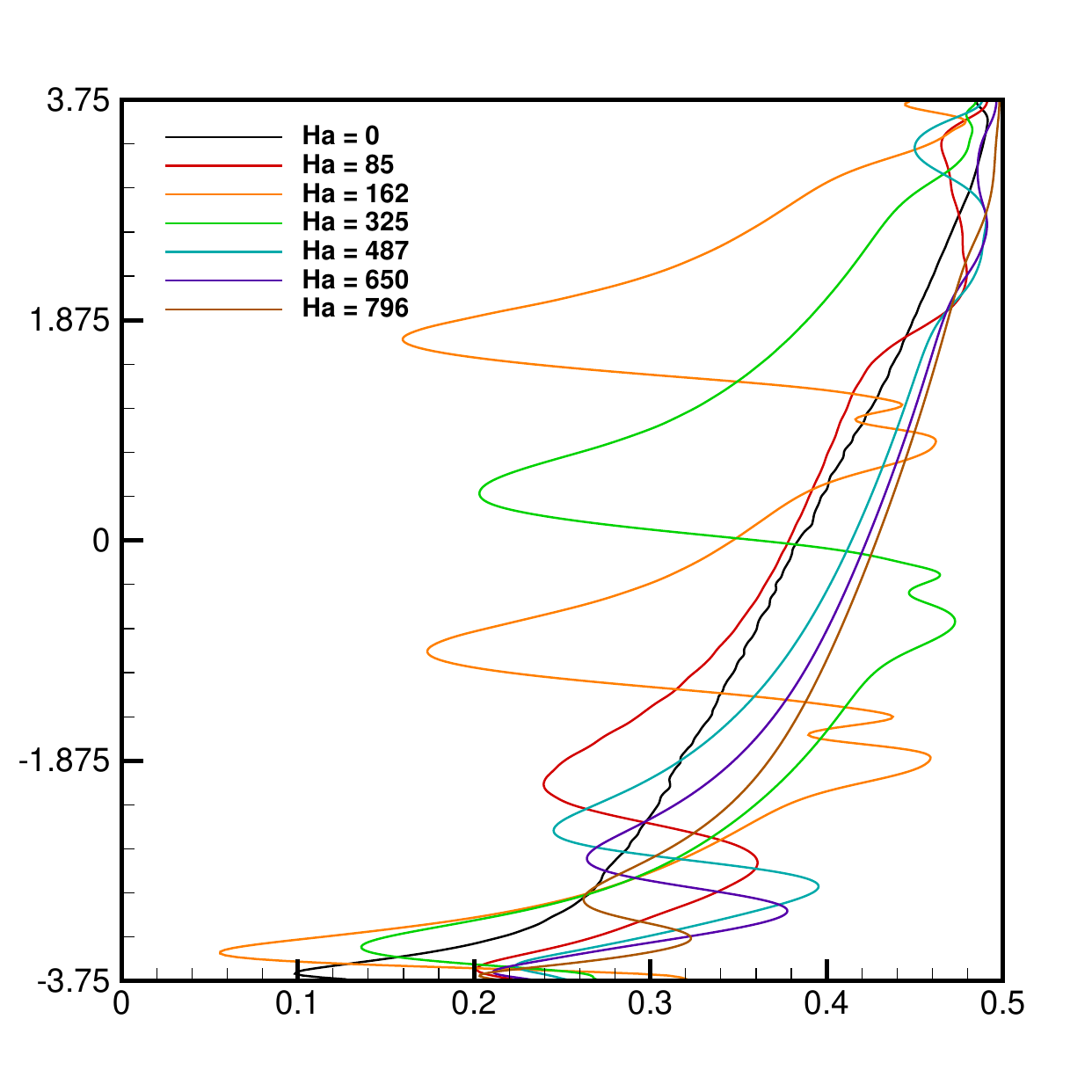}
    \axisput{50}{2}{T}
    \axisput{2}{49}{\rotatebox[origin=c]{90}{Z}}
\end{overpic}\\
\begin{overpic}[width=0.41\textwidth,clip=]{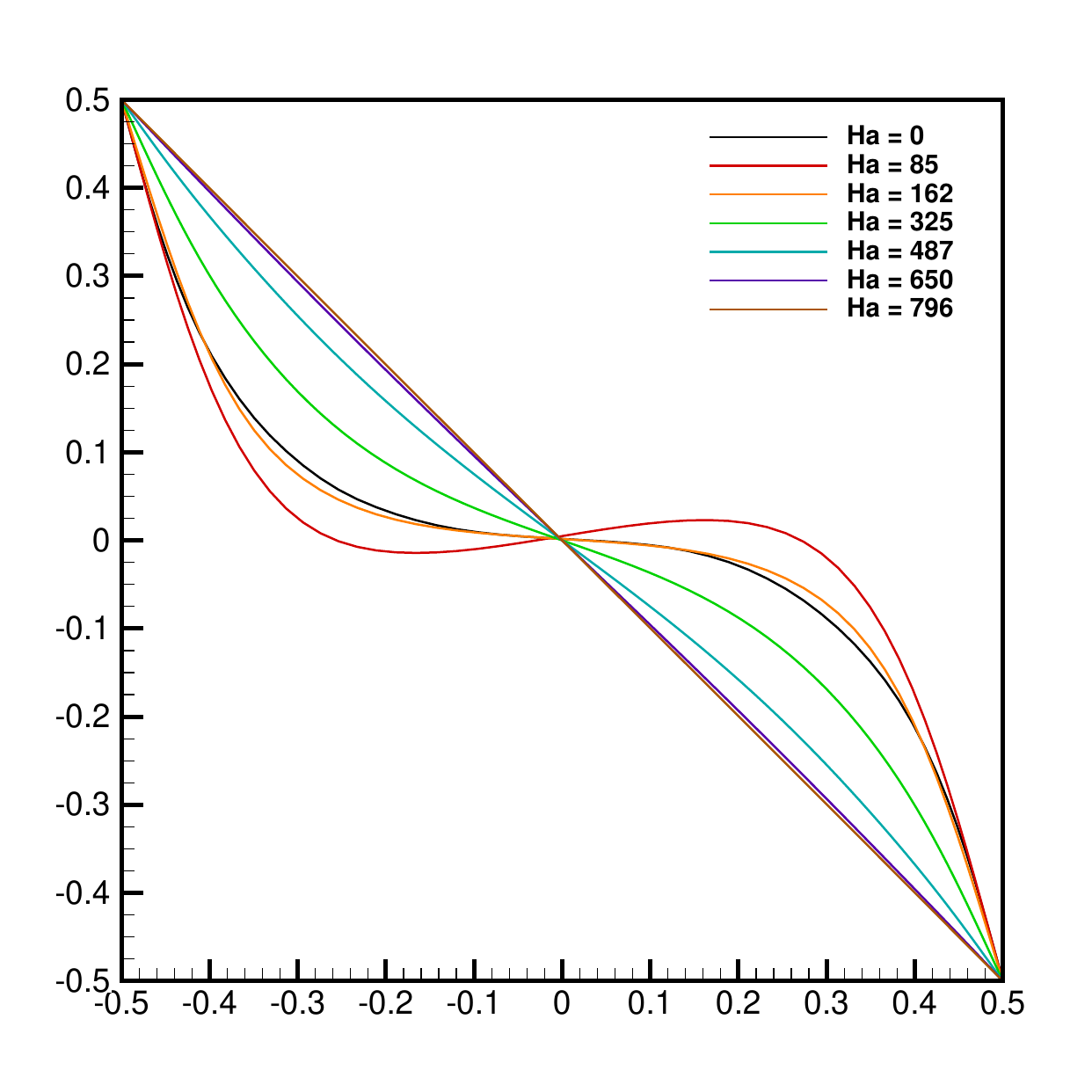}
    \titleput{2}{95}{(b)}
    \axisput{50}{2}{X}
    \axisput{2}{49}{\rotatebox[origin=c]{90}{T}}
\end{overpic}\hfil\hfil
\begin{overpic}[width=0.41\textwidth,clip=]{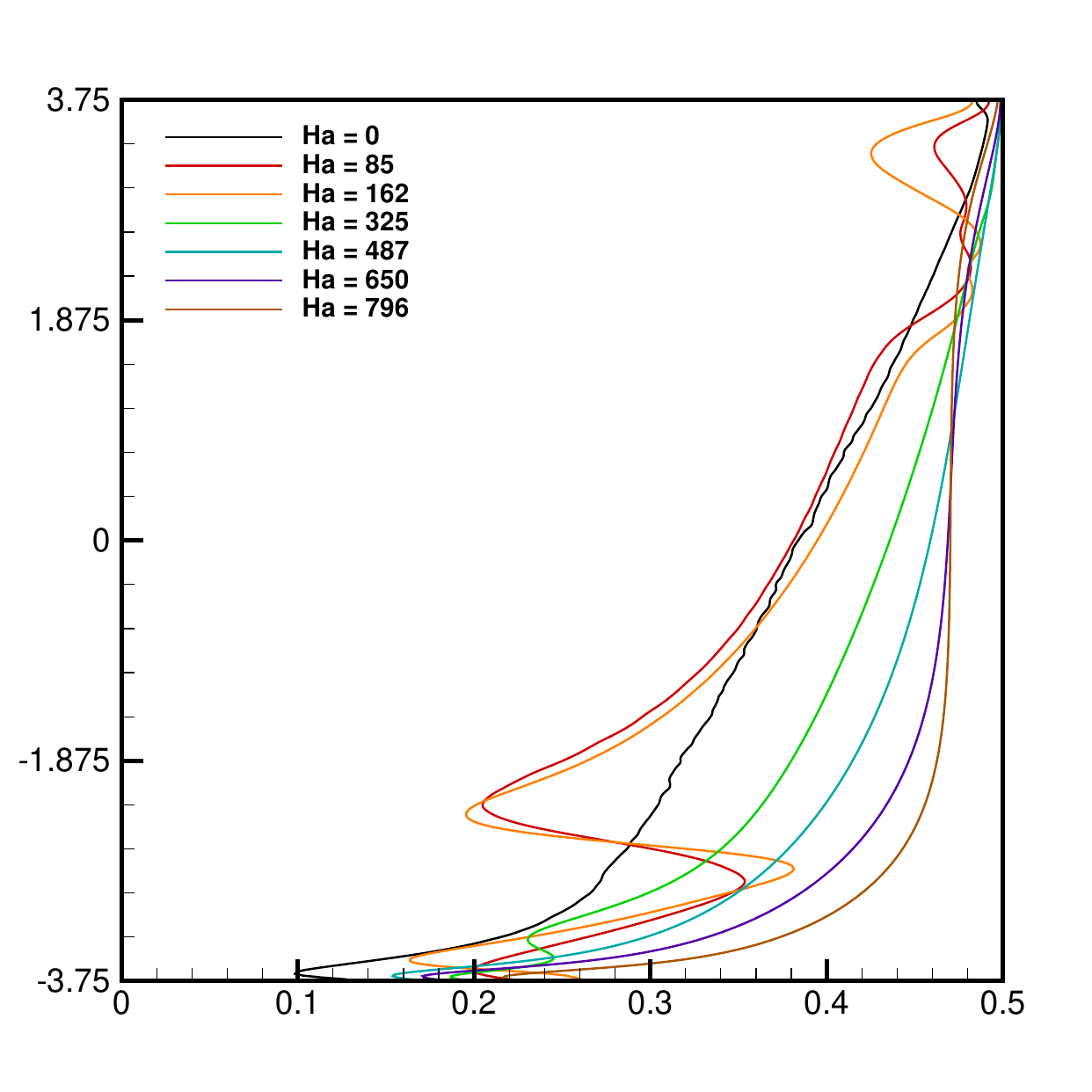}
    \axisput{50}{2}{T}
    \axisput{2}{49}{\rotatebox[origin=c]{90}{Z}}
\end{overpic}\\
\begin{overpic}[width=0.41\textwidth,clip=]{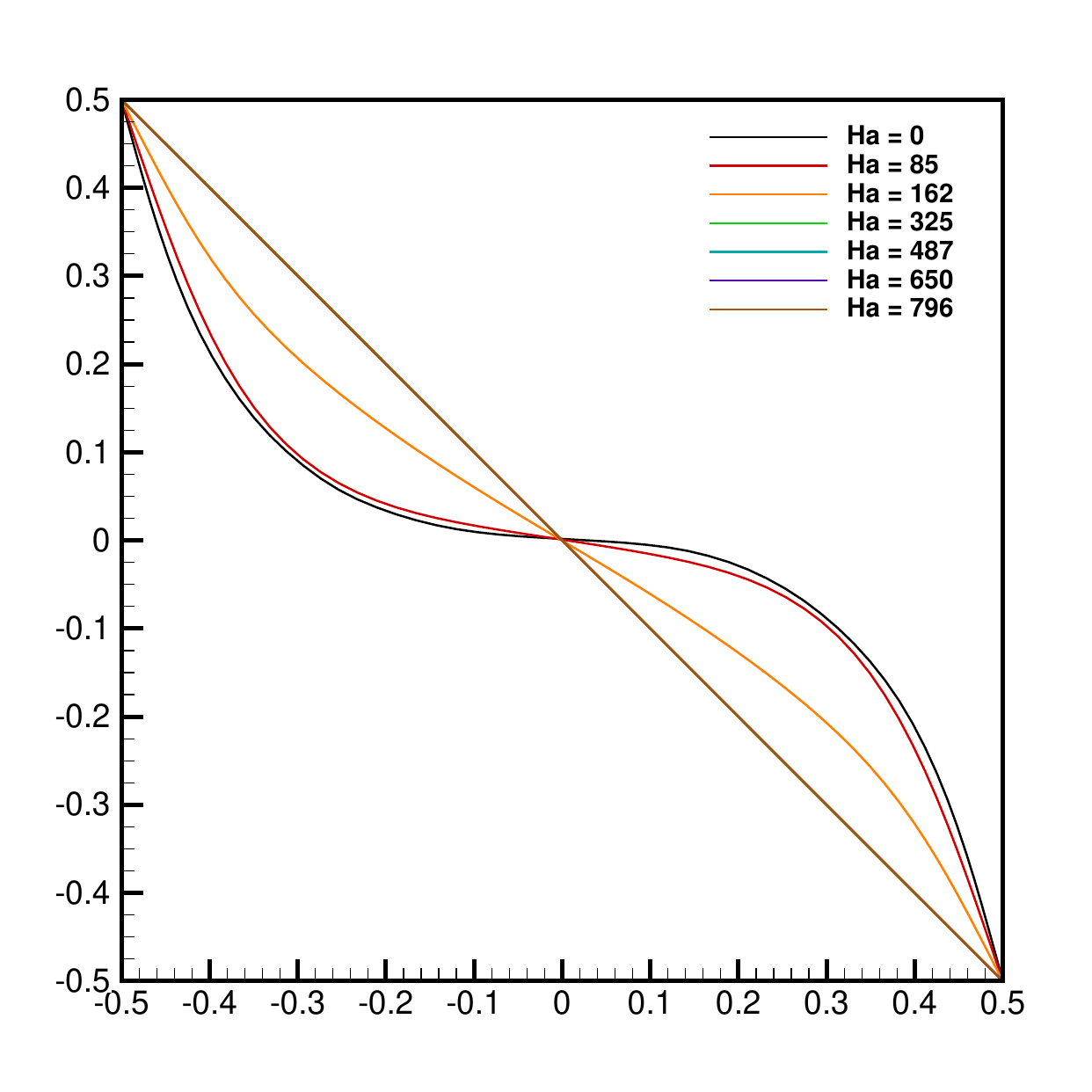}
    \titleput{2}{95}{(c)}
    \axisput{50}{2}{X}
    \axisput{2}{49}{\rotatebox[origin=c]{90}{T}}
\end{overpic}\hfil\hfil
\begin{overpic}[width=0.41\textwidth,clip=]{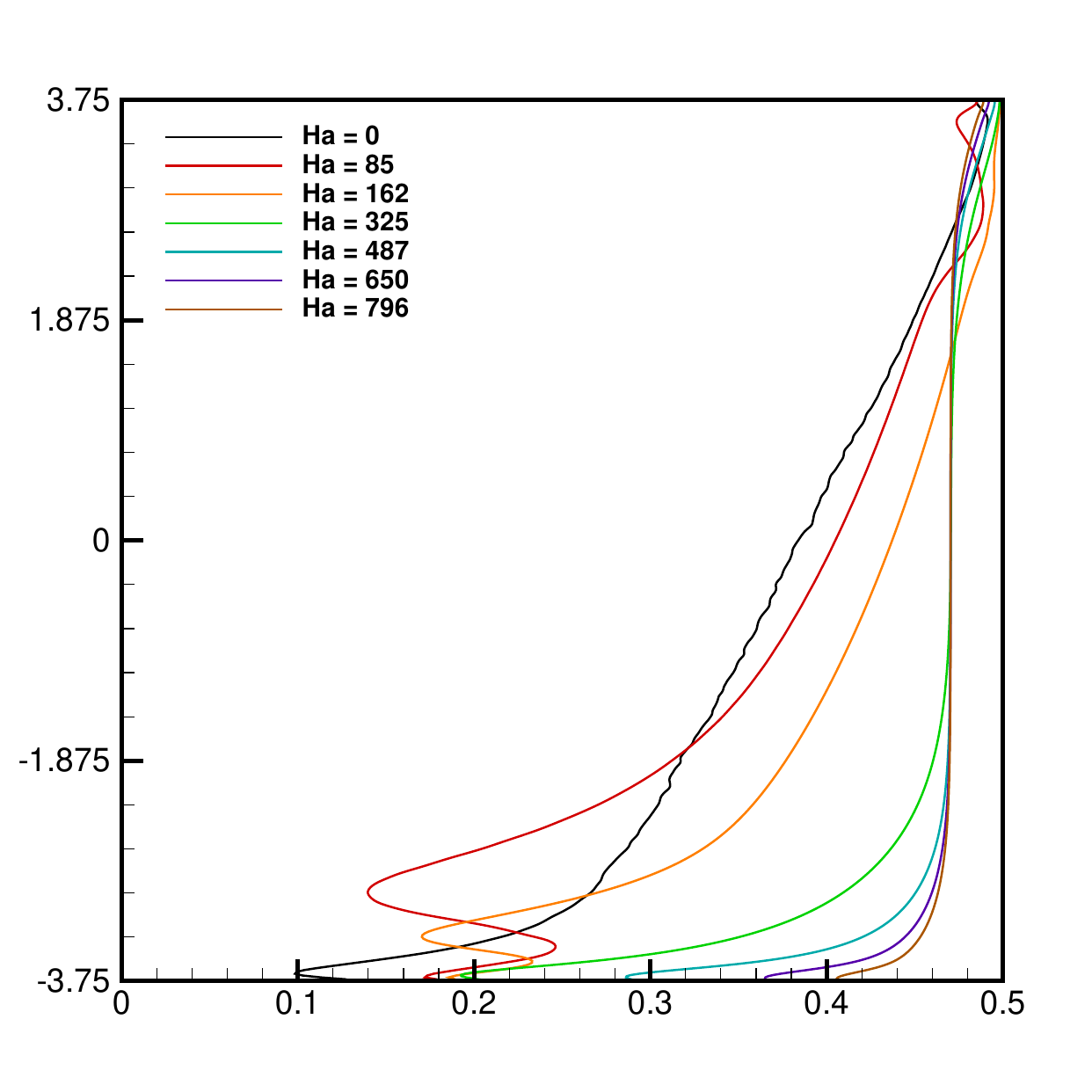}
    \axisput{50}{2}{T}
    \axisput{2}{49}{\rotatebox[origin=c]{90}{Z}}
\end{overpic}
\caption{Timed-averaged profiles of temperature along the lines \xdataline~(left) and \zdataline~(right) shown in Fig.~\ref{fig:doamin}b. Results obtained for $C_w=0$ \emph{(a)} and the configuration A with $C_w = 0.01$ \emph{(b)} and $C_w = 0.1$ \emph{(c)} are shown for various $\Ha$.}
\label{fig:t_ac}
\end{figure*}

The direct effect of an imposed magnetic field on the flow is two-fold. Firstly, the flow becomes  anisotropic and, at high $\Ha$, quasi-two-dimensional (see Fig.~\ref{fig:uz_ac_xy}a). Another effect is the suppression of velocity fluctuations leading to stabilization of the near-wall jets and large-scale circulation eddies. Indirectly, the imposed magnetic field also affects the pattern of the large-scale circulation, which explains the change of values of $\Nu$ and $E$ visible in Tables \ref{tbl:Nu_ac}, \ref{tbl:E_ac} and Fig.~\ref{fig:NU_E_ac}. The change of the pattern is presented in Figure 12 of Ref.~\onlinecite{krasnov2023tensor} and further illustrated by the vertical profiles of velocity and temperature in Figs.~\ref{fig:uz_ac_z}a and \ref{fig:t_ac}a. The mean flow has one large-scale circulation vortex at $\Ha=0$ and $\Ha=85$. Three circulation vortices appear at $\Ha=162$, which causes strong increase of $\Nu$ and drop of $E$. Two circulation vortices are found at $\Ha=325$. Circulation patterns with just one large eddy and gradually decreasing strength of the mean flow are found at $\Ha\ge 487$. 

A comment is in order concerning the two flow regimes, with one and two major circulation eddies, found in Ref.~\onlinecite{krasnov2023tensor} at $\Ha=400$ depending on the size of the grid. Only one grid identical to the finer grid of Ref.~\onlinecite{krasnov2023tensor} is used in this study. Accordingly, only the two-eddy solution with higher $\Nu$ and lower $E$ is reported here at $\Ha=400$.

\subsubsection{All walls are electrically conducting}
\label{sec:results:allWalls:c>0}
When all walls are electrically conducting, the effect of an imposed magnetic field on the flow is greatly simplified. It can be summarized as a straightforward suppression. One aspect of that is the suppression of velocity fluctuations. It can be seen in Fig.~\ref{fig:time_NU_E_ac} and Tables \ref{tbl:Nu_ac} and \ref{tbl:E_ac} that at all the considered values of $C_w$ there is a threshold value of $\Ha$ above which the fluctuations disappear in a fully developed flow. Even at the lowest $C_w=0.01$, the flow is steady-state at $\Ha\ge 400$. The threshold decreases with increasing $C_w$ ($\Ha\ge 240$ at $C_w=0.1$ and $\Ha\ge 162$ at $C_w=1$ and 50). When the fully developed flow remains unsteady, the amplitude of velocity fluctuations monotonically decreases with increasing $C_w$ or $\Ha$. 

Another aspect of the magnetic suppression is the transformation of the mean flow. The large-scale circulation consists of only one major eddy for all values $\Ha>0$ and $C_w>0$ (see Fig.~\ref{fig:cont_ac_vert} and Fig.~\ref{fig:uz_ac_z} for an illustration). Accordingly, an increase of $\Nu$ observed in the flow with $C_w=0$ at moderate values of $\Ha$ and explained earlier by formation of a multi-eddy circulation pattern does not appear in flows with electrically insulating walls. Both $\Nu$ and $E$ decrease monotonically with increasing $\Ha$ or $C_w$. At $C_w\ge 0.1$ and large $\Ha$, $E\sim 10^{-5}-10^{-3}$ and $\Nu$ is close to the pure conduction value 1.0 (see Tables \ref{tbl:Nu_ac} and \ref{tbl:E_ac} and Fig.~\ref{fig:NU_E_ac}). This effect is further illustrated by the velocity and temperature profiles in Figs.~\ref{fig:uz_ac_xy}-\ref{fig:t_ac}. We see in Figs.~\ref{fig:uz_ac_xy} and \ref{fig:uz_ac_z} that velocity of near-wall jets decreases as either $\Ha$ or $C_w$ increases. For example, changing $C_w$ from 0 to 0.01 reduces the maximum jet velocity by about 10\% at $\Ha=85$ and by about 50\% and 70\% at $\Ha=487$ and 796, respectively. Weakened convective heat transfer results in temperature distributions approaching pure conduction distributions with exception of the regions near the top and bottom walls, where a horizontal flow persists (see Figs.~\ref{fig:cont_ac_vert} and \ref{fig:t_ac}). 

One apparent exception from the just described transformation is the case of the flows with $C_w=0.01$ and $\Ha$ increasing from 85 to 162. Fig.~\ref{fig:uz_ac_z}b shows increase of the maximum vertical velocity in the core of the jet. Analysis of three-dimensional velocity distributions, however, demonstrates the accompanying  decrease of the jet's width (see Fig.~\ref{fig:uz_ac_xy}b), so the values of $\Nu$ are nearly the same at $\Ha=85$ and 162, and $E$ is substantially smaller at $\Ha=162$ (see Tables  \ref{tbl:Nu_ac} and \ref{tbl:E_ac}).

An important observation can be made based on the velocity distributions shown in Figs.~\ref{fig:cont_ac_vert} and \ref{fig:cont_ac_hor} and the velocity profiles in the right-hand side columns of Fig.~\ref{fig:uz_ac_xy}b,c. The assumption of quasi-two-dimensionality often used in the analysis of high-$\Ha$ MHD flows is generally inaccurate for the flows considered here. One sees in Fig.~\ref{fig:uz_ac_xy}c, that even in the case with the highest $\Ha = 796$ the velocity variation in the core flow along the magnetic field line (the $y$-direction) is non-negligible. Similar behavior is demonstrated by the profiles of the mean velocity at all $C_w>0$. The curvature is higher at moderately high values of $\Ha$, such as $\Ha=400$ or 325 (see Fig.~\ref{fig:uz_ac_xy}).  The phenomenon can be attributed to the large ratio between the $y$- and $x$-dimensions of the horizontal cross-section of the upward and downward jets. Similar behavior is observed in flows within narrow ducts with magnetic fields in the transverse direction parallel to the longer side (see, e.g., Ref.~\onlinecite{Krasnov:2010} for an illustration).

\subsection{Configuration B: only hot and cold walls are electrically conducting}
\label{sec:results:HCWalls}

In the flows considered in this section, only the hot and cold walls may have $C_w>0$. The other two vertical walls and the top and bottom walls of the box are electrically perfectly insulating. The configuration is partially motivated by the interesting physics of the magnetic field effect in such flows, which we will present below. The other part of the motivation is practical. In laboratory or technological convection flows, the hot and cold walls have to be made of metal, typically copper or aluminum. Unless they are intentionally electrically insulated (e.g., by a layer of epoxy, as in the experiments\cite{Authie:2002,Authie:2003}), their electric conductance in non-negligible.

\begin{figure}
\centering
\begin{overpic}[width=0.48\textwidth,clip=]{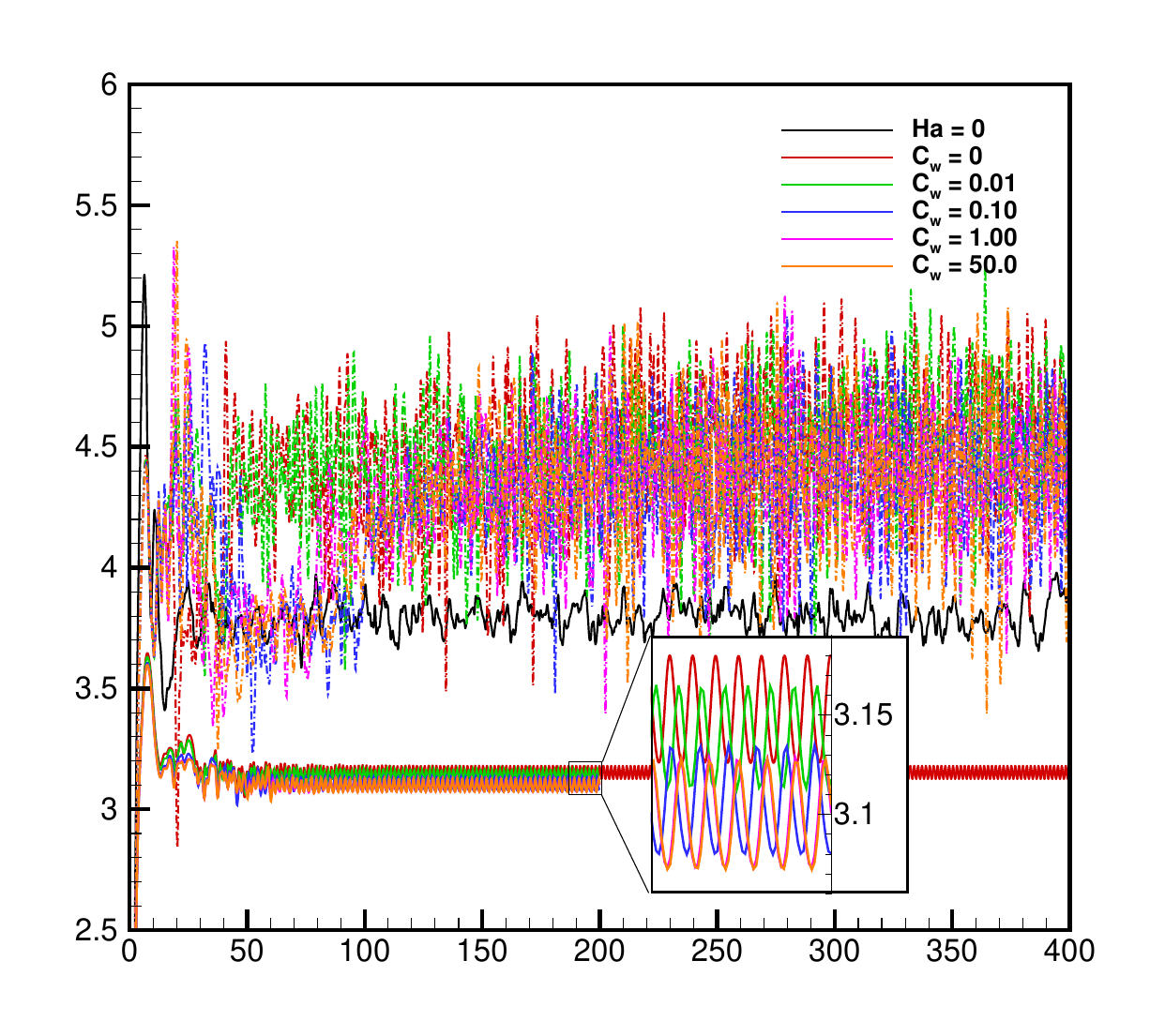}
    \titleput{1}{85}{(a)}
    \axisput{48}{2}{time}
    \axisput{1}{42}{\rotatebox[origin=c]{90}{$\Nu (t)$}}
\end{overpic}
\begin{overpic}[width=0.48\textwidth,clip=]{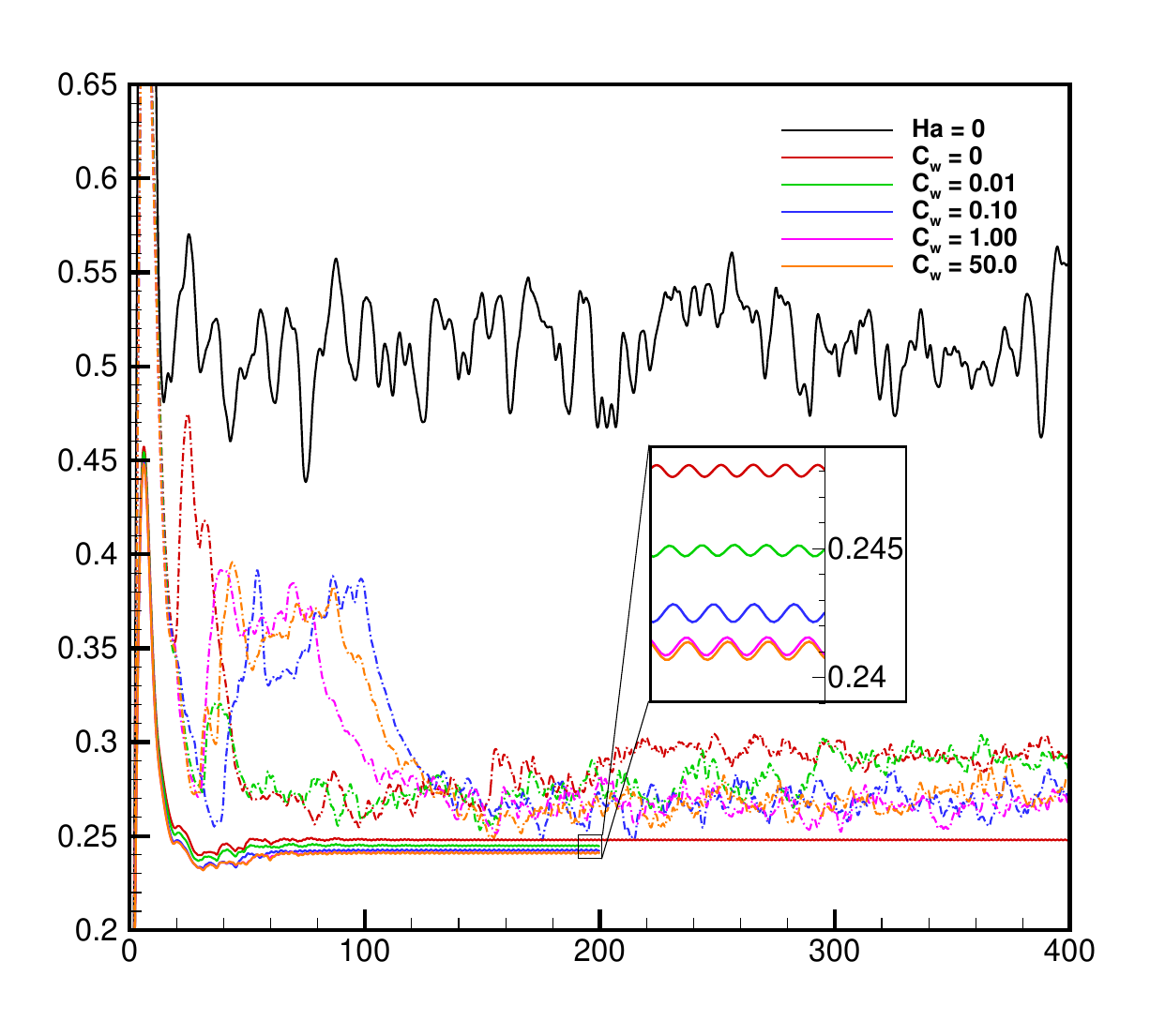}
    \titleput{1}{85}{(b)}
    \axisput{48}{2}{time}
    \axisput{1}{42}{\rotatebox[origin=c]{90}{$E (t)$}}
\end{overpic}
\caption{\emph{(a)} $\Nu(t)$ and \emph{(b)} $E(t)$  (see  \eqref{eqn:Nu} and \eqref{eqn:TKE}) for $\Ha = 325$ (dashed line) and $\Ha = 650$ (solid line) for the configuration B of the wall conductivity conditions. Black solid line represents the case  $\Ha = 0$.}
\label{fig:time_NU_E_tc}
\end{figure}

Fig.~\ref{fig:time_NU_E_tc} shows time signals $\Nu(t)$ \eqref{eqn:Nu} and $E(t)$ \eqref{eqn:TKE} during the entire simulations for flows at moderately high and high values of $\Ha$. It can be seen that unlike the flows with all walls electrically conducting, the fluctuations are only mildly affected by the magnetic field. The flows remain time-dependent at all $0\le C_w\le 50$ and all, but the largest value of $\Ha$. At a given value of $\Ha$ the amplitude of the fluctuations is not significantly affected by $C_w$. This is illustrated for $\Ha=325$ and 650 in Fig.~\ref{fig:time_NU_E_tc}, and is true for all the explored values of $\Ha$.

\begin{figure}
\begin{overpic}[width=0.48\textwidth,clip=]{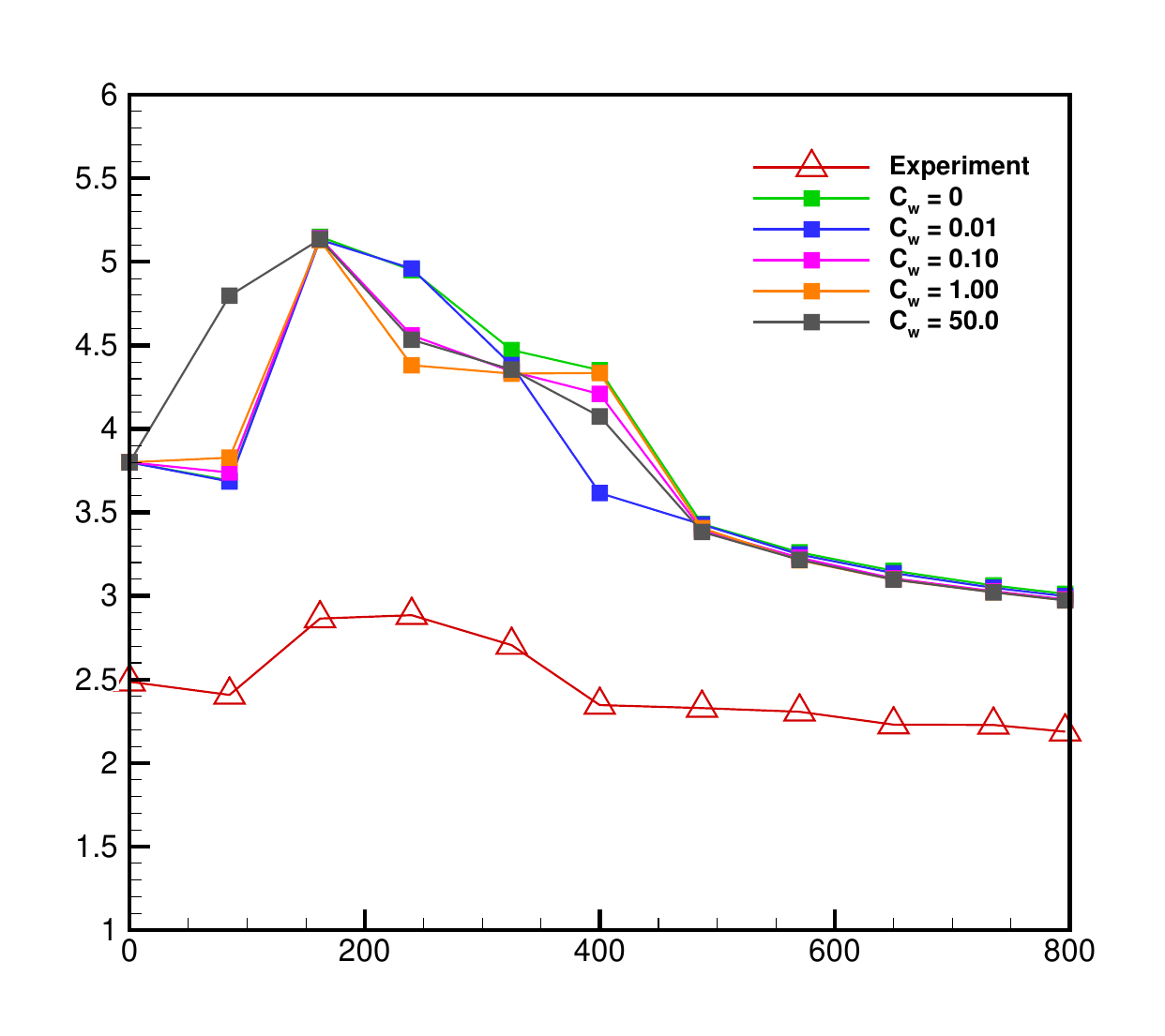}
    \titleput{2}{84}{(a)}
    \axisput{49}{2}{$Ha$}
    \axisput{2}{44}{\rotatebox[origin=c]{90}{$Nu$}}
\end{overpic}
\begin{overpic}[width=0.48\textwidth,clip=]{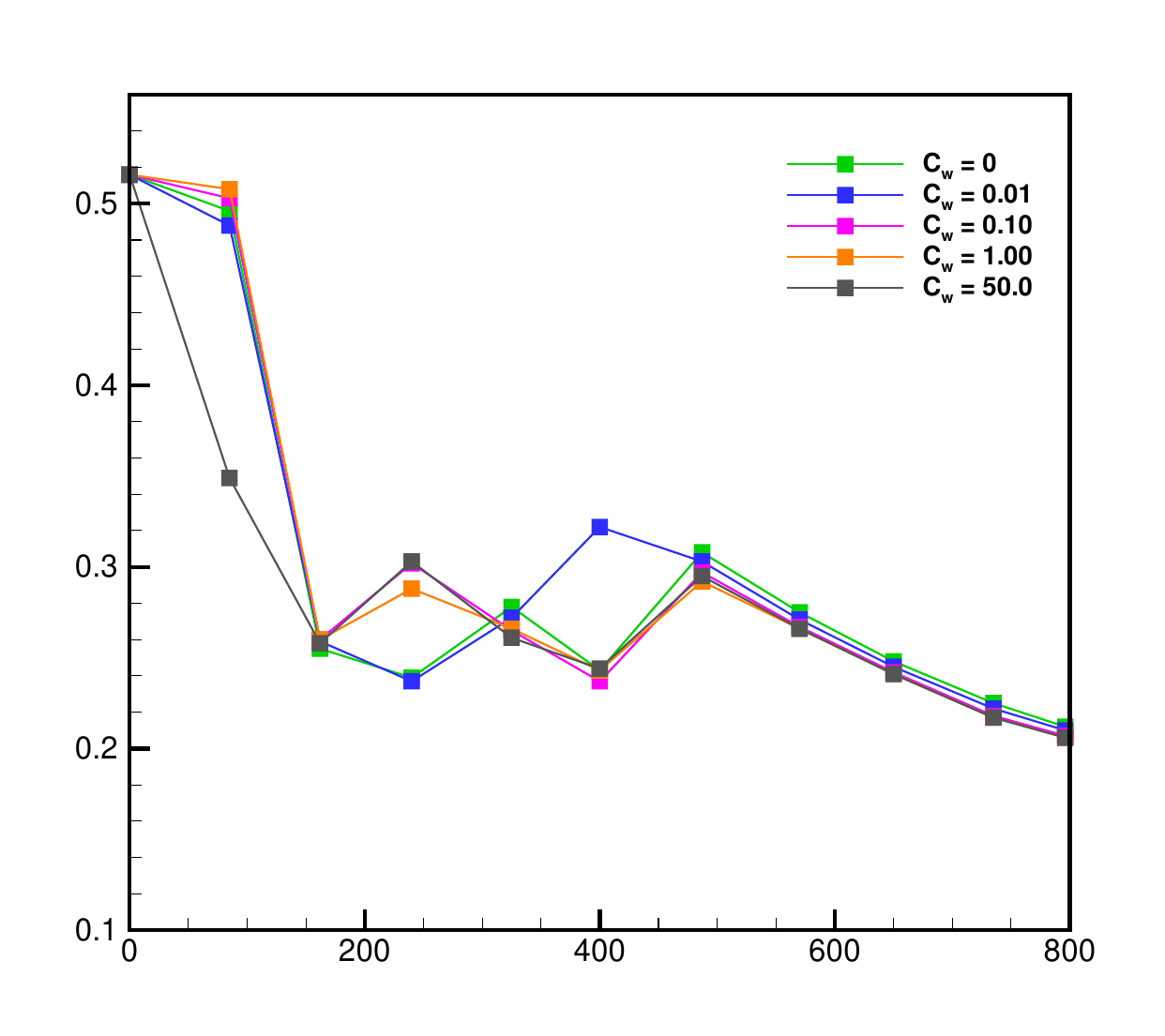}
    \titleput{2}{84}{(b)}
    \axisput{49}{2}{$Ha$}
    \axisput{2}{44}{\rotatebox[origin=c]{90}{$E$}}
\end{overpic}
\caption{Time averaged values of $\Nu$ and $E$ for configuration B (only hot and cold walls are electrically conducting). Data for the case of all walls electrically insulating and the experimental data\cite{Authie:2003} (available only for $\Nu$) are shown for comparison.
}
\label{fig:NU_E_tc}
\end{figure}

\begin{figure*}
\centering
\begin{overpic}[width=0.115\textwidth,clip=]{cont_h0_0_xz_no-title_final}
    \titleput{1}{95}{(a)}
    \axisput{15.5}{1.5}{X}
    \axisput{2}{48}{\rotatebox[origin=c]{90}{Z}}
\end{overpic}
\begin{overpic}[width=0.115\textwidth,clip=]{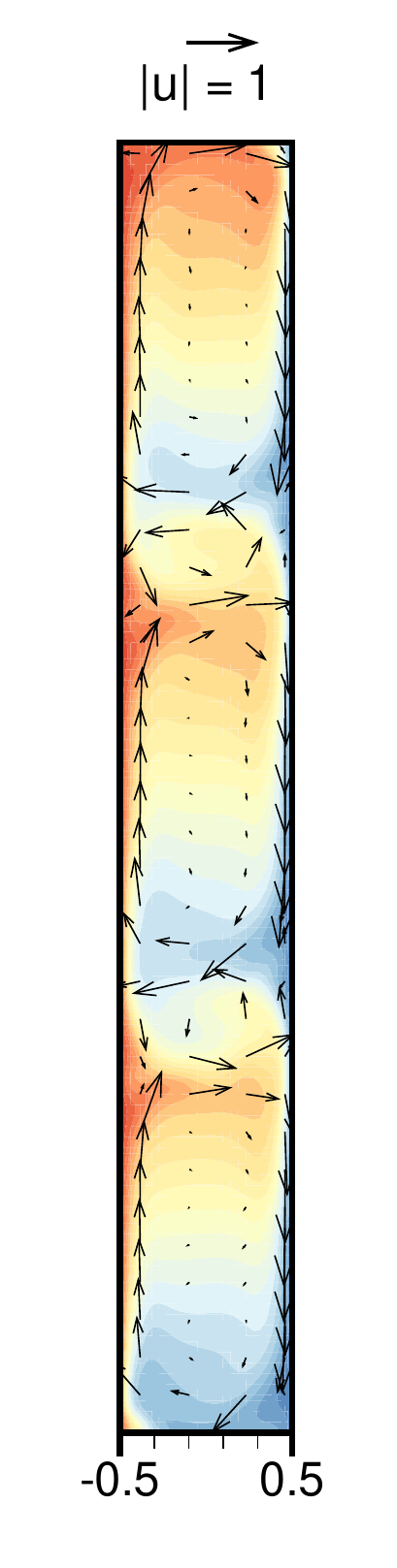}
    \titleput{1}{95}{(b)}
    \axisput{11.5}{1.5}{X}
\end{overpic}
\begin{overpic}[width=0.115\textwidth,clip=]{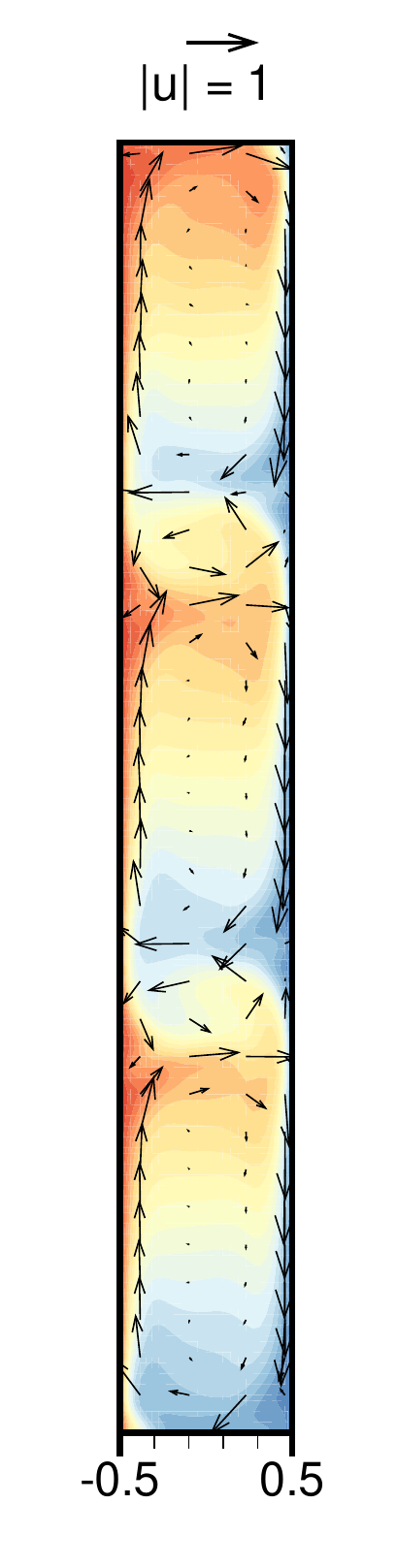}
    \titleput{1}{95}{(c)}
    \axisput{11.5}{1.5}{X}
\end{overpic}
\begin{overpic}[width=0.115\textwidth,clip=]{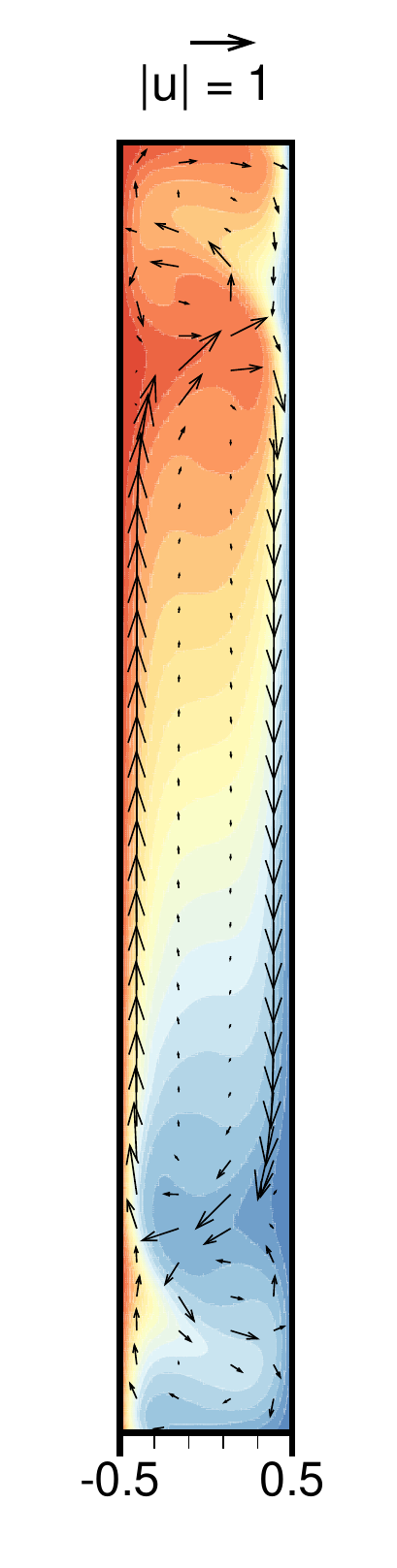}
    \titleput{1}{95}{(d)}
    \axisput{11.5}{1.5}{X}
\end{overpic}
\begin{overpic}[width=0.115\textwidth,clip=]{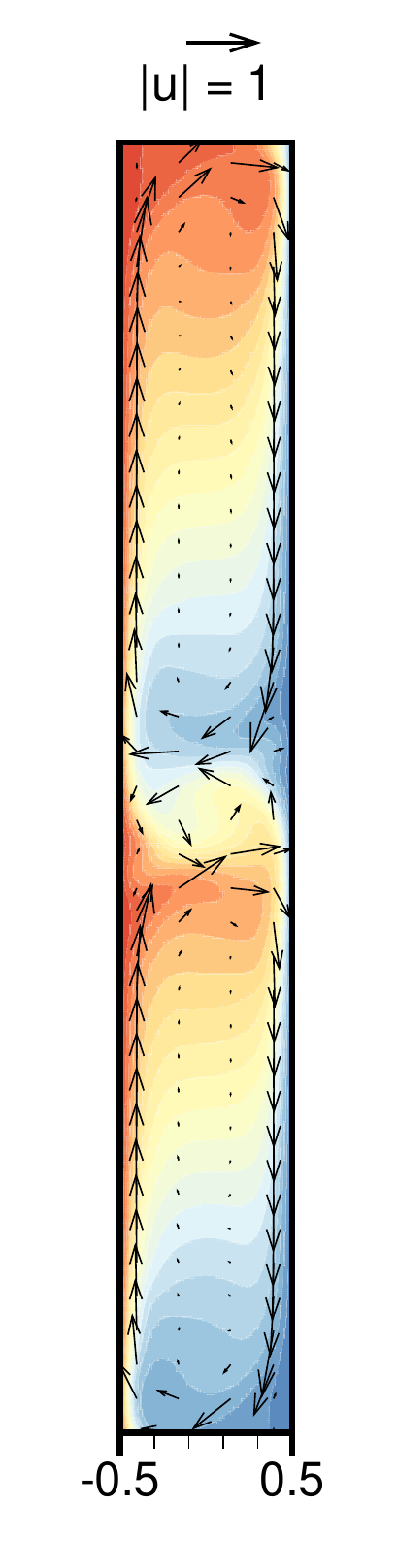}
    \titleput{1}{95}{(e)}
    \axisput{11.5}{1.5}{X}
\end{overpic}
\begin{overpic}[width=0.115\textwidth,clip=]{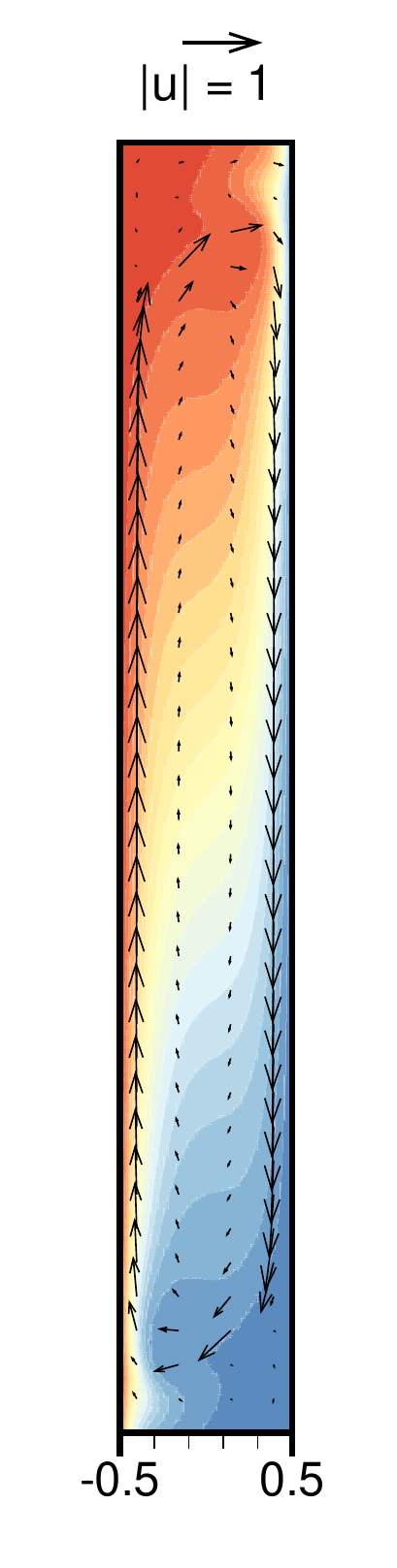}
    \titleput{1}{95}{(f)}
    \axisput{11.5}{1.5}{X}
\end{overpic}
\begin{overpic}[width=0.23\textwidth,clip=]{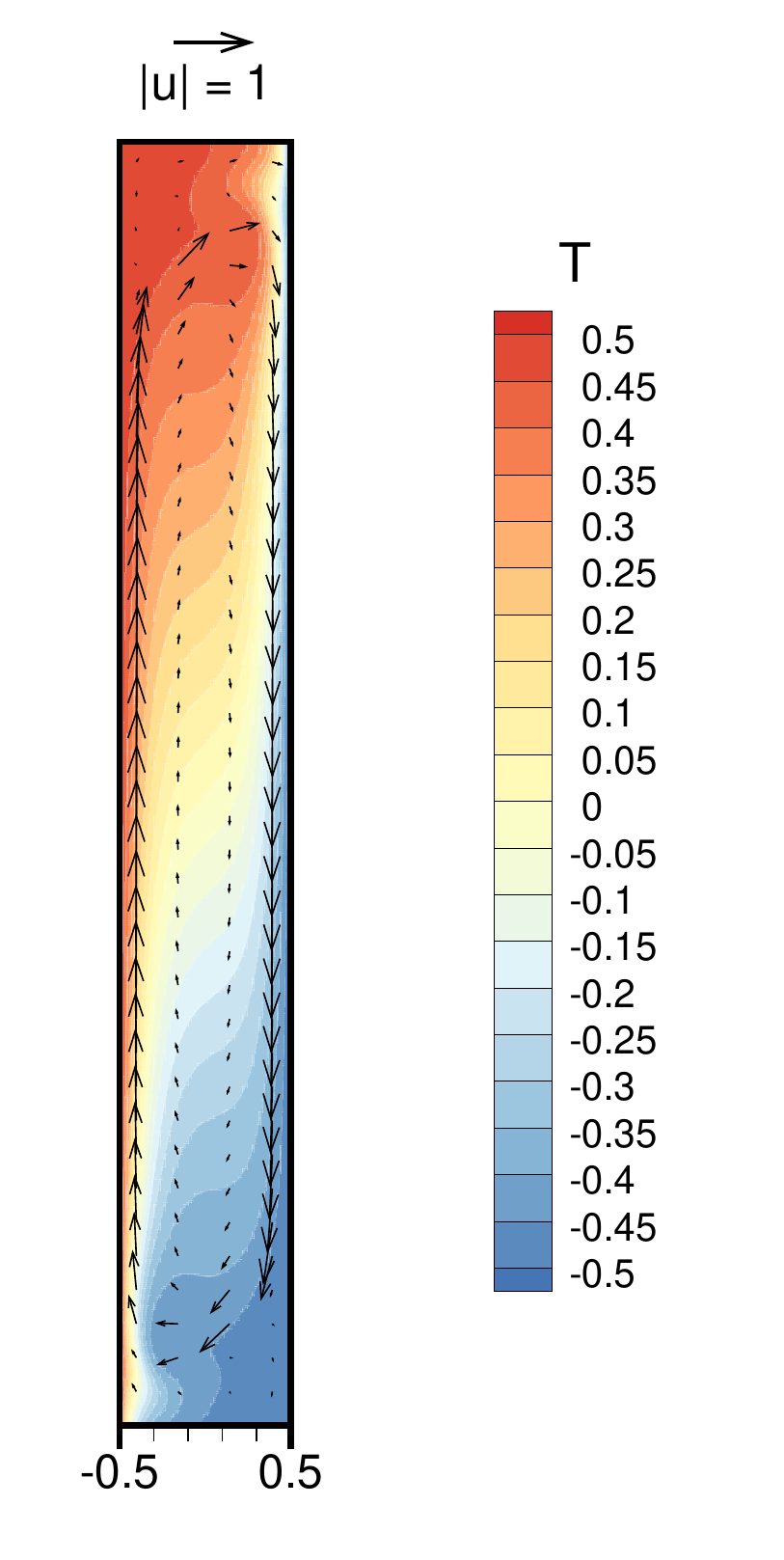}
    \titleput{-0.5}{95}{(g)}
    \axisput{11.5}{1.5}{X}
\end{overpic}
\begin{overpic}[width=0.115\textwidth,clip=]{cont_h0_0_yz_no-title_final}
    \axisput{15.5}{1.5}{Y}
    \axisput{2}{48}{\rotatebox[origin=c]{90}{Z}}
\end{overpic}
\begin{overpic}[width=0.115\textwidth,clip=]{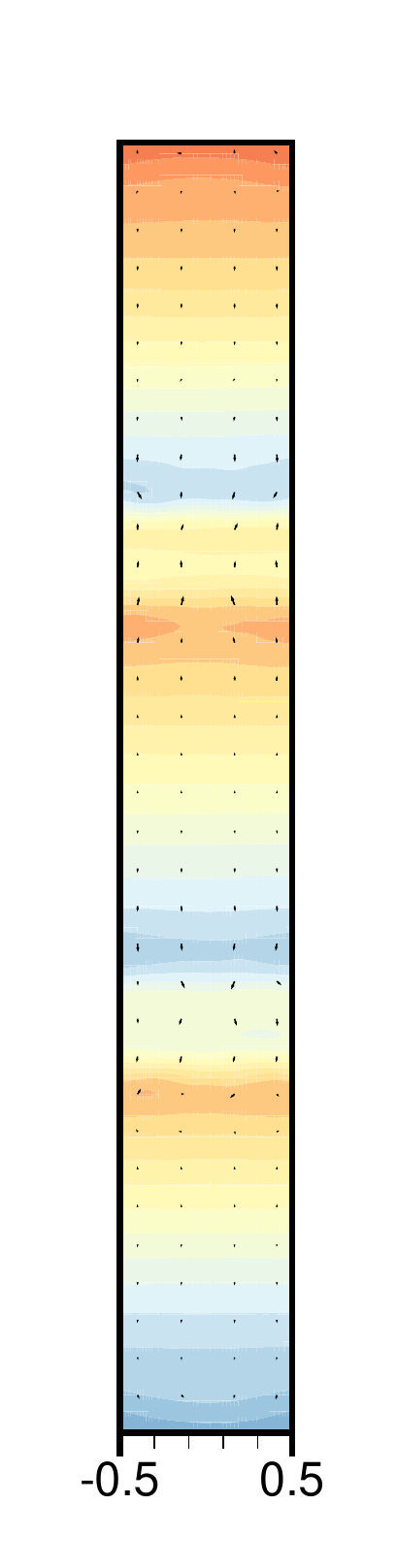}
    \axisput{11.5}{1.5}{Y}
\end{overpic}
\begin{overpic}[width=0.115\textwidth,clip=]{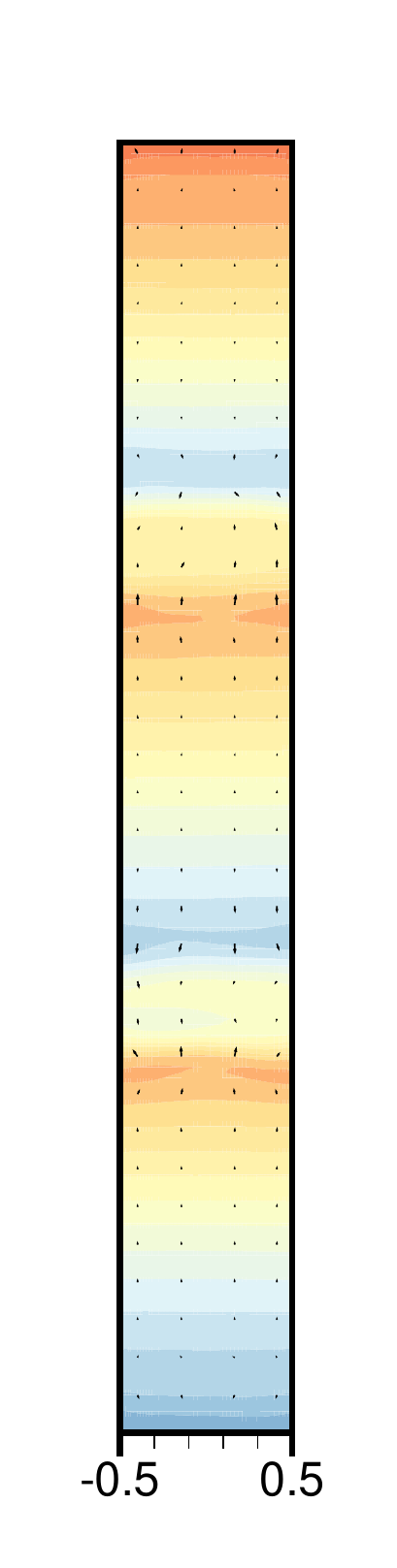}
    \axisput{11.5}{1.5}{Y}
\end{overpic}
\begin{overpic}[width=0.115\textwidth,clip=]{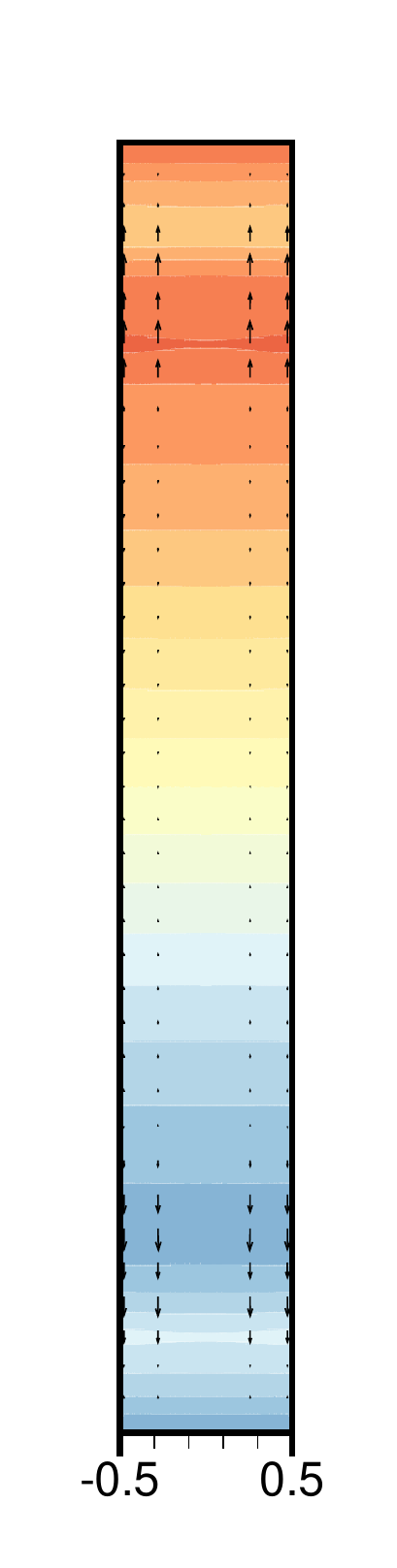}
    \axisput{11.5}{1.5}{Y}
\end{overpic}
\begin{overpic}[width=0.115\textwidth,clip=]{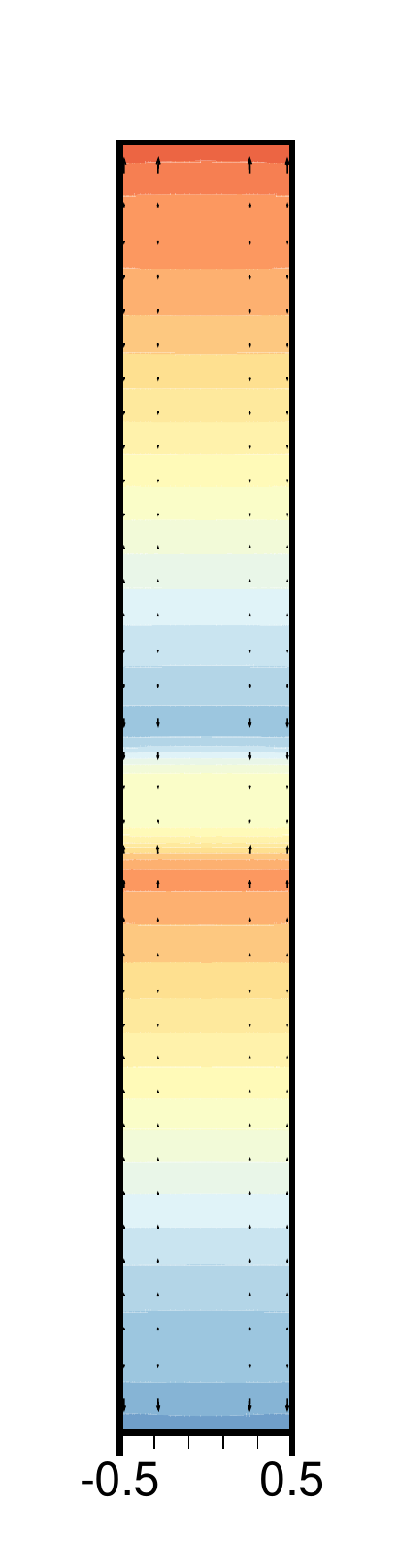}
    \axisput{11.5}{1.5}{Y}
\end{overpic}
\begin{overpic}[width=0.115\textwidth,clip=]{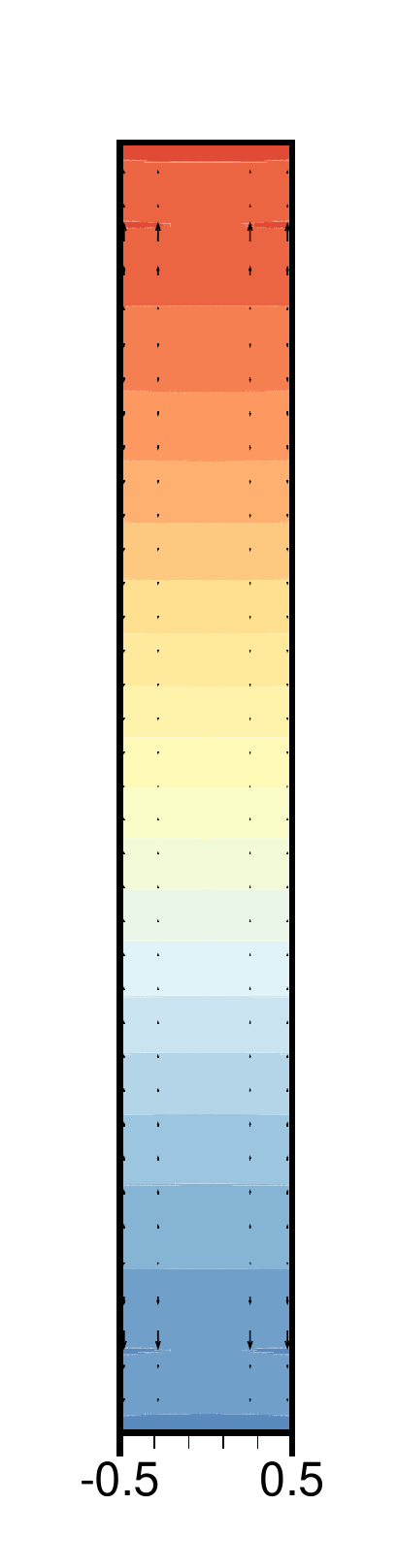}
    \axisput{11.5}{1.5}{Y}
\end{overpic}
\begin{overpic}[width=0.23\textwidth,clip=]{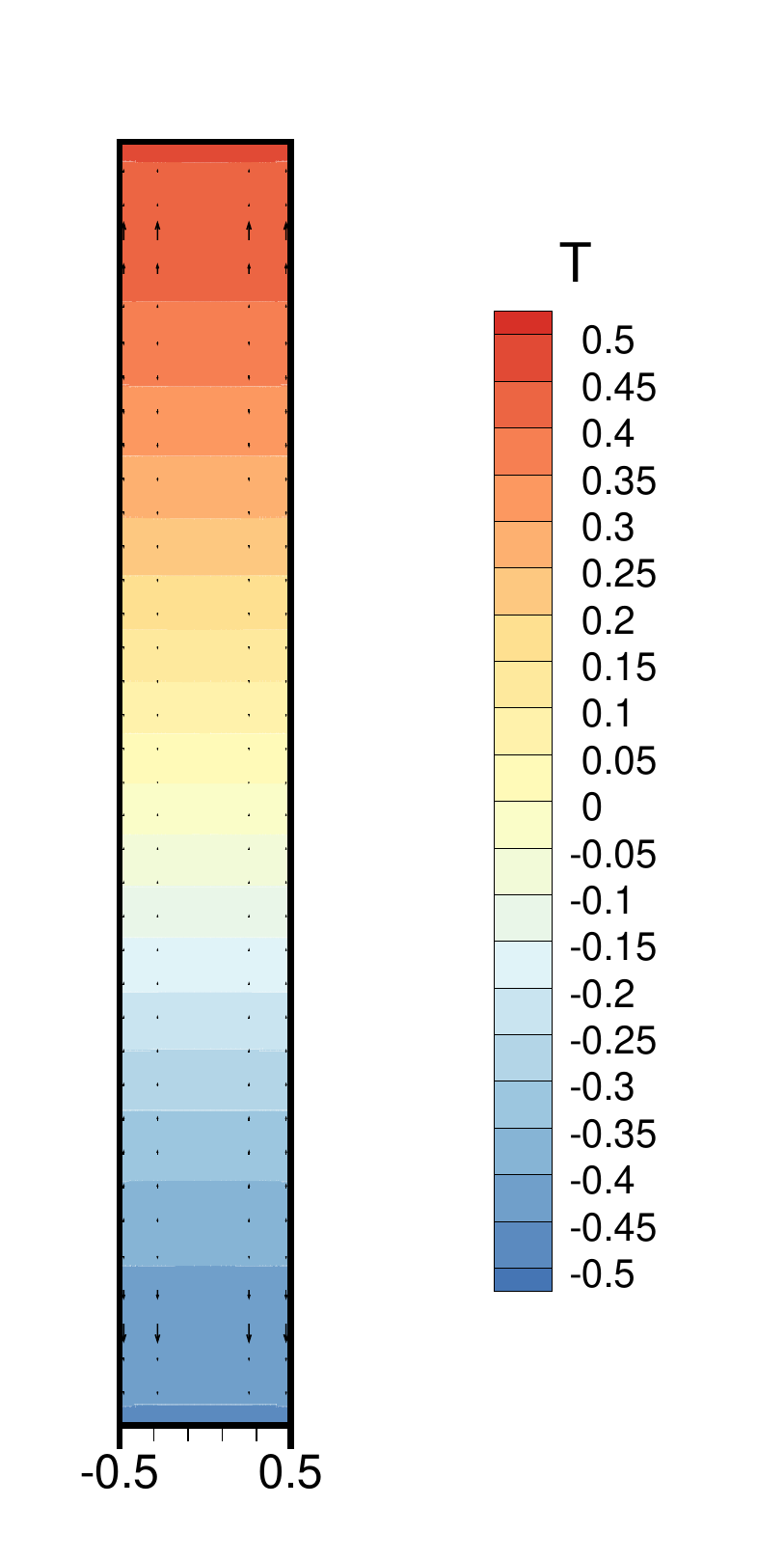}
    \axisput{11.5}{1.5}{Y}
\end{overpic}
\caption{Time-averaged distributions of $T$ and a projection of the velocity on the vertical mid-plane cross-sections $y = 0$ and $x = 0$ (see Fig.~\ref{fig:doamin}c) for configuration B. Results for the flows with $Ha = 0$ \emph{(a)}, $Ha = 162$, $C_w = 0.01$ \emph{(b)}, $Ha = 162$, $C_w = 1.0$ \emph{(c)}, $Ha = 400$, $C_w = 0.01$ \emph{(d)}, $Ha = 400$, $C_w = 1.0$ \emph{(e)}, $Ha = 796$, $C_w = 0.01$ \emph{(f)} and $Ha = 796$, $C_w = 1.0$ \emph{(g)} are shown. The top row shows cross-section $y = 0$ perpendicular to the magnetic field direction. The bottom row shows the cross-section $x = 0$ parallel to the magnetic field direction.}
\label{fig:cont_tc_vert}
\end{figure*}

\begin{figure*}
\centering
\begin{overpic}[width=0.28\textwidth,clip=]{cont_h0_0_xy_no-title_final}
    \titleput{0}{99}{(a)}
    \axisput{49}{0.5}{X}
    \axisput{0}{49}{\rotatebox[origin=c]{90}{Y}}
\end{overpic}
\begin{overpic}[width=0.28\textwidth,clip=]{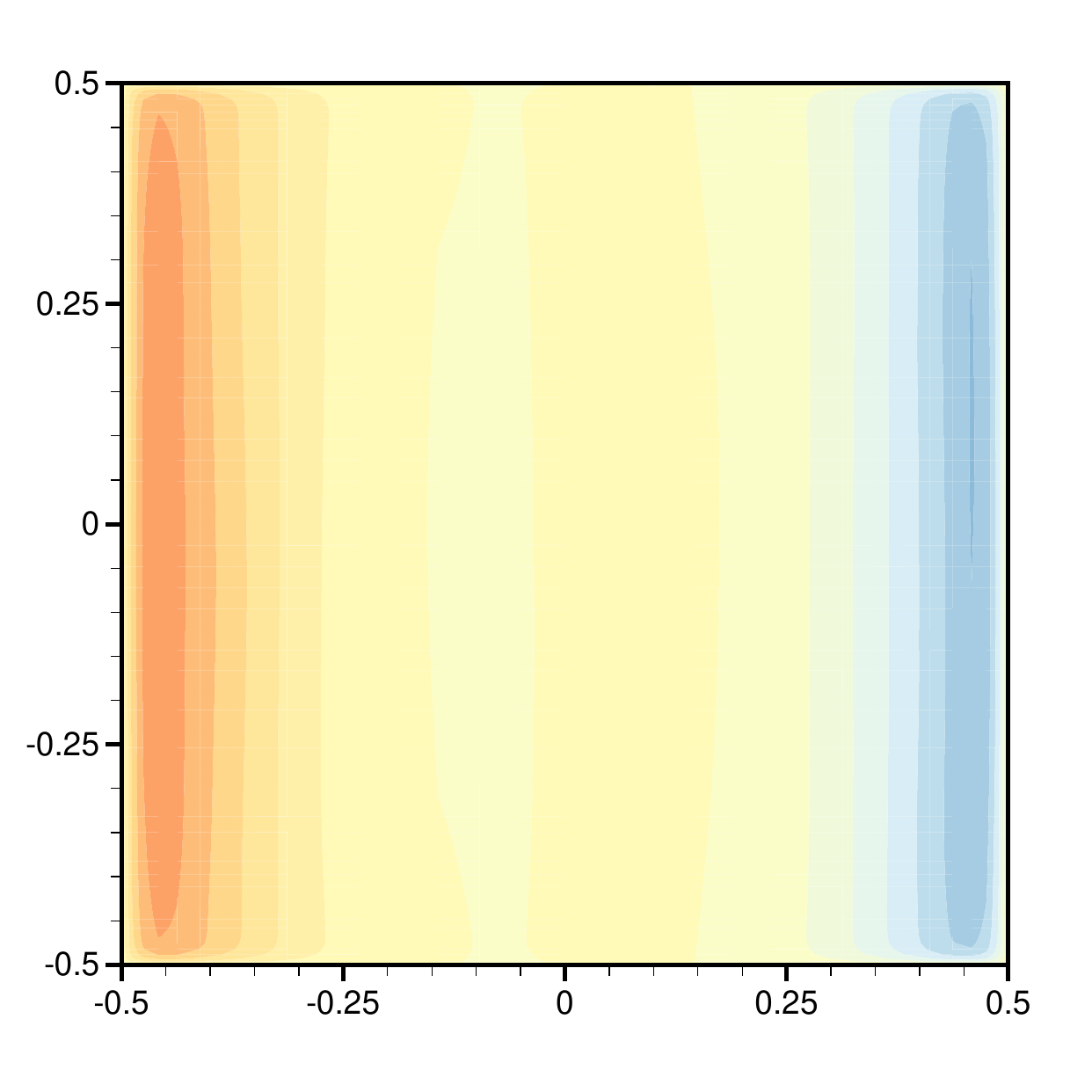}
    \titleput{0}{99}{(b)}
    \axisput{49}{0.5}{X}
    \axisput{0}{49}{\rotatebox[origin=c]{90}{Y}}
\end{overpic}
\begin{overpic}[width=0.315\textwidth,clip=]{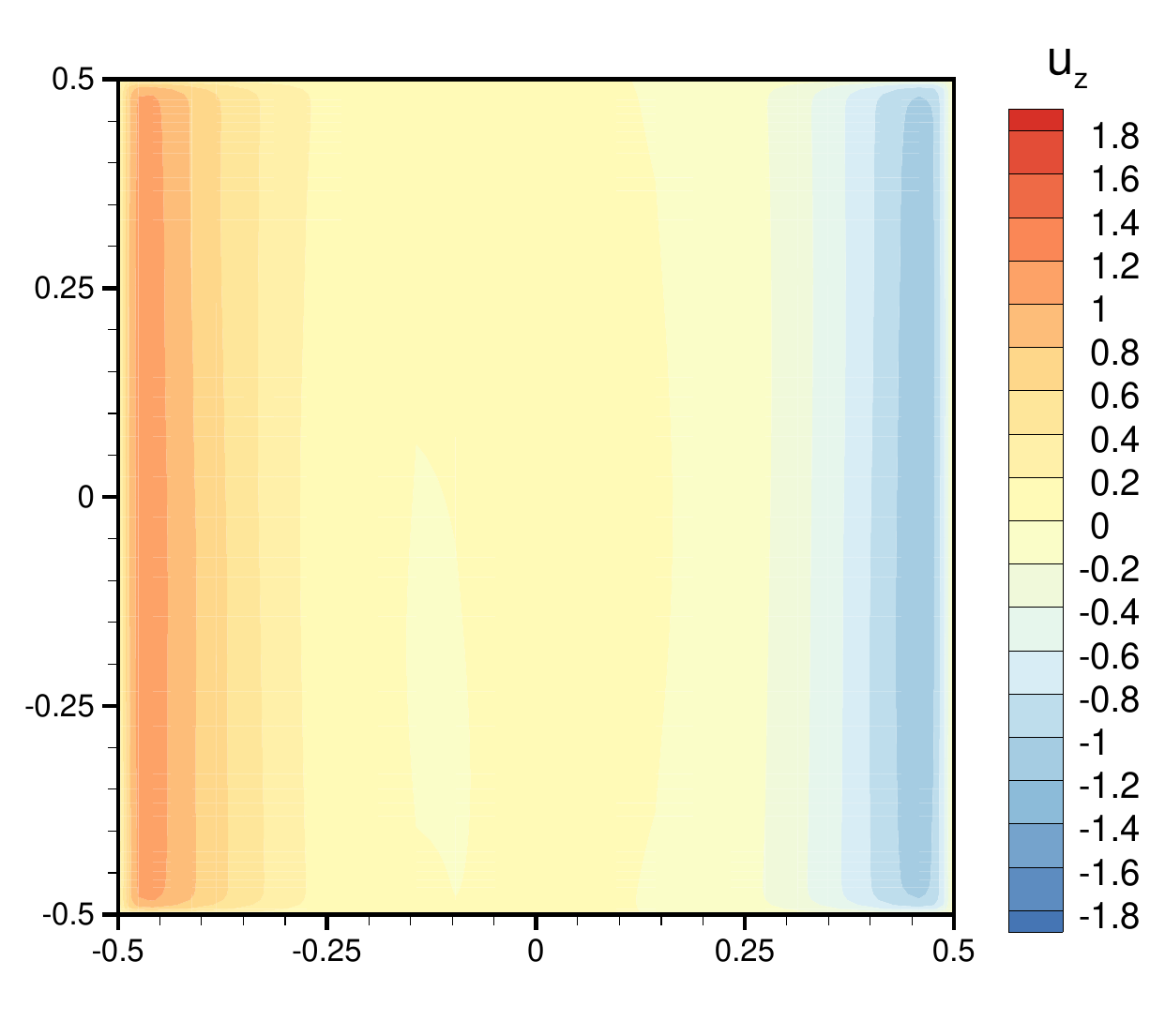}
    \titleput{1}{88}{(c)}
    \axisput{43.5}{0.25}{X}
    \axisput{0}{43.5}{\rotatebox[origin=c]{90}{Y}}
\end{overpic}\vskip1em
\begin{overpic}[width=0.28\textwidth,clip=]{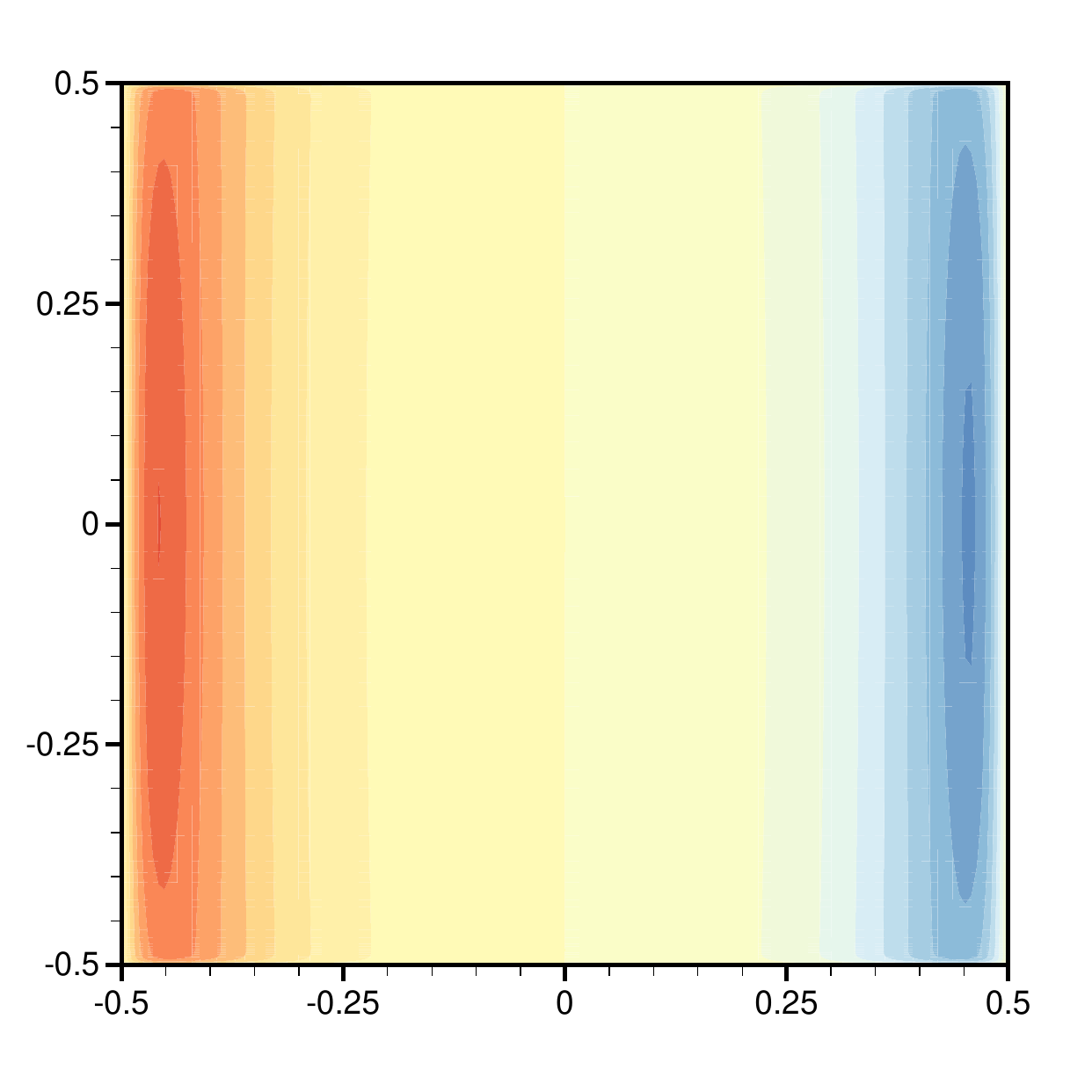}
    \titleput{0}{99}{(d)}
    \axisput{49}{0.5}{X}
    \axisput{0}{49}{\rotatebox[origin=c]{90}{Y}}
\end{overpic}
\begin{overpic}[width=0.28\textwidth,clip=]{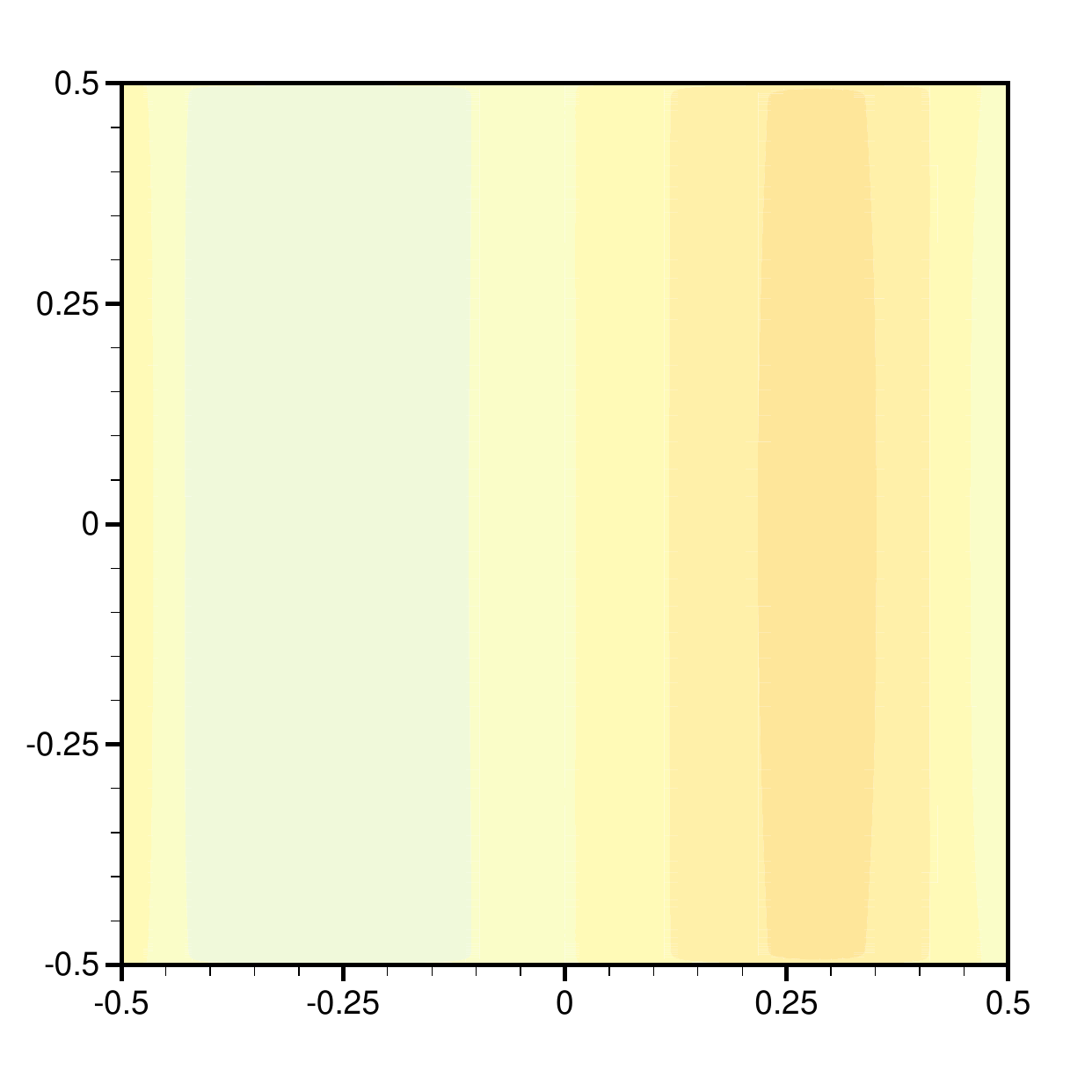}
    \titleput{0}{99}{(e)}
    \axisput{49}{0.5}{X}
    \axisput{0}{49}{\rotatebox[origin=c]{90}{Y}}
\end{overpic}
\begin{overpic}[width=0.315\textwidth,clip=]{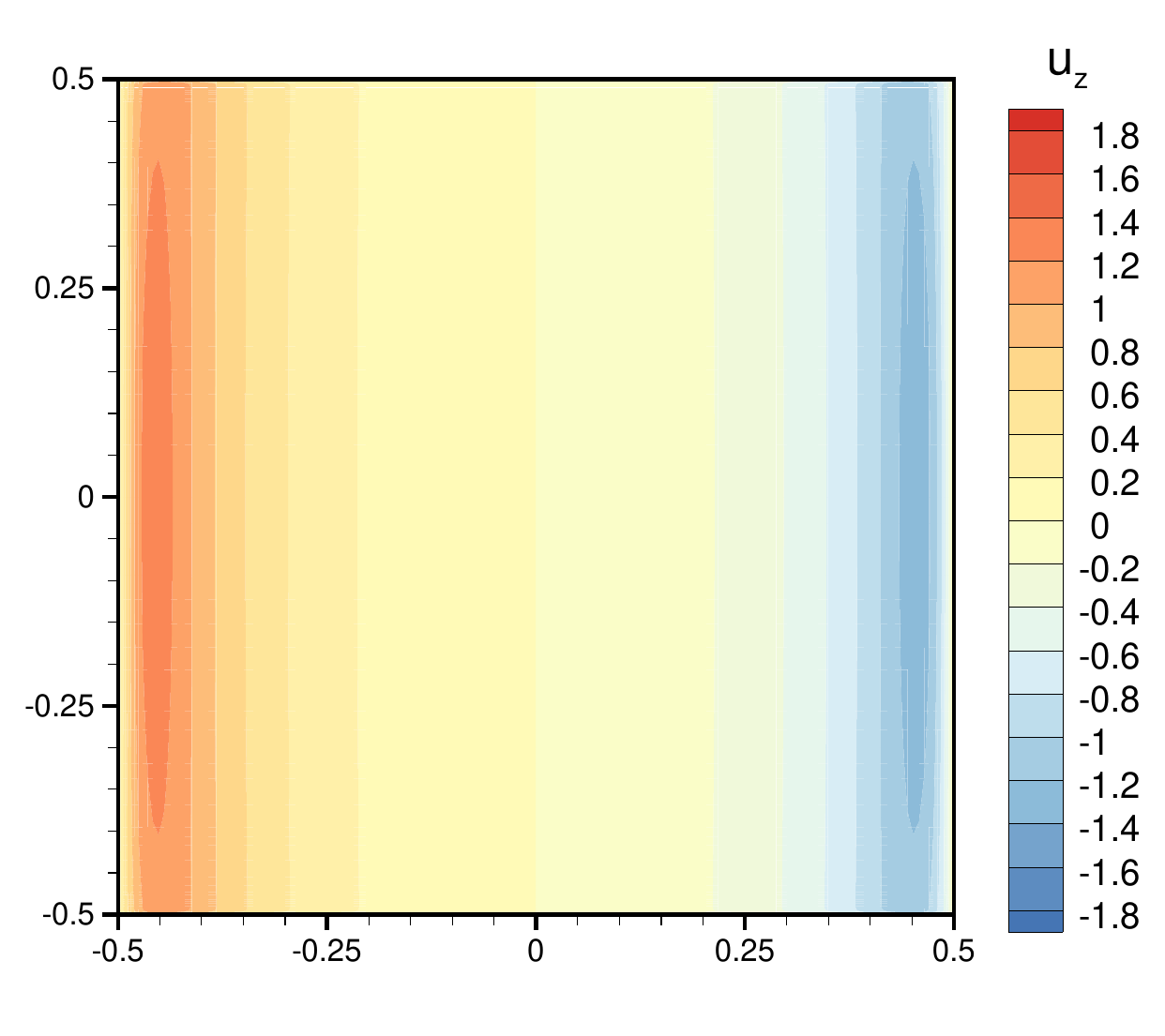}
    \titleput{1}{88}{(f)}
    \axisput{43.5}{0.25}{X}
    \axisput{0}{43.5}{\rotatebox[origin=c]{90}{Y}}
\end{overpic}
\caption{Time-averaged distributions of vertical velocity $u_z$ on the horizontal mid-plane cross-section $z = 0$ (see Fig.~\ref{fig:doamin}c) for configuration B. Results for the flows with $Ha = 0$ \emph{(a)}, $Ha = 162$, $C_w = 0.01$ \emph{(b)}, $Ha = 162$, $C_w = 1.0$ \emph{(c)}, $Ha = 400$, $C_w = 0.01$ \emph{(d)}, $Ha = 400$, $C_w = 1.0$ \emph{(e)} and $Ha = 796$, $C_w = 0.01$ \emph{(f)} are shown.}
\label{fig:cont_tc_hor}
\end{figure*}

\begin{figure*}
\begin{overpic}[width=0.41\textwidth,clip=]{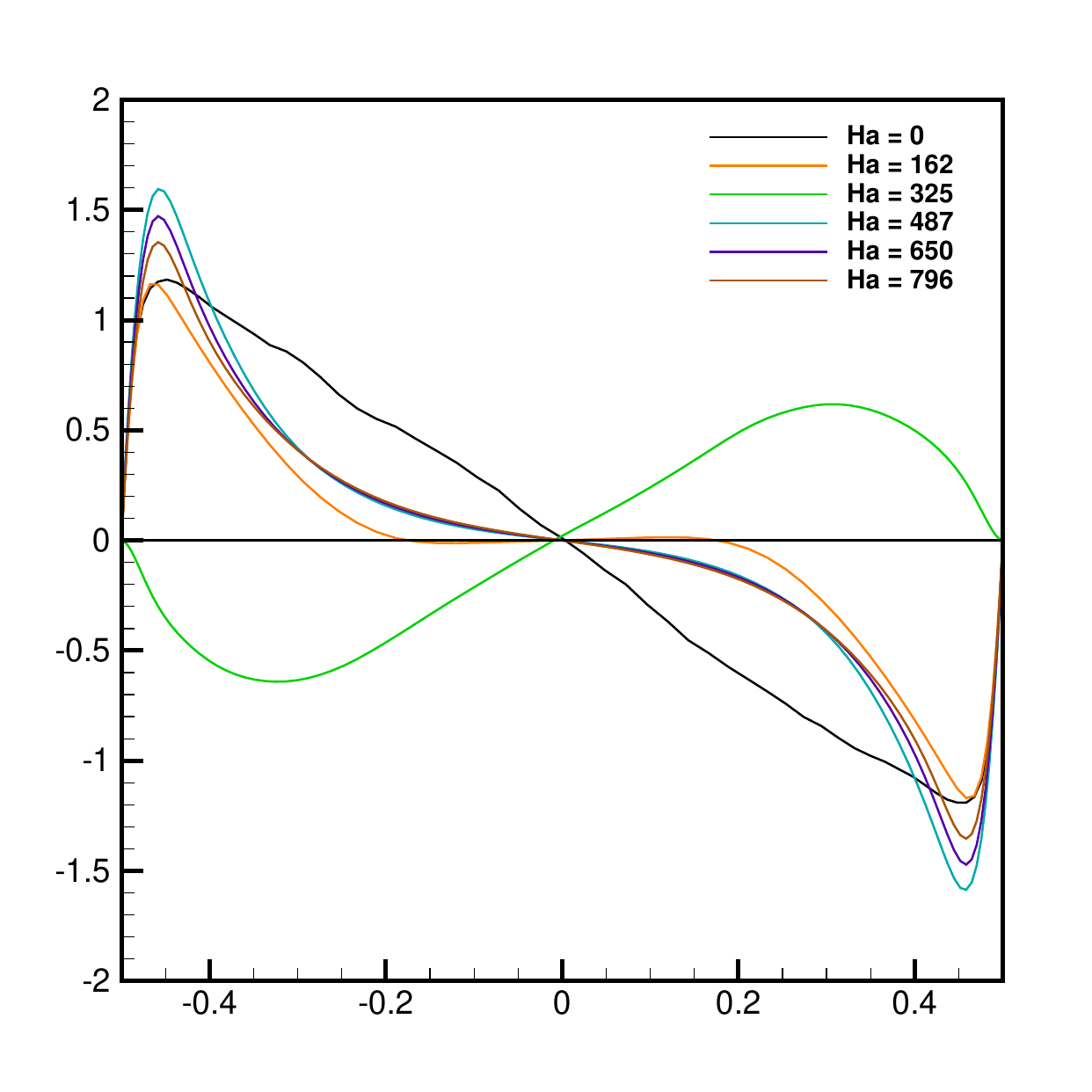}
    \titleput{2}{94}{(a)}
    \axisput{50}{2}{X}
    \axisput{2}{50}{\rotatebox[origin=c]{90}{$u_z$}}
\end{overpic}\hfil\hfil
\begin{overpic}[width=0.41\textwidth,clip=]{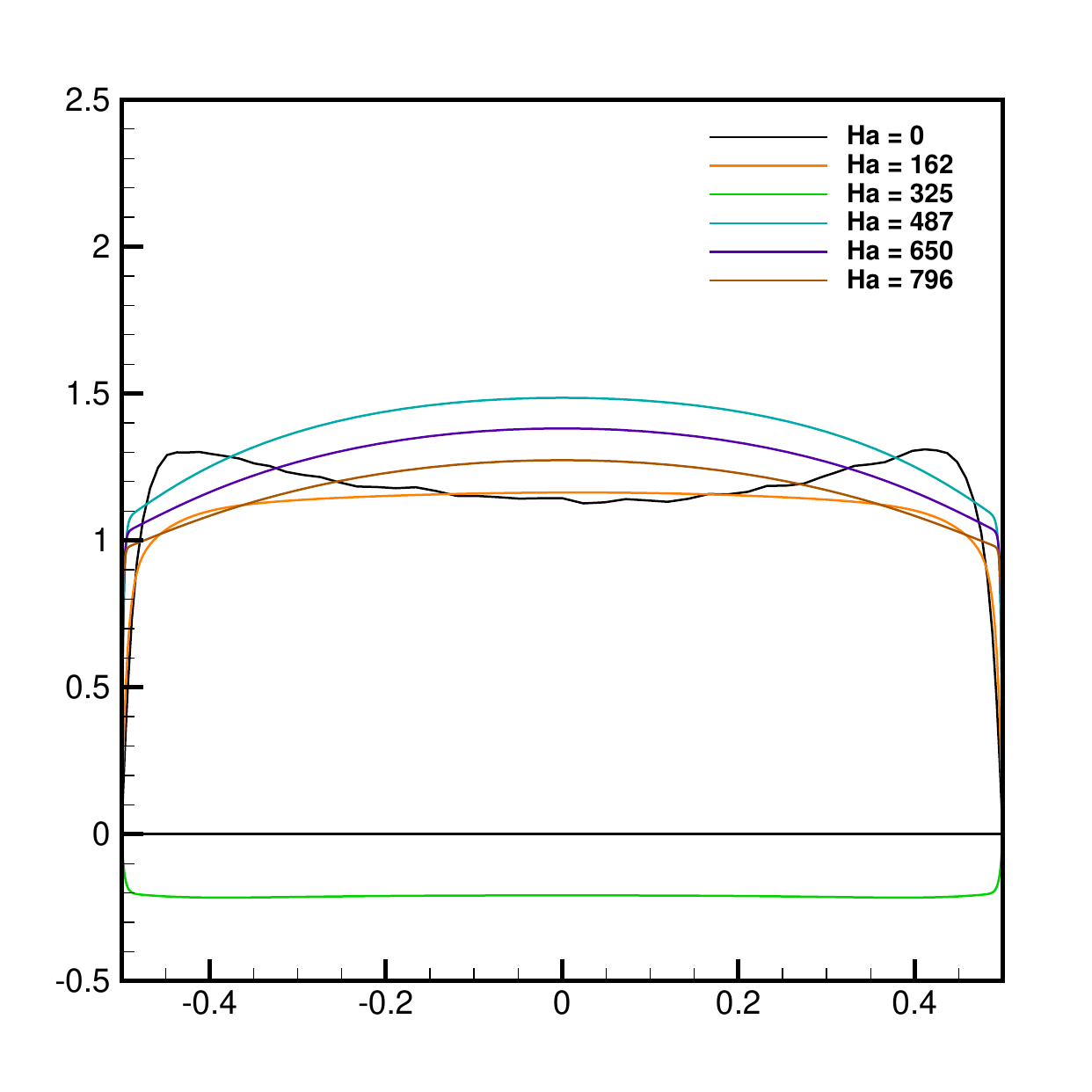}
    \axisput{50}{2}{Y}
    \axisput{2}{50}{\rotatebox[origin=c]{90}{$u_z$}}
\end{overpic}\\
\begin{overpic}[width=0.41\textwidth,clip=]{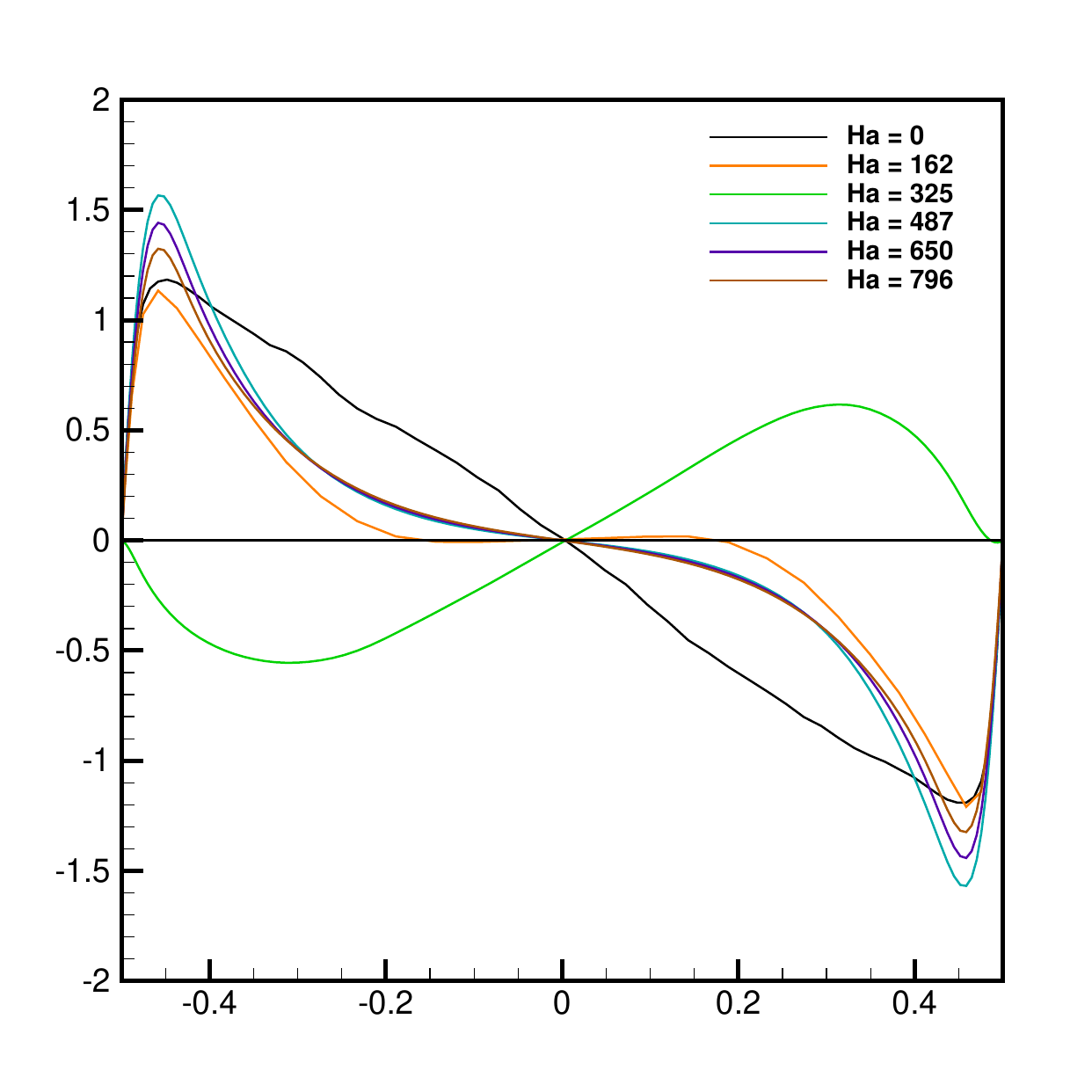}
    \titleput{2}{94}{(b)}
    \axisput{50}{2}{X}
    \axisput{2}{50}{\rotatebox[origin=c]{90}{$u_z$}}
\end{overpic}\hfil\hfil
\begin{overpic}[width=0.41\textwidth,clip=]{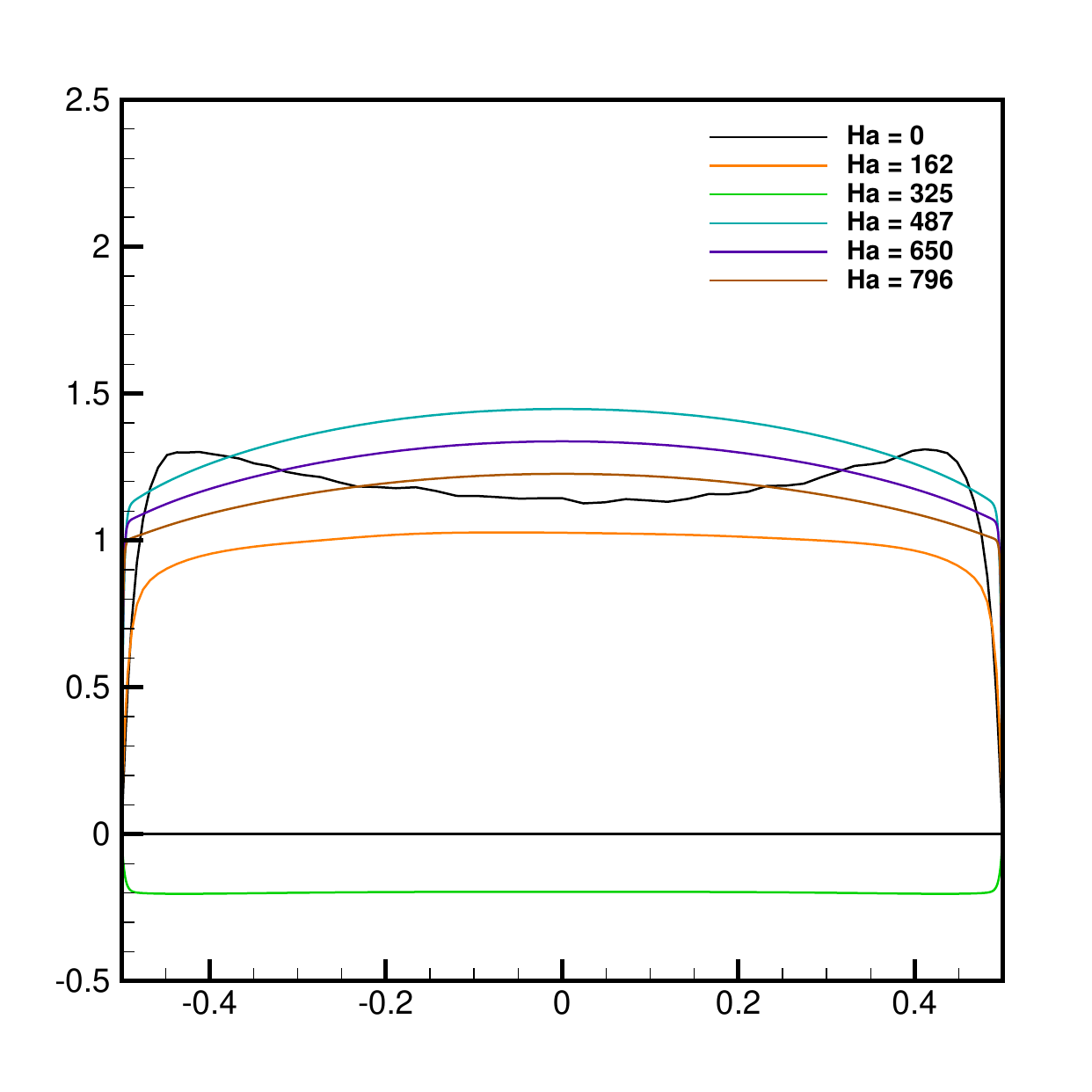}
    \axisput{50}{2}{Y}
    \axisput{2}{50}{\rotatebox[origin=c]{90}{$u_z$}}
\end{overpic}\\
\begin{overpic}[width=0.41\textwidth,clip=]{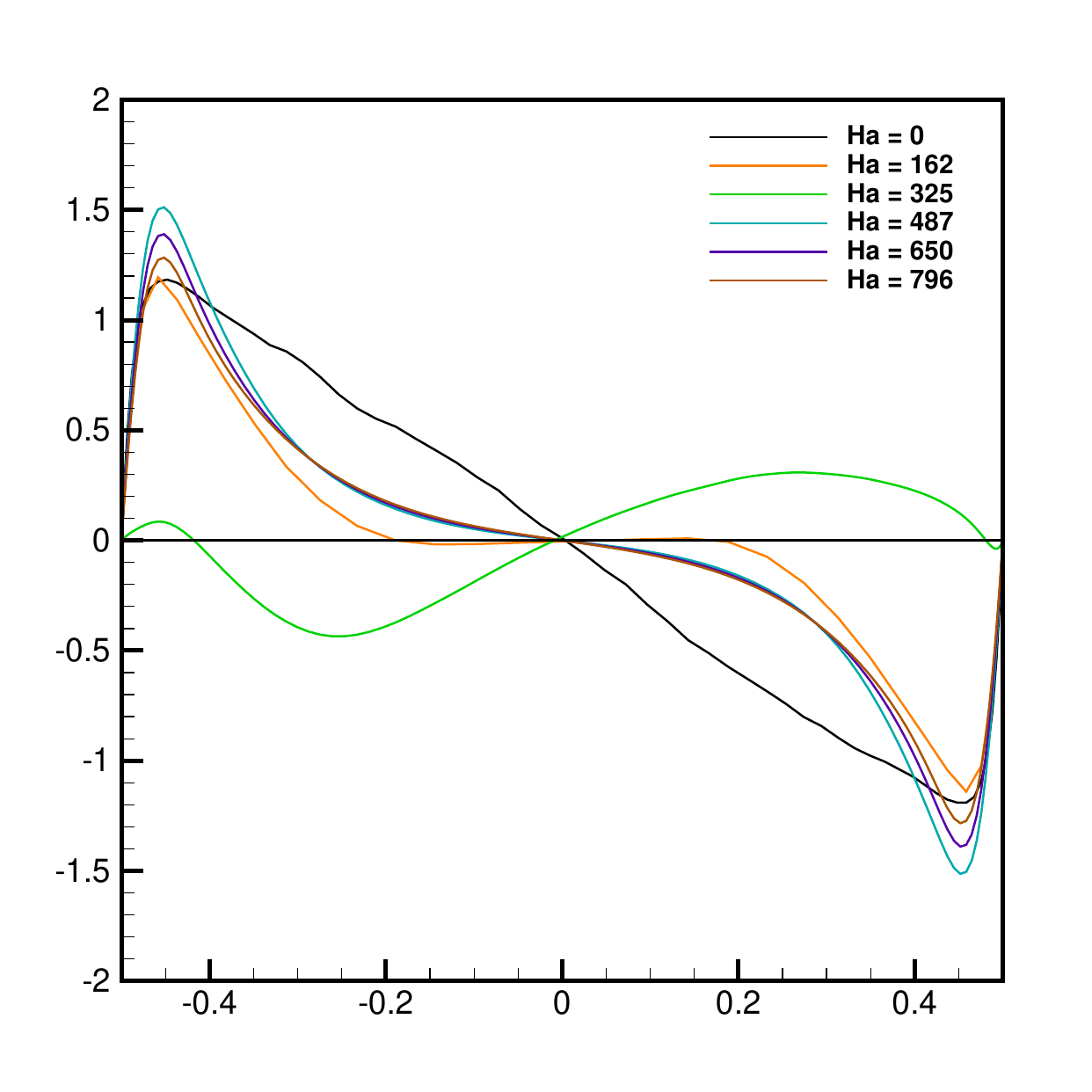}
    \titleput{2}{94}{(c)}
    \axisput{50}{2}{X}
    \axisput{2}{50}{\rotatebox[origin=c]{90}{$u_z$}}
\end{overpic}\hfil\hfil
\begin{overpic}[width=0.41\textwidth,clip=]{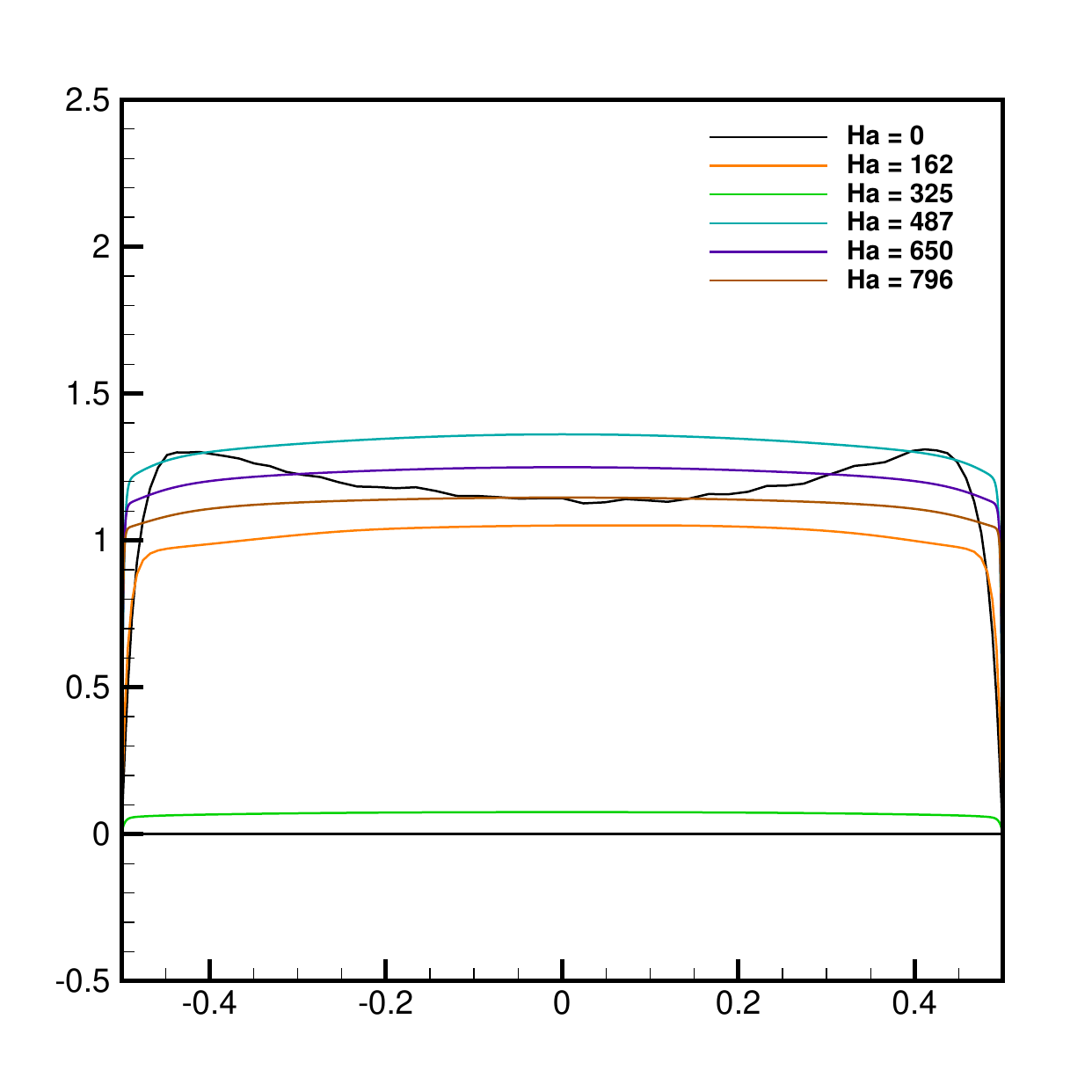}
    \axisput{50}{2}{Y}
    \axisput{2}{50}{\rotatebox[origin=c]{90}{$u_z$}}
\end{overpic}
\caption{Timed-averaged profiles of vertical velocity along the lines \xdataline~(left) and \ydataline~(right) shown in Fig.~\ref{fig:doamin}b. Results obtained for $C_w=0$ \emph{(a)} and the configuration B with $C_w = 0.01$ \emph{(b)} and $C_w = 0.1$ \emph{(c)} are shown for various $\Ha$.}
\label{fig:uz_tc_xy}
\end{figure*}

\begin{figure*}
\centering
\begin{overpic}[width=0.41\textwidth,clip=]{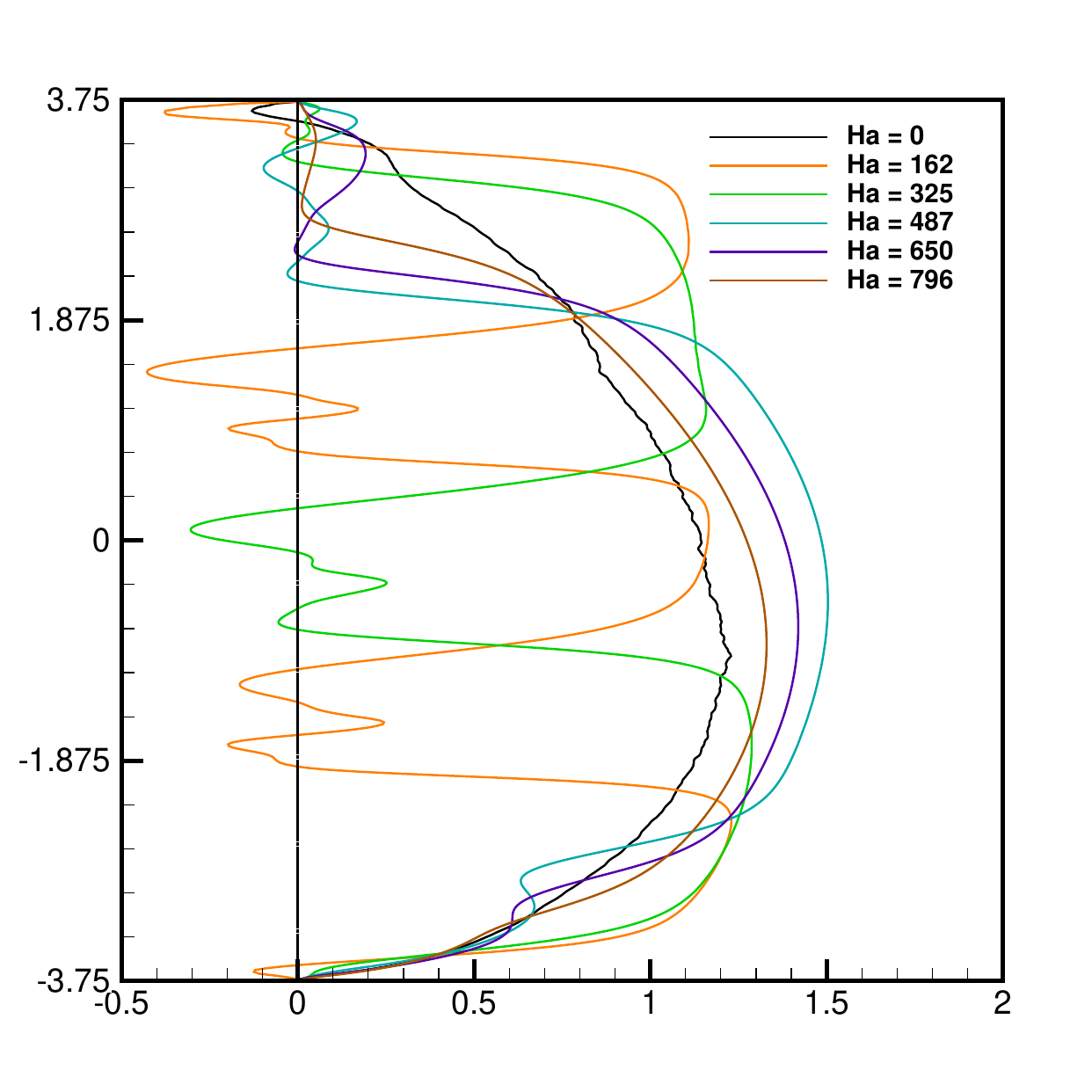}
    \titleput{2}{95}{(a)}
    \axisput{50}{2}{$u_z$}
    \axisput{2}{49}{\rotatebox[origin=c]{90}{Z}}
\end{overpic}\hfil\hfil
\begin{overpic}[width=0.41\textwidth,clip=]{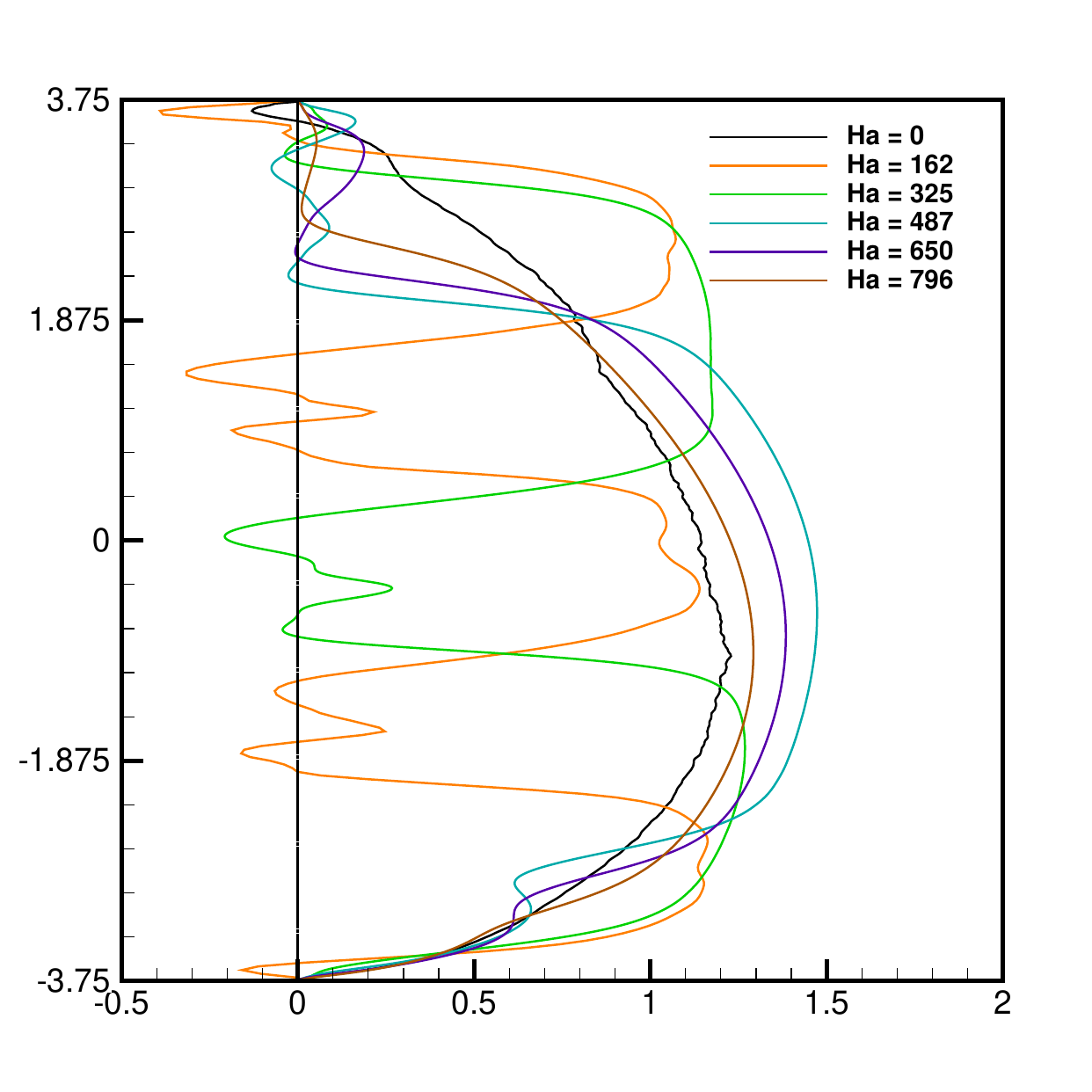}
    \titleput{2}{95}{(b)}
    \axisput{50}{2}{$u_z$}
    \axisput{2}{49}{\rotatebox[origin=c]{90}{Z}}
\end{overpic}\\
\begin{overpic}[width=0.41\textwidth,clip=]{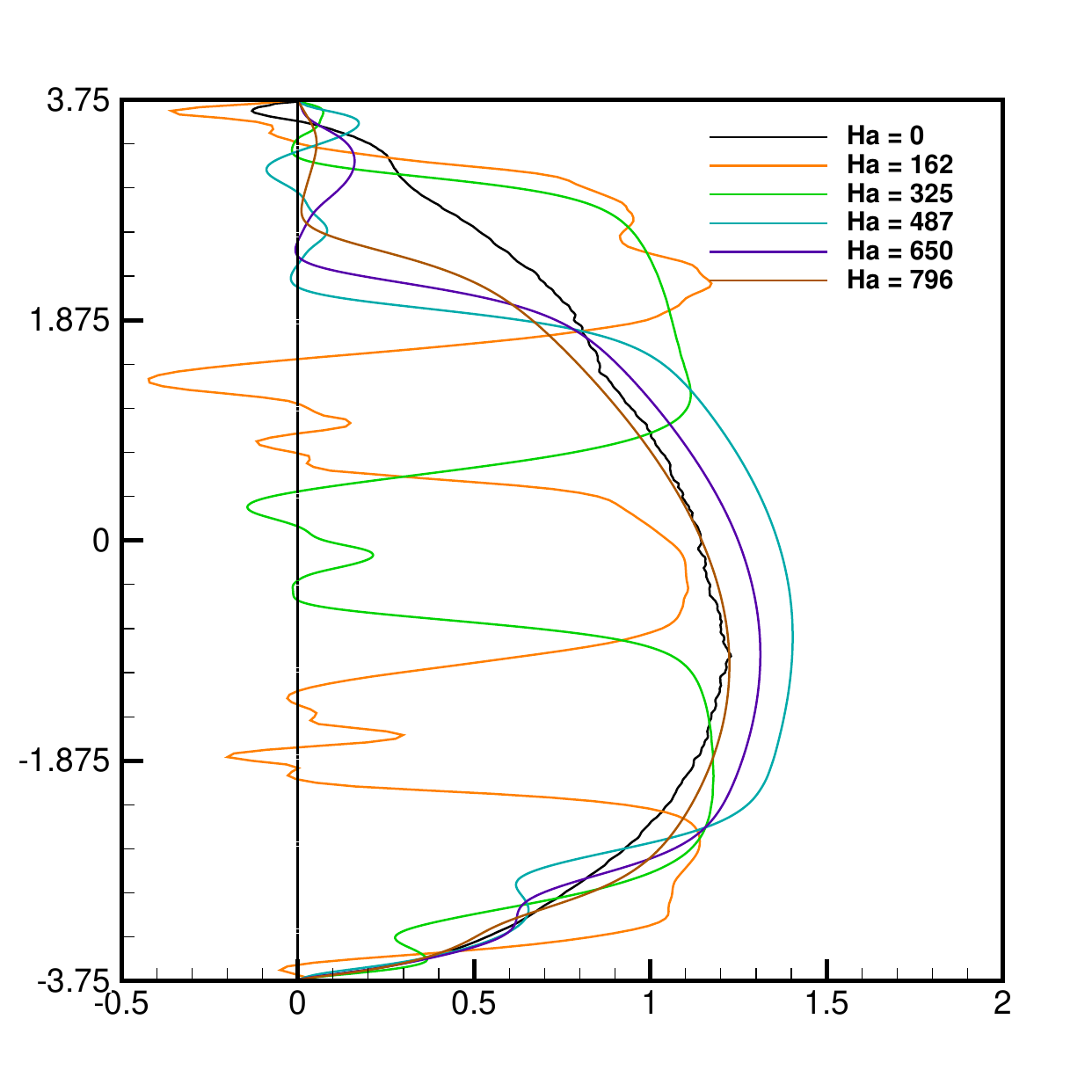}
    \titleput{2}{95}{(c)}
    \axisput{50}{2}{$u_z$}
    \axisput{2}{49}{\rotatebox[origin=c]{90}{Z}}
\end{overpic}\hfil\hfil
\begin{overpic}[width=0.41\textwidth,clip=]{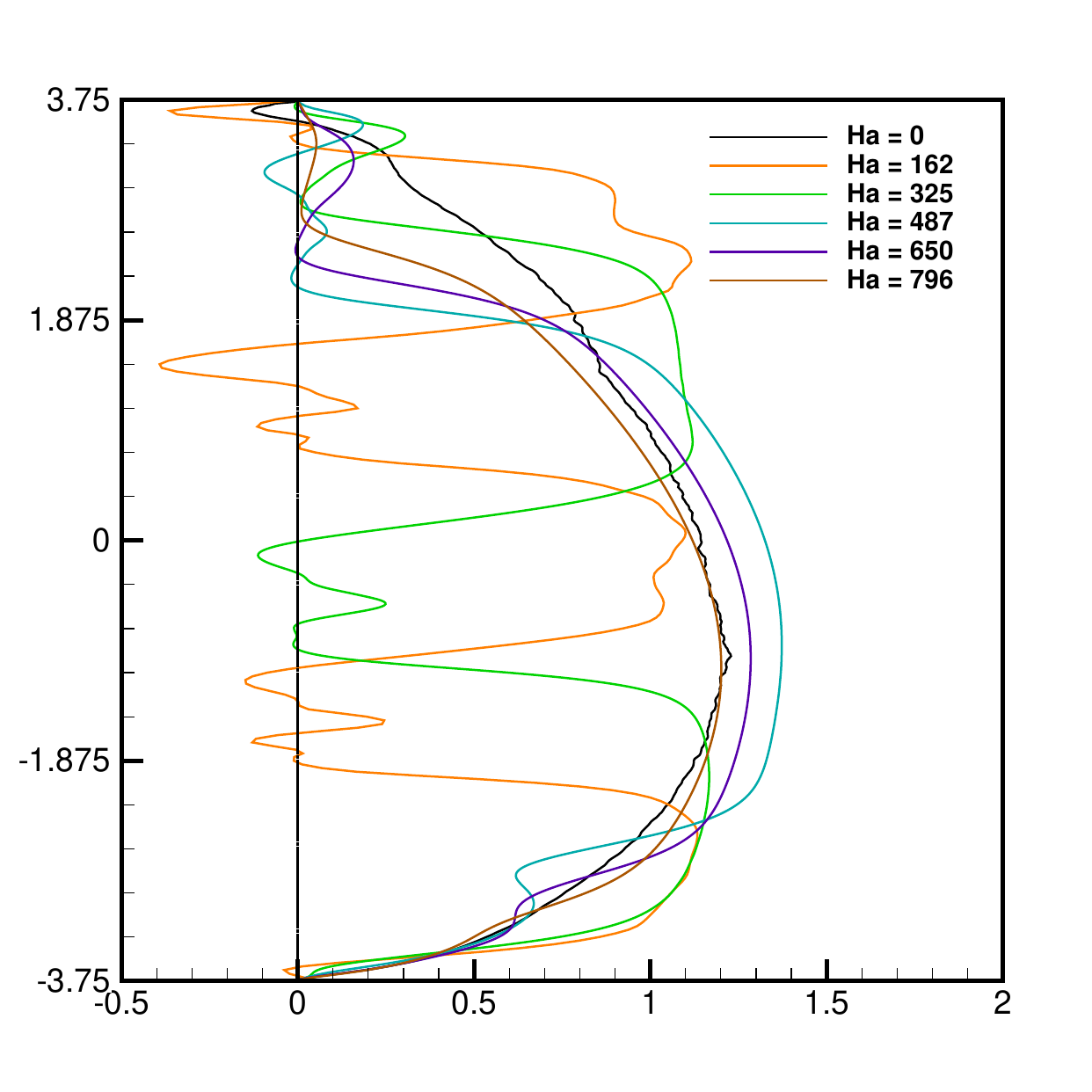}
    \titleput{2}{95}{(d)}
    \axisput{50}{2}{$u_z$}
    \axisput{2}{49}{\rotatebox[origin=c]{90}{Z}}
\end{overpic}
\caption{Timed-averaged profiles of vertical velocity along the line \zdataline~shown in Fig.~\ref{fig:doamin}b. Results obtained for $C_w=0$ \emph{(a)} and the configuration B with  $C_w = 0.01$ \emph{(b)} $C_w = 0.1$ \emph{(c)}, and $C_w = 1.0$ \emph{(d)} are shown for various $\Ha$. A black dotted line indicates $u_z=0$.}
\label{fig:uz_tc_z}
\end{figure*}

\begin{figure*}
\begin{overpic}[width=0.41\textwidth,clip=]{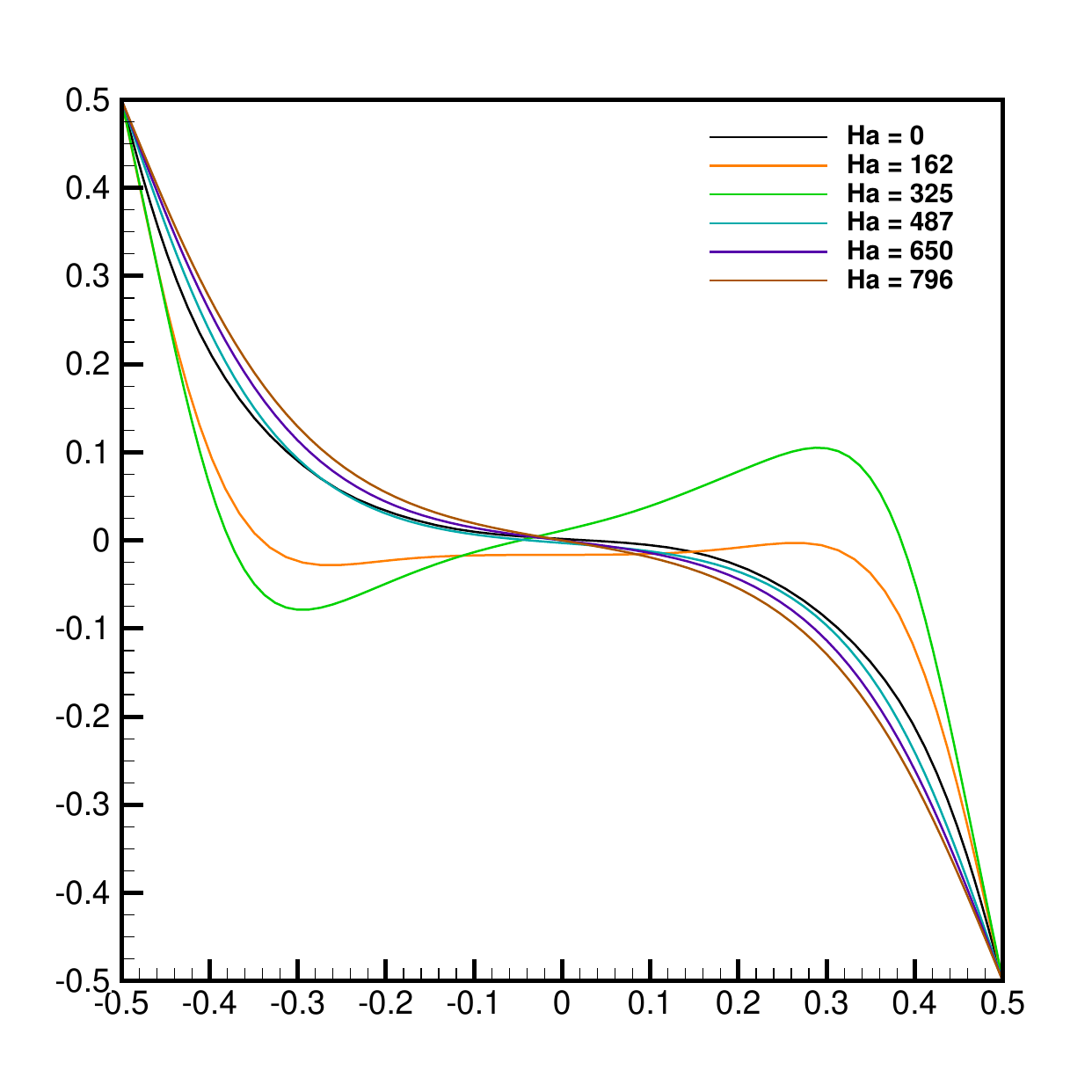}
    \titleput{2}{95}{(a)}
    \axisput{50}{2}{X}
    \axisput{2}{49}{\rotatebox[origin=c]{90}{T}}
\end{overpic}\hfil\hfil
\begin{overpic}[width=0.41\textwidth,clip=]{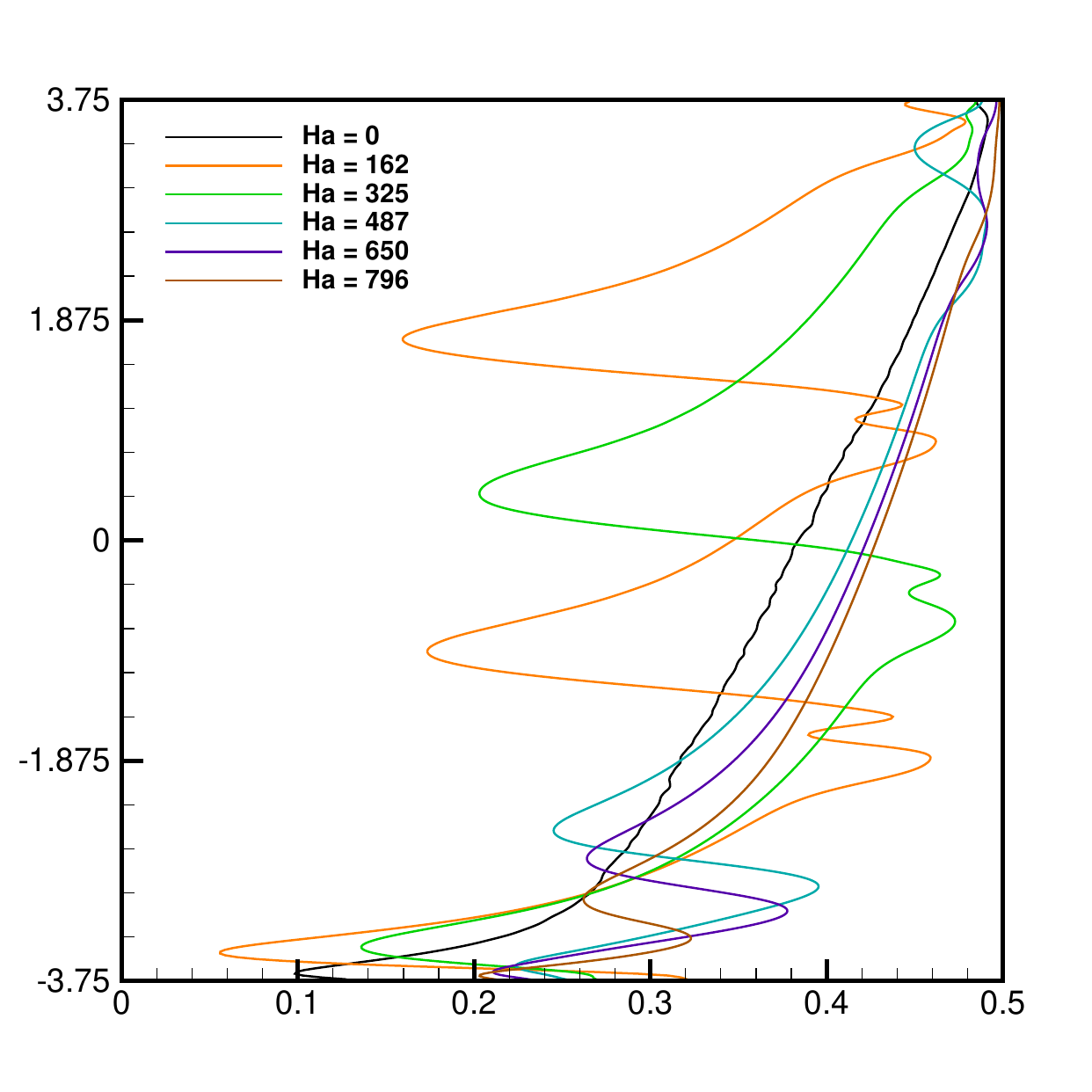}
    \axisput{50}{2}{T}
    \axisput{2}{49}{\rotatebox[origin=c]{90}{Z}}
\end{overpic}\\
\begin{overpic}[width=0.41\textwidth,clip=]{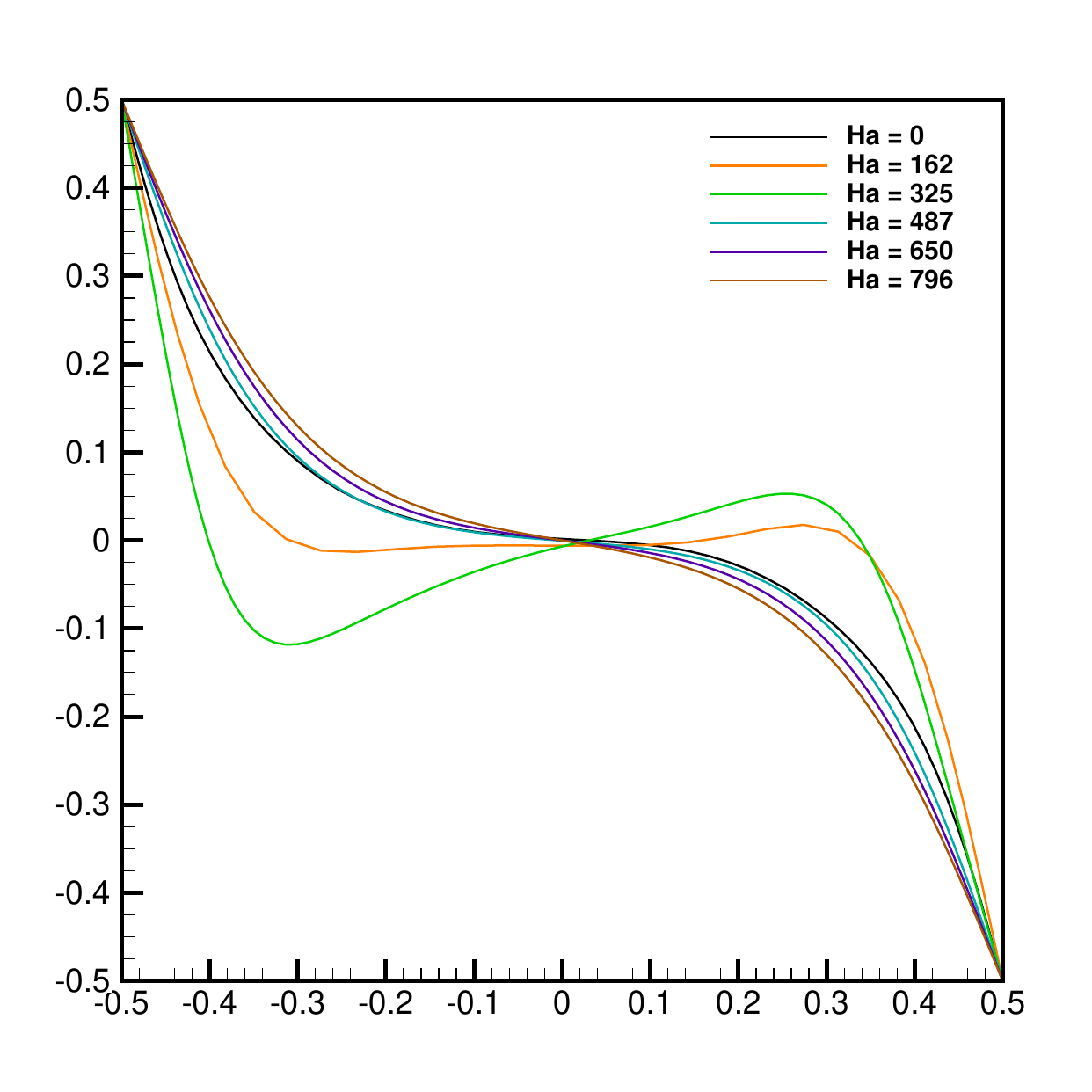}
    \titleput{2}{95}{(b)}
    \axisput{50}{2}{X}
    \axisput{2}{49}{\rotatebox[origin=c]{90}{T}}
\end{overpic}\hfil\hfil
\begin{overpic}[width=0.41\textwidth,clip=]{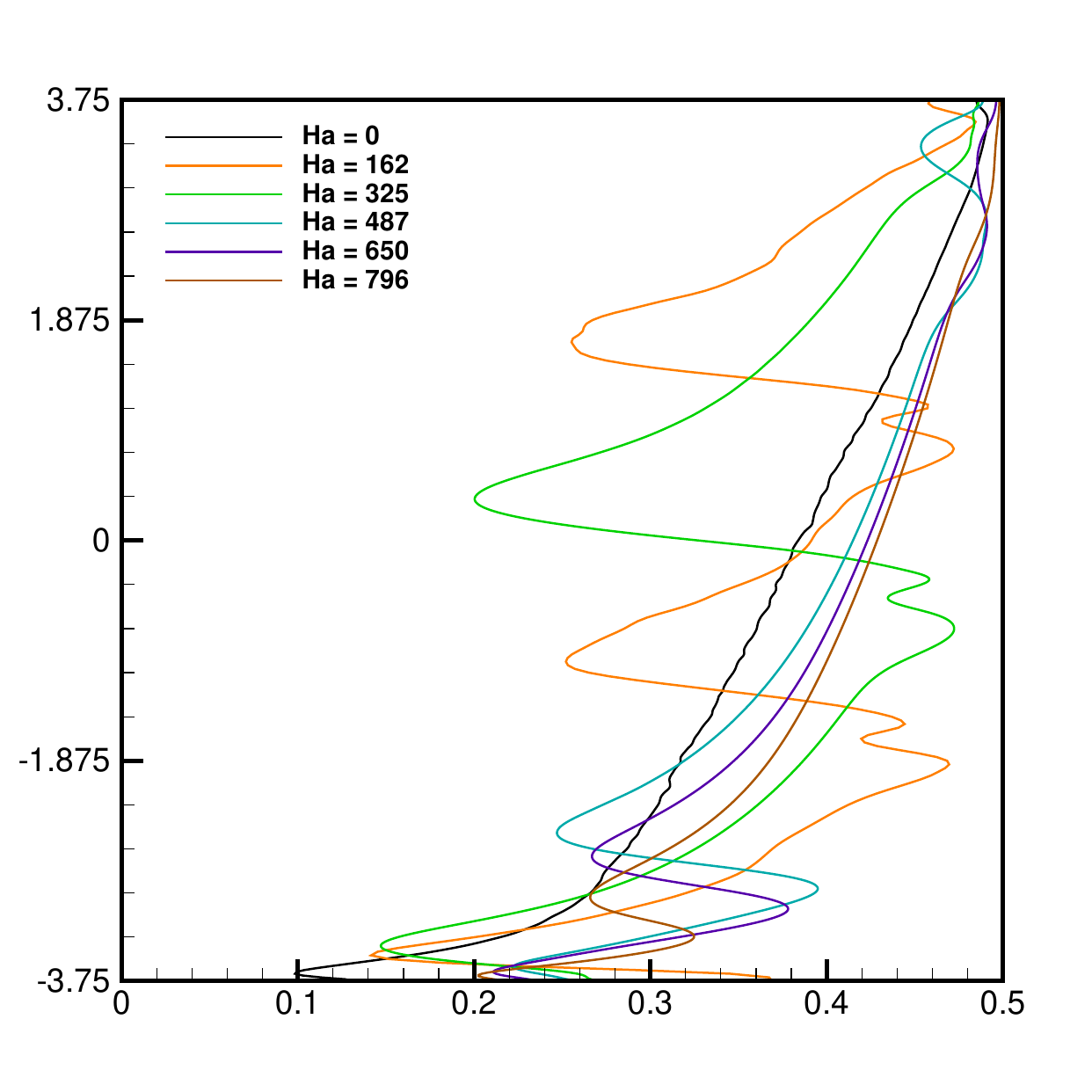}
    \axisput{50}{2}{T}
    \axisput{2}{49}{\rotatebox[origin=c]{90}{Z}}
\end{overpic}\\
\begin{overpic}[width=0.41\textwidth,clip=]{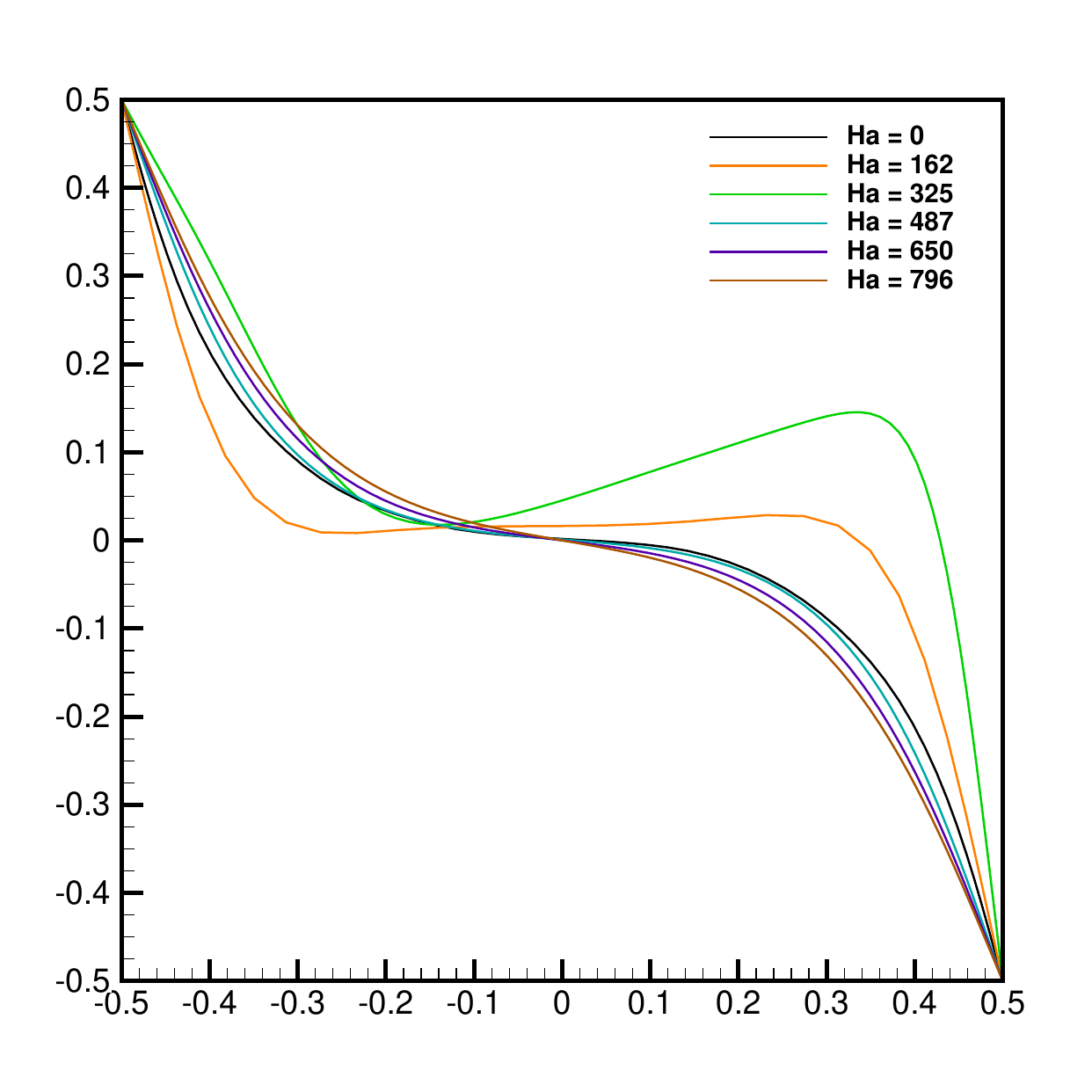}
    \titleput{2}{95}{(c)}
    \axisput{50}{2}{X}
    \axisput{2}{49}{\rotatebox[origin=c]{90}{T}}
\end{overpic}\hfil\hfil
\begin{overpic}[width=0.41\textwidth,clip=]{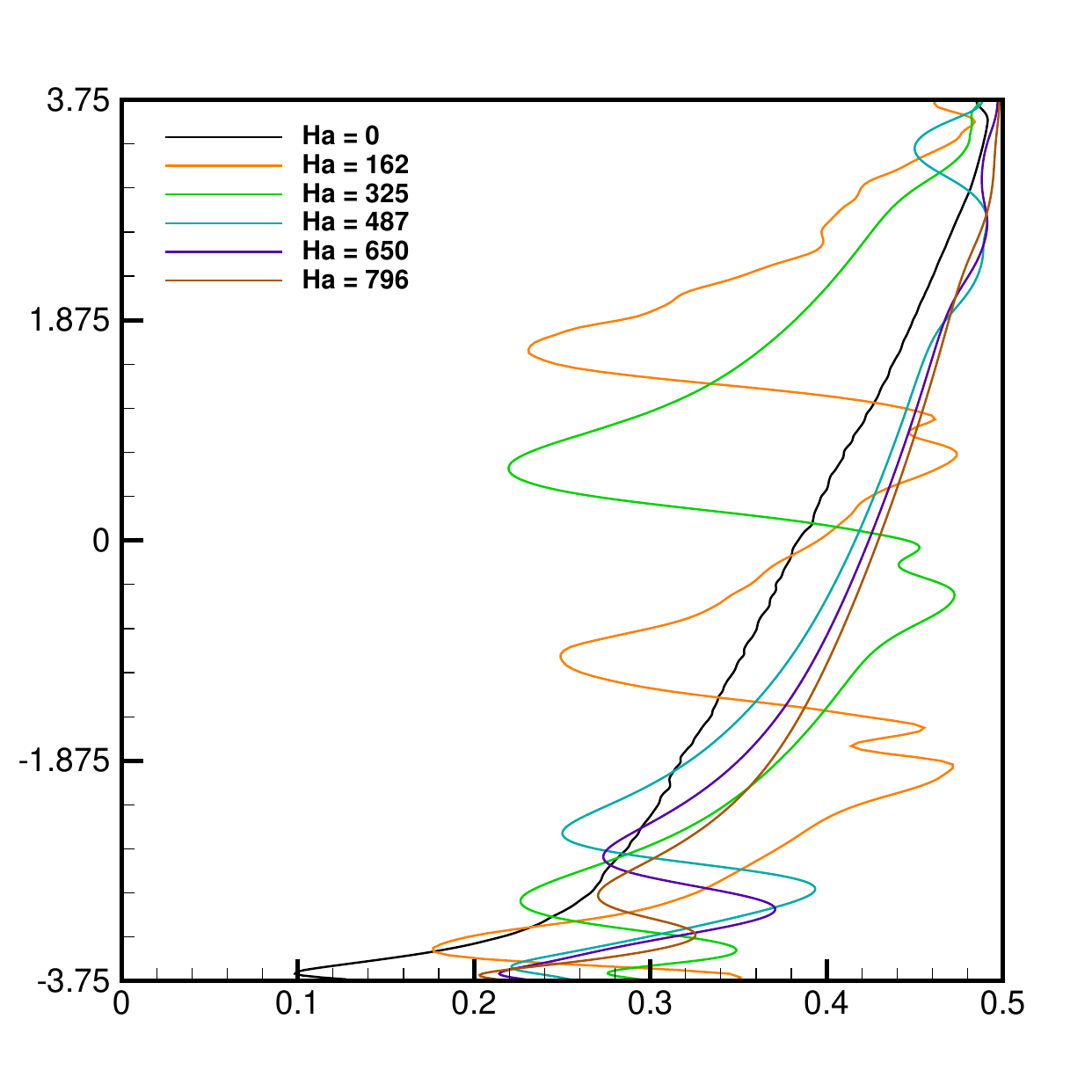}
    \axisput{50}{2}{T}
    \axisput{2}{49}{\rotatebox[origin=c]{90}{Z}}
\end{overpic}
\caption{Timed-averaged profiles of temperature along the lines \xdataline~(left) and \zdataline~(right) shown in Fig.~\ref{fig:doamin}b. Results obtained for $C_w=0$ \emph{(a)} and the configuration B with $C_w = 0.01$ \emph{(b)} and $C_w = 0.1$ \emph{(c)} are shown for various $\Ha$.}
\label{fig:t_tc}
\end{figure*}

The mean flow data calculated by time-averaging over the last 100 time units of each simulation are presented in tables \ref{tbl:Nu_tc}-\ref{tbl:LSC_tc} and figures \ref{fig:NU_E_tc}-\ref{fig:t_tc}. Tables \ref{tbl:Nu_tc} and \ref{tbl:E_tc} and Fig.~\ref{fig:NU_E_tc} show time-averaged values of $\Nu$ and $E$. We observe that the effect of wall conductivity is much weaker than in the case of configuration A. At all considered values of $C_w>0$, the variation of $\Nu$ and $E$ with $\Ha$ is qualitatively the same as in the flow with $C_w=0$. In particular, we see increase of $\Nu$ at moderate $\Ha$ followed by a monotonic decrease at high $\Ha$. The values of $E$ are reduced strongly when the magnetic field is introduced (at $\Ha=85$ and 162), and change slowly at moderate and high $\Ha$. 

The quantitative impact of $C_w$ on $\Nu$ and $E$ is also weak, with exception of several data points, where it can be explained by the modification of the flow structure described below. Practically no effect of $C_w$ on $\Nu$ and $E$ is observed at $\Ha\ge 487$.
 
The main feature of the structure of the mean flow, namely the number of major eddies in the large-scale circulation, is reported in Table \ref{tbl:LSC_tc}. Minor vortices, which appear in the corners of the domain and at confluence of two major eddies are not included in the count. We see that similarly to the case $C_w=0$, multi-eddy circulations appear at moderate values of $\Ha$. The cases when the number of eddies in the flow is different from the numbers observed at the same $\Ha$ for other values of $C_w$ (e.g., two eddies at $C_w=50$, $\Ha=85$ or one eddy at $C_w=0.01$, $\Ha=400$) correlate with significant deviations of $\Nu$ and $E$.

\begin{table}
    \centering
    \begin{tabular}{c|c|C{1.2cm}C{1.2cm}C{1.2cm}C{1.2cm}C{1.2cm}}
        \multicolumn{2}{c}{} & \multicolumn{5}{c}{$C_w$} \\ \cline{3-7}
        \multicolumn{2}{c}{} & 0 & 0.01 & 0.1 & 1 & 50 \\ \cline{3-7}
        \multirow{10}{*}{$\Ha$} & 0 & \multicolumn{5}{c}{$\overline{3.80}$} \\
         & 85 & $\overline{3.69}$ & $\overline{3.68}$ & $\overline{3.74}$ & $\overline{3.83}$ & $\overline{4.80}$ \\ 
         & 162 & $\overline{5.15}$ & $\overline{5.13}$ & $\overline{5.14}$ & $\overline{5.13}$ & $\overline{5.14}$ \\ 
         & 240 & $\overline{4.95}$ & $\overline{4.96}$ & $\overline{4.56}$ & $\overline{4.38}$ & $\overline{4.53}$ \\ 
         & 325 & $\overline{4.47}$ & $\overline{4.38}$ & $\overline{4.34}$ & $\overline{4.33}$ & $\overline{4.36}$ \\ 
         & 400 & $\overline{4.35}$ & $\overline{3.62}$ & $\overline{4.21}$ & $\overline{4.34}$ & $\overline{4.08}$ \\ 
         & 487 & $\overline{3.43}$ & $\overline{3.43}$ & $\overline{3.39}$ & $\overline{3.41}$ & $\overline{3.38}$ \\ 
         & 570 & $\overline{3.26}$ & $\overline{3.25}$ & $\overline{3.23}$ & $\overline{3.21}$ & $\overline{3.22}$ \\ 
         & 650 & $\overline{3.15}$ & $\overline{3.14}$ & $\overline{3.11}$ & $\overline{3.10}$ & $\overline{3.10}$ \\ 
         & 735 & $\overline{3.06}$ & $\overline{3.05}$ & $\overline{3.03}$ & $\overline{3.02}$ & $\overline{3.02}$ \\ 
         & 796 & ${3.01}$ & ${3.00}$ & ${2.98}$ & ${2.97}$ & ${2.97}$ \\ 
    \end{tabular}
    \caption{Time averaged values of $\Nu$ \eqref{eqn:meanNuE} computed for the configuration B (hot and cold walls have the same conductance ratio $C_w$). Values marked with an overline indicate cases in which $\Nu(t)$ fluctuates significantly in a fully developed flow. The signal at $\Ha=796$ is steady-state after an initial transient.}
    \label{tbl:Nu_tc}
\end{table}

\begin{table*}
    \centering
    \begin{tabular}{c|c|C{2.2cm}C{2.2cm}C{2.2cm}C{2.2cm}C{2.2cm}}
        \multicolumn{2}{c}{} & \multicolumn{5}{c}{$C_w$} \\ \cline{3-7}
        \multicolumn{2}{c}{} & 0 & 0.01 & 0.1 & 1 & 50 \\ \cline{3-7}
        \multirow{10}{*}{$\Ha$} & 0 & \multicolumn{5}{c}{$\overline{5.16 \times 10^{-1}}$} \\
         & 85 & $\overline{4.96 \times 10^{-1}}$ & $\overline{4.88 \times 10^{-1}}$ & $\overline{5.03 \times 10^{-1}}$ & $\overline{5.08 \times 10^{-1}}$ & $\overline{3.49 \times 10^{-1}}$ \\
         & 162 & $\overline{2.55 \times 10^{-1}}$ & $\overline{2.59 \times 10^{-1}}$ & $\overline{2.60 \times 10^{-1}}$ & $\overline{2.60 \times 10^{-1}}$ & $\overline{2.58 \times 10^{-1}}$ \\
         & 240 & $\overline{2.39 \times 10^{-1}}$ & $\overline{2.37 \times 10^{-1}}$ & $\overline{3.02 \times 10^{-1}}$ & $\overline{2.88 \times 10^{-1}}$ & $\overline{3.03 \times 10^{-1}}$ \\
         & 325 & $\overline{2.78 \times 10^{-1}}$ & $\overline{2.72 \times 10^{-1}}$ & $\overline{2.65 \times 10^{-1}}$ & $\overline{2.66 \times 10^{-1}}$ & $\overline{2.61 \times 10^{-1}}$ \\
         & 400 & $\overline{2.43 \times 10^{-1}}$ & $\overline{3.22 \times 10^{-1}}$ & $\overline{2.37 \times 10^{-1}}$ & $\overline{2.43 \times 10^{-1}}$ & $\overline{2.44 \times 10^{-1}}$ \\
         & 487 & $\overline{3.08 \times 10^{-1}}$ & $\overline{3.03 \times 10^{-1}}$ & $\overline{2.97 \times 10^{-1}}$ & $\overline{2.92 \times 10^{-1}}$ & $\overline{2.95 \times 10^{-1}}$ \\
         & 570 & $\overline{2.75 \times 10^{-1}}$ & $\overline{2.71 \times 10^{-1}}$ & $\overline{2.67 \times 10^{-1}}$ & $\overline{2.66 \times 10^{-1}}$ & $\overline{2.66 \times 10^{-1}}$ \\
         & 650 & $\overline{2.48 \times 10^{-1}}$ & $\overline{2.45 \times 10^{-1}}$ & $\overline{2.42 \times 10^{-1}}$ & $\overline{2.41 \times 10^{-1}}$ & $\overline{2.41 \times 10^{-1}}$ \\
         & 735 & $\overline{2.25 \times 10^{-1}}$ & $\overline{2.22 \times 10^{-1}}$ & $\overline{2.18 \times 10^{-1}}$ & $\overline{2.17 \times 10^{-1}}$ & $\overline{2.17 \times 10^{-1}}$ \\
         & 796 & ${2.12 \times 10^{-1}}$ & ${2.10 \times 10^{-1}}$ & ${2.07 \times 10^{-1}}$ & ${2.06 \times 10^{-1}}$ & ${2.06 \times 10^{-1}}$ \\
    \end{tabular}
    \caption{Time averaged values of $E$ \eqref{eqn:meanNuE} computed for the configuration B (hot and cold walls have the same conductance ratio $C_w$). Values marked with an overline indicate cases in which $\Nu(t)$ fluctuates significantly in a fully developed flow. The signal at $\Ha=796$ is steady-state after an initial transient.}
    \label{tbl:E_tc}
\end{table*}

\begin{table}
    \centering
    \begin{tabular}{c|c|C{1cm}C{1cm}C{1cm}C{1cm}C{1cm}}
        \multicolumn{2}{c}{} & \multicolumn{5}{c}{$C_w$} \\ \cline{3-7}
        \multicolumn{2}{c}{} & 0 & 0.01 & 0.1 & 1 & 50 \\ \cline{3-7}
        \multirow{10}{*}{$\Ha$} & 0 & \multicolumn{5}{c}{turbulent (1)} \\
         & 85 & 1 & 1 & 1 & 1 & 2 \\
         & 162 & 3 & 3 & 3 & 3 & 3 \\
         & 240 & 3 & 3 & 2 & 2 & 2 \\
         & 325 & 2 & 2 & 2 & 2 & 2 \\
         & 400 & 2 & 1 & 2 & 2 & 2 \\
         & 487 & 1 & 1 & 1 & 1 & 1 \\
         & 570 & 1 & 1 & 1 & 1 & 1 \\
         & 650 & 1 & 1 & 1 & 1 & 1 \\
         & 735 & 1 & 1 & 1 & 1 & 1 \\
         & 796 & 1 & 1 & 1 & 1 & 1 \\
    \end{tabular}
    \caption{The number of major eddies in the large-scale circulation for the flows of configuration B.}
    \label{tbl:LSC_tc}
\end{table}

The structure of the mean flow in the case of configuration B is illustrated by the mid-plane distributions in Figs.~\ref{fig:cont_tc_vert} and \ref{fig:cont_tc_hor} and profiles of velocity and temperature in Figs.~\ref{fig:uz_tc_xy}-\ref{fig:t_tc}. We see that the effect of the Lorentz force on the flow structure is, in general, much weaker than in the case of configuration A. The width and maximum velocity of near-wall jets are practically unaffected by $C_w$ in the entire range $0\le C_w\le 50$ (only a slight decrease of jet velocity can identified). The effect of $\Ha$ at a given $C_w$ is noticeable, but much weaker than for flows with all walls conducting (cf. Figs.~\ref{fig:uz_tc_xy}, \ref{fig:uz_tc_z} and \ref{fig:uz_ac_xy}, \ref{fig:uz_ac_z}). Strong jets and large-scale circulation are observed even at the largest $C_w=50$, $\Ha=796$. Convective heat transfer associated with the strong circulation leads to mean temperature distributions varying only slightly with $C_w$ and $\Ha$ and remaining far different from the pure conduction profiles (see Fig.~\ref{fig:t_tc}). 

The profiles of $u_z$ in the right column of Fig.~\ref{fig:uz_tc_z} illustrate that the flow approaches a quasi-two-dimensional state at high $\Ha$. Curvature of the core flow profile is still visible, which, as before, can be attributed to large horizontal aspect ratio of the jet, but is much lower than in the case of configuration A. Increasing $C_w$ results in a lower curvature and a more uniform vertical velocity profile in the core flow. The temperature distributions are nearly two-dimensional at high $\Ha$. This indicates that a two-dimensional approximation can be used for the flow.

\begin{figure*}
\begin{overpic}[width=0.46\textwidth,clip=]{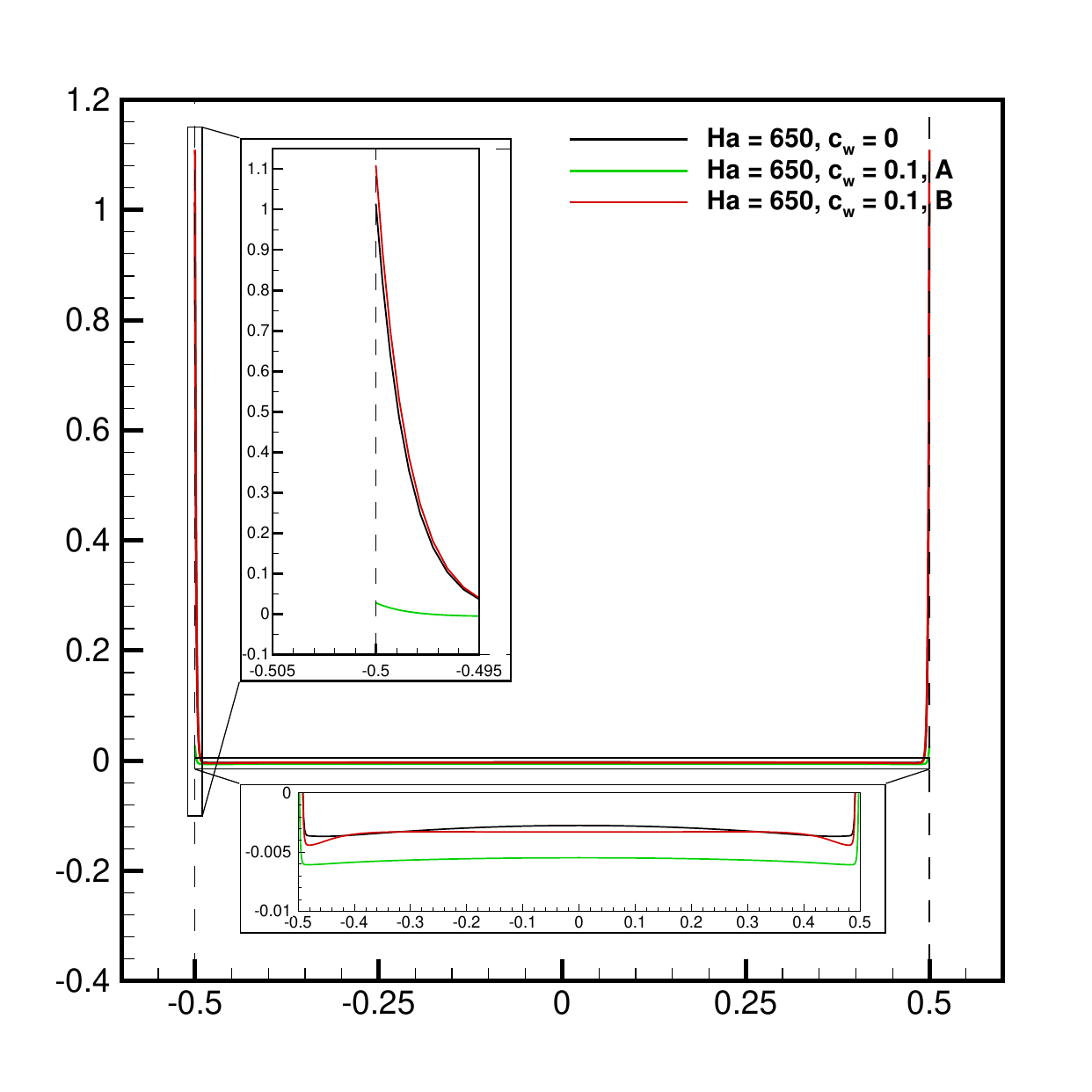}
    \titleput{2}{100}{(a)}
    \axisput{50}{2.5}{Y}
    \axisput{0}{50}{\rotatebox[origin=c]{90}{$j_x$ or $F_z$}}
\end{overpic}
\begin{overpic}[width=0.5175\textwidth,clip=]{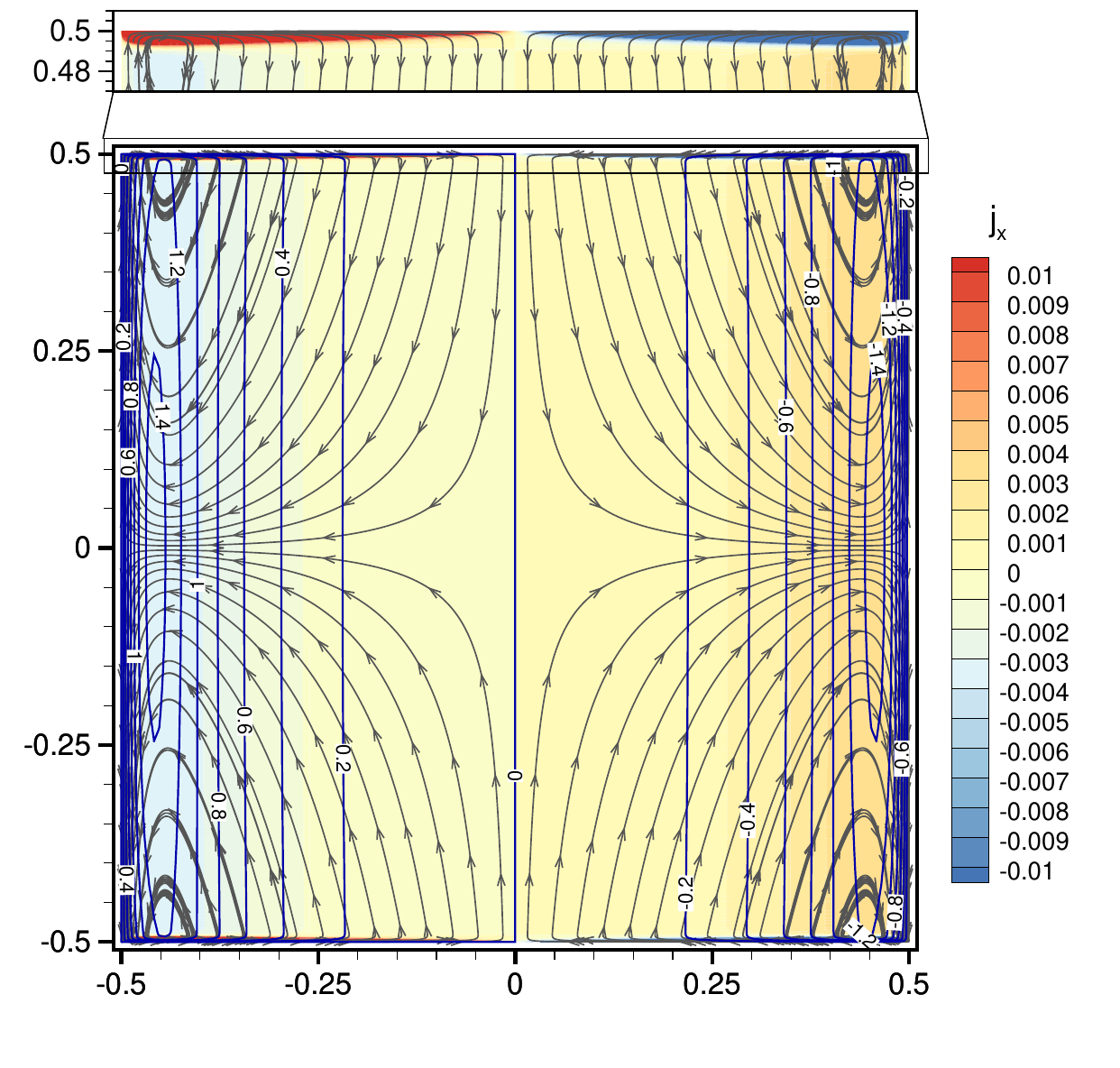}
    \titleput{0}{96}{(b)}
    \axisput{44.75}{2.25}{X}
    \axisput{2.5}{45}{\rotatebox[origin=c]{90}{Y}}
\end{overpic}
\vskip 1em
\begin{overpic}[width=0.46\textwidth,clip=]{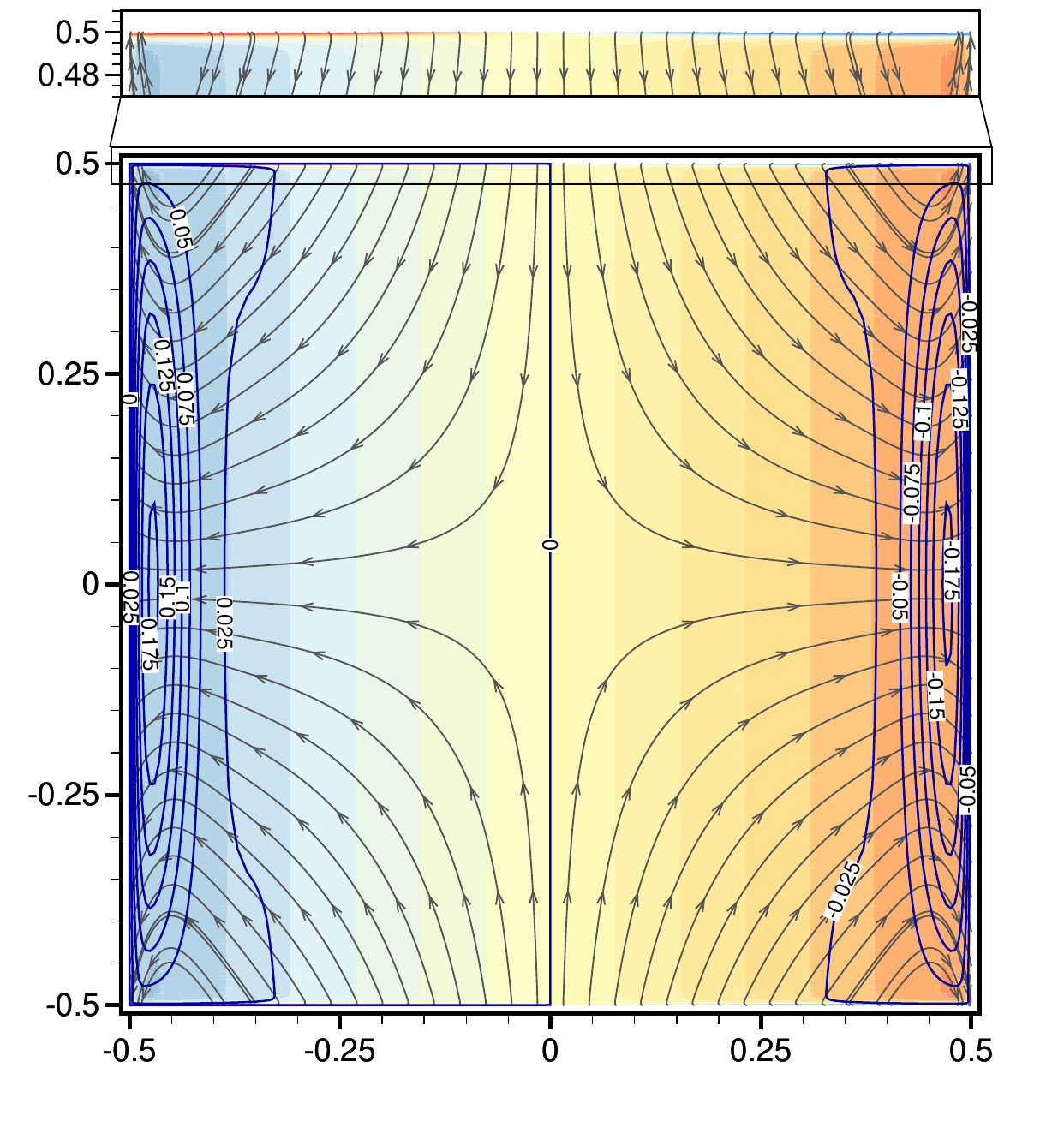}
    \titleput{0}{103}{(c)}
    \axisput{47}{2.25}{X}
    \axisput{3}{47.5}{\rotatebox[origin=c]{90}{Y}}
\end{overpic}
\begin{overpic}[width=0.5175\textwidth,clip=]{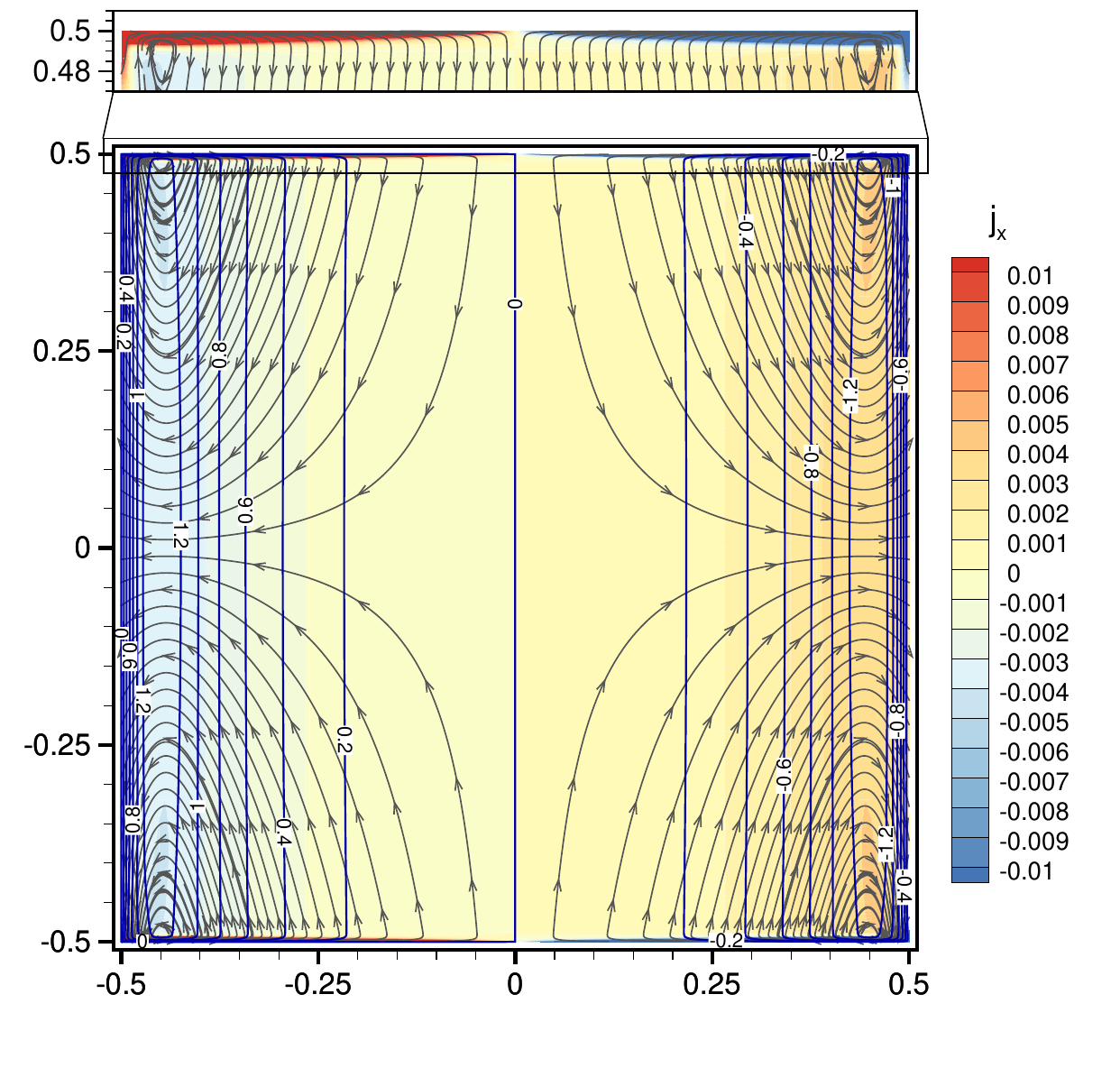}
    \titleput{1}{98}{(d)}
    \axisput{44.75}{2.25}{X}
    \axisput{2.5}{45}{\rotatebox[origin=c]{90}{Y}}
\end{overpic}
\caption{\emph{(a)} Timed-averaged profiles of $j_x=F_z$ along the line \ydataline~parallel to the magnetic field (see Fig.~\ref{fig:doamin}b). Results obtained at $Ha = 650$ and three wall configurations: $C_w = 0$, configuration A with $C_w = 0.1$, and configuration B with $C_w = 0.1$  are shown. The two zoom-ins show the distributions within the Hartmann layer at the wall $y =-0.5$ and in the core flow. \emph{(b-d)} Time-averaged distributions of the current paths (gray streamlines), the magnitude of $j_x=F_z$ (color), and the vertical velocity $v_z$ (blue isolines) in the horizontal mid-plane cross-section $z=0$ (see Fig.~\ref{fig:doamin}c). Results for the flows at $Ha = 650$ with $C_w=0$ \emph{(b)}, configuration A with $C_w = 0.1$ \emph{(c)} and configuration B with $C_w = 0.1$ \emph{(d)} are shown. }
\label{fig:cur_y}
\end{figure*}

\section{Discussion and concluding remarks}
\label{sec:conclusions}

A parametric study of the effect of electric conductivity of the walls on magnetoconvection in a tall box with vertical hot and cold walls is carried out. Two configurations of electric boundary conditions are explored. In configuration A, all walls of the box have the same finite electric conductance ratio $C_w$. In configuration B, only the hot and cold sidewalls are electrically conducting, with the same values of $C_w$. The other two sidewalls and the top and bottom walls are electrically insulating. Numerical simulations are completed at $\Pra=0.025$, $\Ray=7.5\times 10^5$, $0\le \Ha\le 796$, $0\le C_w \le 50$. The new results obtained for flows with conducting walls are compared with those for insulating walls and with the experimental data of Ref.~\onlinecite{Authie:2002} The focus of the analysis is on the effect of the magnetic field and wall conductivity on the structure of the mean flow and the integral properties, such as the Nusselt number $\Nu$ and the volume-averaged kinetic energy $E$.

The main finding of the study is the significant impact of the configuration of wall conductivities on the MHD transformation of the flow. When all walls are electrically conducting (configuration A), the transformation can be described as a straightforward suppression. Increasing the strength of the magnetic field leads to the damping of velocity fluctuations, reduction in the strength and simplification of the shape of large-scale circulation, and a decrease in $\Nu$ and $E$. The effect is stronger at higher $C_w$. At $C_w\ge 0.1$ and high $\Ha$, the heat transfer is reduced to nearly pure conduction.

The effect of the magnetic field on the flow is entirely different when only the sidewalls parallel to the magnetic field are electrically conducting (configuration B). The suppression of the flow is much weaker. The flow retains strong large-scale circulation and high values of $\Nu$ and  $E$ in the entire range $0\le \Ha\le 796$. Velocity fluctuations persist in fully developed flows, except in the case of the highest considered value $\Ha=796$. Interestingly, the variation of the electric conductivity of the hot and cold walls in the range between perfect insulation $C_w=0$ and nearly perfect conduction $C_w=50$ does not have a significant impact on flow properties. The few exceptions are examples of a subtle effect, where a change in $C_w$ alters the number of loops in the large-scale circulation. 

To explain the profound difference between the MHD responses of the two wall configurations, we consider the typical time-averaged distributions of induced electric currents and Lorentz forces. The focus is on the vertical component $F_z$ of the force, which directly affects near-wall jets. This component has the same value as the component $j_x$ of the current. The flows at $Ha=650$ are considered for three sets of wall conductivities: $C_w=0$; $C_w=0.1$ in configuration A; and $C_w=0.1$ in configuration B.

In Fig.~\ref{fig:cur_y}a, profiles of time-averaged $j_x$ are shown along the line $x=-0.47$, $z=0$, parallel to the magnetic field and located within the upward jet near the hot wall. The fields of time-averaged $j_x$, vertical velocity $v_z$, and current paths in the horizontal mid-plane cross-section $z=0$ are depicted for each of the three aforementioned cases in Figs.~\ref{fig:cur_y}b-d.

The explanation illustrated by Fig.~\ref{fig:cur_y} is based on the effect of the conductivity of the walls at $y=\pm 0.5$ perpendicular to the magnetic field on the distribution of the induced electric currents. If the walls are electrically insulating (as in the case of $C_w=0$ and in configuration B with $C_w=0.1$), thin Hartmann boundary layers form at them (see the insert at $y=-0.5$ in Fig.~\ref{fig:cur_y}a and Figs.~\ref{fig:cur_y}b,d). All current loops must close within these layers, implying strong currents and significant electric resistance to them. The cumulative result is the increase of the total electric resistance of the fluid domain to the induced electric currents. This does not occur in the case of configuration A, when the walls at $y=\pm 0.5$ are electrically conducting, so the current loops may close through them (see Fig.~\ref{fig:cur_y}c).

We observe in Fig.~\ref{fig:cur_y} that wall conductivity does not have a significant effect on the pattern of induced currents in the core flow, including the zones of the upward and downward jets. However, it does significantly modify the current strength. The insert in Fig.~\ref{fig:cur_y}a and the distributions of $j_x$ in Figs.~\ref{fig:cur_y}b-d show that the magnitude of $j_x$ and, consequently, $F_z$ in the jet's area is approximately twice as large in the case of configuration A compared to $C_w=0$ or configuration B. Since the component $F_z$ opposes the jet's vertical velocity, the damping of the jets by the Lorentz force is significantly stronger in the case of configuration B, i.e., when the walls perpendicular to the magnetic field are electrically conducting.

Two other conclusions can be drawn from the results presented in this paper. One concerns the applicability of the two-dimensional approximation, generally possible for high-$\Ha$ MHD flows, in which  velocity and temperature distributions are nearly uniform in the direction of the magnetic field within the core flow,  to magnetoconvection.  It has been long understood\cite{ZikanovASME:2021} that applicability of the approximation requires that the imposed temperature gradient is perpendicular to the magnetic field. Our results show that the accuracy of the approximation can be low even when this requirement is met. We see in Figs.~~\ref{fig:uz_ac_z} and ~\ref{fig:uz_tc_z} that vertical velocity $u_z$ varies significantly in the field direction in the core flow even at the highest $\Ha$ considered here.

Another comment concerns the disagreement between the data of the experiment\cite{Authie:2003} proposed as a benchmark for MHD numerical models\cite{Smolentsev:2015verification} and our computational data (see Figs.~\ref{fig:NU_E_ac}a and \ref{fig:NU_E_tc}a). A similar disagreement was observed in Ref.~\onlinecite{Authie:2003} and attributed in Ref.~\onlinecite{Smolentsev:2015verification} to either imperfection of the experiment or insufficient accuracy of the numerical model in Ref.~\onlinecite{Authie:2003}. The fact that a similar disagreement is found in our simulations on a much finer grid allows us to exclude the latter possibility. The data in Fig.~\ref{fig:NU_E_tc}a also show that the disagreement cannot be caused by imperfect electric insulation of the hot and cold walls in the experiment. The most plausible remaining explanation appears to be the difficulty of maintaining the condition of constant uniform temperature of hot and cold walls in the experimental setup.

\begin{acknowledgments}
The authors would like to thank the University Computing Center at the Technical University Ilmenau and the Advanced Research Computing at the University of Michigan for providing computing time on their HPC clusters.
\end{acknowledgments}

\section*{Author Declarations}
The authors have no conflicts to disclose.

\section*{Author Contributions}
\textbf{Ali Akhtari}: Conceptualization (equal); Investigation (equal); Methodology (equal); Software (equal); Visualization (equal); Writing – original draft (equal); Writing – review \& editing (equal).
\textbf{Oleg Zikanov}: Conceptualization (equal); Methodology (equal); Software (equal); Supervision (equal); Visualization (equal); Writing – review \& editing (equal).
\textbf{Dmitry Krasnov}: Conceptualization (equal); Methodology (equal); Software (equal); Supervision (equal); Writing – review \& editing (equal).

\section*{Data Availability}
The data that support the findings of this study are available from the corresponding author upon reasonable request.

\bibliography{mhd}

\end{document}